\documentclass{emulateapj}
\usepackage{multirow}
\usepackage{epsfig}
 \usepackage{url}
\usepackage{bmpsize}
\usepackage{amsmath}
\usepackage[toc,page]{appendix}
\usepackage[caption=false]{subfig}

\begin{document}

\title{
The Entire Virial Radius of the Fossil Cluster RXJ1159+5531: \\
I. Gas Properties
}

\author{Yuanyuan Su\altaffilmark{1$\dagger$}
, David Buote\altaffilmark{1},
 Fabio Gastaldello\altaffilmark{2} and Fabrizio Brighenti\altaffilmark{3}}

\affil{$^1$Department of Physics and Astronomy, University of 
California, Irvine, 4129 Frederick Reines Hall, Irvine, CA 92697, USA}

\affil{$^2$INAF-IASF-Milano, Via E. Bassini 15, I-20133 Milano, Italy}
\affil{$^3$Dipartimento di Fisica e Astronomia, Universita di Bologna, via Ranzani 1, 40126 Bologa, Italy}
\altaffiltext{$\dagger$}{Email: yuanyuas@uci.edu}

\begin{abstract}

Previous analysis of the fossil-group/cluster RXJ1159+5531 with X-ray observations from a central {\sl Chandra} pointing and an offset-North {\sl Suzaku} pointing indicate a radial intracluster medium (ICM) entropy profile at the virial radius ($R_{\rm vir}$) consistent with predictions from gravity-only cosmological simulations, in contrast to other cool-core clusters.  To examine the generality of these results, we present three new {\sl Suzaku} observations that, in conjunction with the North pointing, provide complete azimuthal coverage out to $R_{\rm vir}$. With two new {\sl Chandra} ACIS-I observations overlapping the North {\sl Suzaku} pointing, we have resolved $\gtrsim$50\% of the cosmic X-ray background there. We present radial profiles of the ICM density, temperature, entropy, and pressure obtained for each of the four directions. We measure only modest azimuthal scatter in the ICM properties at $R_{\rm 200}$ between the {\sl Suzaku} pointings: 7.6\% in temperature and 8.6\% in density, while the systematic errors can be significant. The temperature scatter, in particular, is lower than that studied at $R_{\rm 200}$ for a small number of other clusters observed with {\sl Suzaku}. 
These azimuthal measurements verify that RXJ1159+5531 is a regular, highly relaxed system. The well-behaved entropy profiles we have measured for RXJ1159+5531 disfavor the weakening of the accretion shock as an explanation of the entropy flattening found in other cool-core clusters but is consistent with other explanations such as gas clumping, electron-ion non-equilibrium, non-thermal pressure support, and cosmic ray acceleration. Finally, we mention that the large-scale galaxy density distribution of RXJ1159+5531 seems to have little impact on its gas properties near $R_{\rm vir}$.

\end{abstract}

\section{\bf Introduction}

\smallskip

Galaxy clusters represent the final stage of the hierarchical formation, and they are potent laboratories for testing models of structure formation.
They are especially valuable when considering global quantities computed within their virial radii ($R_{\rm vir}$), 
since the outskirts of galaxy clusters contain most of their baryons, dark matter, and metal content.
The entropy profile is a sensitive indicator of non-gravitational heating of the intracluster medium (ICM), since entropy is conserved in an adiabatic process. 
In studies of galaxy clusters, entropy is conventionally defined as $K(r)=T(r)/n(r)^{2/3}$ where $T$ and $n$ are gas temperature and density.  
The deviation of the entropy profile from that expected from pure gravitational collapse $K(r)\propto r^{1.1}$ (Voit et al.\ 2005) reflects the role of non-gravitational processes in cluster formation. 
Cluster outskirt near the virial radius are the front lines of cluster formation and are still growing. 
Primordial gas and sub-structures have been continuously accreted into the ICM along large filamentary structures, leaving a shock-heated region near the $R_{\rm vir}$.

To date, a dozen of massive galaxy clusters have been observed with {\sl Suzaku} out to $R_{\rm vir}$. These studies found that the gas properties near $R_{\rm vir}$ disagree with the predictions of gravity-only cosmological simulations (e.g. Su et al.\ 2013, Bautz et al.\ 2009; George et al.\ 2009; Simionescu et al.\ 2011; Hoshino et al.\ 2010; Walker et al.\ 2012a; Walker et al.\ 2013, see Reiprich et al.\ 2013 for a review). 
In particular, outside of $R_{500}$\footnote{$R_{\Delta}$ is the radius within which the average density is $\Delta$ times the cosmological critical density. $R_{200}\approx R_{\rm vir}$ and $R_{500}\approx 0.6\,R_{\rm vir}$. We adopt $R_{200}$ as $R_{\rm vir}$ the fiducial virial radius in our study for two reasons. First, $R_{200}$ is most commonly used as in X-ray cluster studies. Second, we are able to obtain interesting constraints on the ICM properties out to that radius with all four {\sl Suzaku} observations. Since $R_{108}$ was previously employed by Humphrey et al.\ (2012) to study the ICM properties of RXJ1159+5531, we also make some comparisons of the ICM properties at that radius.} the observed entropies are significantly less than the gravity-only prediction. This entropy deficit cannot be explained by feedback, the effect of which is expected to be small at such large radii, and should instead produce an excess of entropy over that produced by only gravitational evolution.

In the case of the Perseus Cluster, Simionescu et al.\ (2011) also measured an enclosed gas fraction within $R_{200}$ that exceeds the cosmic value by 50\%. 
Simionescu et al.\ (2011), in particular, advocate a clumpy ICM to reconcile the observations with structure formation models, since clumped gas emit more efficiently than uniformly distributed gas.
The clumping factor, defined as 
$$C=\frac{{\langle n_{\rm e}}^2\rangle}{{\langle n_{\rm e}\rangle}^2},$$ 
with $C\geq 1$, is used to describe the deviation from uniformly distributed gas.
In the case of the Perseus Cluster, a clumping factor of $\sim$16 is required at $R_{200}$. C$\sim$7 and $\sim$9 are required for Abell~1835 (Bonamente et al.\ 2013) and PKS~0745-191 (Walker et al.\ 2012b) respectively to reconcile the observed entropy profile and the expected $r^{1.1}$ power-law model from gravity-only simulations. In contrast, significantly smaller clumping factors are predicted for simulated clusters. For example, Nagai \& Lau (2011) obtained $C=$1.3--2 at the $R_{\rm vir}$. Similarly, Vazza et al.\ (2013) and Zhuravleva et al.\ (2013) report $C\leq$3 in the cluster outskirts. 

Apparently, other factors besides gas clumping need to be taken into account to explain the observed entropy profiles. 
Hoshino et al.\ (2010) and Akamatsu et al.\ (2011) attribute the entropy flattening
to electrons and ions being out of thermal equilibrium due to recent accretion shocks (Hoshino et al.\ 2010, Akamatsu et al.\ 2011); however, some simulations suggest such effects should not be significant (Wong \& Sarazin 2009). 
Femiano \& Lapi (2014) proposed that the rapid radial decrement of the temperature caused by non-gravitational effects is responsible for the observed entropy flattening. Fujita et al.\ (2013) presented a scenario that cosmic ray acceleration could consume the kinetic energy of infalling gas and affect the entropy profile in the cluster outskirts. 
Another explanation is provided by 
Cavaliere et al.\ (2011), who propose a cluster evolutionary model incorporating the effects of merger shocks weakening over time (towards low redshift); Walker et al.\ (2012a) finds these models are consistent with the observed entropy flattening, although their model contains several free parameters. All the above explanations predict that the entropy profile depends on the relaxation state and mass of the cluster.
Most studies have focused on massive, cool-core clusters. 
It is important to extend these studies to lower mass clusters.

Indeed, an intriguing example of such a system is the poor-cluster/fossil-group RXJ1159+5531 (Humphrey et al.\ 2012).
We obtained good constraints on its gas properties all the way out to $R_{\rm 108}$, finding that at its virial entropy profile agrees with the prediction from gravity-only simulations and
its baryon fraction within $R_{\rm 108}$ is fully consistent with the cosmic value (0.15 Planck Collaboration 2013; 0.17 Komatsu et al.\ 2011). 
Fossil groups are empirically defined as systems with 1) a central
dominant galaxy more than two optical magnitudes brighter than the second
brightest galaxy within half a virial radius; and 2) an extended thermal
X-ray halo with $L_{\rm (X,bol)}$ $>$ $10^{42}$ $h_{50}^{-2}$ erg s$^{-1}$ (Jones et al. 2003).
Perhaps as a fossil-group, RXJ1159+5531 may be sufficiently evolved and relaxed (e.g. Harrison et al.\ 2012) that hydrostatic equilibrium is  an accurate approximation, and there is little clumping. 
We note that the entropy profile of ESO3060170, the only other fossil group that has been observed with {\sl Suzaku} out to $R_{\rm vir}$, does flatten, but its deviation occurs at a much larger radius compared with other systems, and the value of the entropy at $R_{200}$ is still consistent with the predictions from gravity-only simulations (Su et al.\ 2013).

Recent {\sl Suzaku} observations of cluster outskirts indicate that azimuthal asymmetries are common, even in many clusters that appear symmetric at small scales (e.g., Eckert et al.\ 2013; Urban et al.\ 2014). 
If RXJ1159+5531 is a highly evolved and relaxed system, we would, in contrast, expect to see little azimuthal variation in these properties as predicted in simulations (Vazza et al.\ 2011).
The {\sl Suzaku} observation offset to the North reported in Humphrey et al.\ (2012) only provides $\sim$ 27\% azimuthal coverage beyond $R_{500}$.   
To achieve a complete azimuthal coverage of this valuable system, we acquired deep {\sl Suzaku} observations in the other three directions.  
Together, these observations allow the entire virial radius to be studied.

In the cluster outskirts, where the X-ray emission is background dominated, the low and stable instrumental background of {\sl Suzaku}
is crucial for constraining the properties of the hot gas. 
However,
the observed flat entropy profile if, due to gas clumping, would result from substructures unresolved by the
point-spread function (PSF) of {\sl Suzaku}. 
As an independent study, Eckert et al.\ (2013) combined {\sl Planck} pressure profile and {\sl ROSAT} density profile and obtained entropy profiles of cool-core clusters in line with the baseline entropy profile.
Moreover, in this regime, the cosmic X-ray background (CXB) dominates the cluster outskirts, in particular for energies above about 2\,keV. 
An accurate characterization of the background is therefore a prerequisite for a reliable measurement of the gas properties. 

The {\sl Chandra} X-ray Observatory has superb spatial resolution (0.5$^{\prime\prime}$),
which makes it ideal to address these issues and to complement the {\sl Suzaku} data. 
In addition, the large field-of-view of the {\sl Chandra} ACIS-I combined with a low background (compared to ACIS-S) make it very suitable to investigate the cluster outskirts and provide an independent test of the {\sl Suzaku} results.  
A growing number of {\sl Suzaku} observations have been awarded {\sl Chandra} follow up observations and have been used to refine the analysis of {\sl Suzaku} data. 
Miller et al.\ (2012) demonstrated that a short snapshot {\sl Chandra} observation is able to reduce the uncertainties in the surface brightness measured by {\sl Suzaku} by 50\% at the outskirts of clusters.
We acquired deep ACIS-I observations, mosaicing the portion of the $R_{500}$--$R_{\rm vir}$ region in the north direction that was observed and reported in Humphrey et al.\ (2012). These deep exposures allow us to 
obtain a better characterization of the CXB and allow us to
reduce the uncertainties of the gas properties in the spectroscopic analysis directly.

Adopting a redshift of $z=0.081$ from the NASA/IPAC Extragalactic Database (NED),
we derive a luminosity distance of 368 Mpc (so $1^{\prime} = 90$ kpc), 
assuming a  cosmology with $H_0=70$ km s$^{-1}$ Mpc$^{-1}$,
$\Omega_{\Lambda}=0.7$ and  $\Omega_m=0.3$.
We studied this system out to $R_{108}=12^\prime$ (1100\,kpc) in the north, south, west, and east directions with {\sl Suzaku} ($R_{500}=580$\,kpc and $R_{200}=871$\,kpc). We determined
its gas properties at large radii in all directions, which we present in this paper. 
We will present the hydrostatic constraints of its total mass distribution and dark matter properties in paper II and its metallicity distributions in paper III. 
We describe the observations and data reduction in \S2 and introduce our mass modeling techniques in \S3.
We report results in \S4 and construct detailed systematic error budgets in \S5.
The implications of our results are discussed in \S6, and our main conclusions are summarized in \S7. Uncertainties reported in this paper are at 1$\sigma$ confidence level unless stated otherwise.

\medskip

\section{\bf Observations and data reduction} 

\begin{table*}
\caption{Observation log}
\begin{center}
\begin{tabular}{lcccccc}\hline \hline 
\colhead{Name}&\colhead{Obs ID}&\colhead{Obs Date}&\colhead{Exposure}&\colhead{R.A.}&\colhead{Dec.}&\colhead{Dectectors}\\
\colhead{}&\colhead{}&\colhead{}&\colhead{(ksec)}&\colhead{}&\colhead{}&\colhead{}\\ \hline
{\sl Suzaku} N&804051010 &2009-05-02&84&11 59 48.72 & 55 36 39.6 & XIS0,1,3\\
\multirow{2}{*}{{\sl Suzaku} S}&807064010&2012-05-27& 81&11 59 51.29&55 24 44.3&XIS0,1,3\\
&807064020&2012-12-18& 21&11 59 55.49 &55 26 24.7&XIS0,1,3\\
{\sl Suzaku} E&809063010&2014-05-29& 96& 12 00 46.70&55 31 16.0&XIS0,1,3 \\
{\sl Suzaku} W&809064010 &2014-05-31&94&11 58 59.83& 55 32 39.1 & XIS0,1,3\\
{\sl Chandra} Cen & 4964 &2004-02-11&76&11 59 51.40 & 55 32 01.0 & ACIS-S \\
{\sl Chandra} NW & 14026 &2012-08-09&50&12 00 41.60& 55 39 55.1 & ACIS-I \\
\multirow{2}{*}{{\sl Chandra} NE} & 14473 &2012-08-12&37&\multirow{2}{*}{11 59 29.10} & \multirow{2}{*}{55 42 13.7} & ACIS-I \\
& 14027 &2012-08-09&13& && ACIS-I \\ \hline
    \end{tabular}
  \end{center}
\end{table*}

 \begin{figure}
 \epsscale{1.2}
        \centering
        \includegraphics[width=0.45\textwidth,natwidth=640,natheight=642]{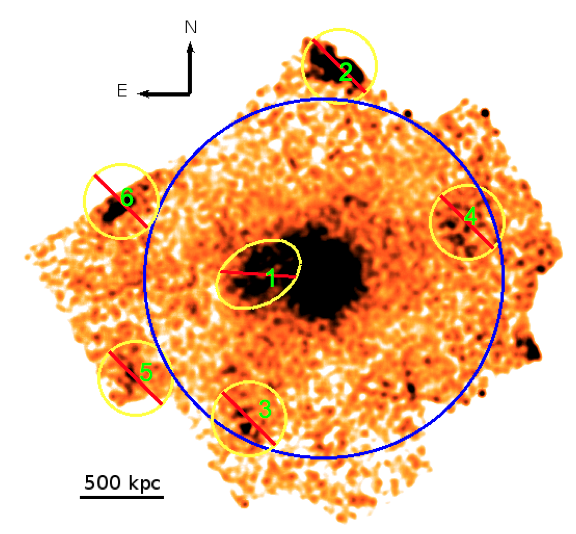}
\figcaption{\label{fig:suzaku}Mosaic of the {\sl Suzaku} XIS1 images of RXJ1159+5531 in the 0.5--4.0 keV energy range with NXB subtracted. The image is corrected for exposure and vignetting. A region of enhanced brightness to the east of the center that contains several bright point sources when compared to the {\sl Chandra} image. Point sources marked in yellow circles were excluded. The blue circle indicates $R_{108}$.}
\end{figure}

\begin{figure}
 \epsscale{1.2}
        \centering
        \plotone{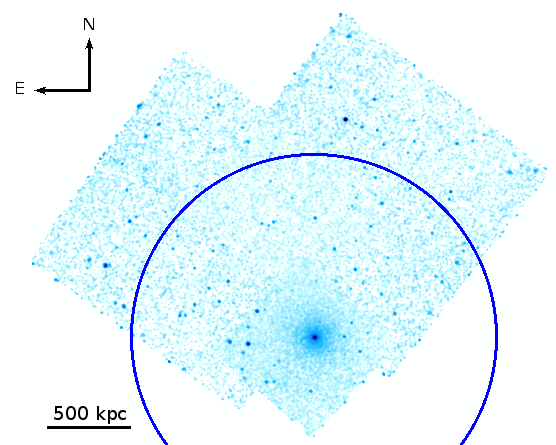}
\figcaption{\label{fig:chandra}Mosaic of the {\sl Chandra} images of RXJ1159+5531 in 0.3-7.0 keV band corrected for exposure variations, containing one central ACIS-S pointing and two ACIS-I offset pointings to the north direction. The blue circle indicates $R_{108}$.}
\end{figure}

RXJ1159+5531 has been observed with {\sl Suzaku} from the center and out to the $R_{\rm vir}$ to north, south, west, and east directions. A mosaic of {\sl Suzaku} pointings is shown in Figure~\ref{fig:suzaku}. 
RXJ1159+5531 has been observed with {\sl Chandra} with one ACIS-S pointing at the center and two ACIS-I offset pointings covering the entire field-of-view of the {\sl Suzaku} north pointing\footnote{There is another 19 ksec ACIS-I pointing at the center of RXJ1159+5531. We did not include it in our analysis in order to to simplify the analysis (in particular, the
background modeling) and avoid additional systematic uncertainty.}. A mosaic of {\sl Chandra} pointings is shown in Figure~\ref{fig:chandra}.
The result of the {\sl Suzaku} analysis of the north pointing combined with the {\sl Chandra} analysis of the central region has been presented in Humphrey et al.\ (2012).  
We processed (and also reprocessed) all these {\sl Suzaku} and {\sl Chandra} observations to guarantee the latest calibrations and consistent reduction process. The observation logs are listed in Table~1. 
Basically, gas properties beyond 0.5 $R_{\rm vir}$ of the north direction are obtained with a joint {\sl Suzaku} and {\sl Chandra} analysis; that of the other three directions were obtained with {\sl Suzaku} observations only.

\medskip

\subsection{\sl Chandra}

We used the CIAO 4.6 and {\sl Heasoft} 6.15 software, and the {\sl Chandra} calibration database ({\sl Caldb}) 4.6.3, to reduce the data.
All data were reprocessed from level 1 events following the standard data reduction threads\footnote{http://cxc.harvard.edu/ciao/threads/index.html}. 
Light curves were extracted from a low surface brightness region of the CCDs using the CIAO script {\tt lc\_clean}. 
High background intervals exceeding 2$\sigma$ above the mean quiescent background rates were excised. Effective exposure times are listed in Table~1. 

Point sources were detected in a 0.3--7.0 keV image with {\tt wavdetect}, supplied with a 1.7 keV exposure map to minimize spurious detections at chip boundaries. The detection threshold was set to 10$^{-6}$ and the scales of {\tt wavdetect} were set to a $\sqrt{2}$ series from 1 to 8. All detected point sources were inspected by eye, and corresponding elliptical regions containing 99\% of the source photons were generated. 
We obtained the 2.0-8.0 keV count rates for each resolved point source. An elliptical annulus centered on each point source was used for ``local" background subtraction, which is just outside the source-extraction region. We further converted the net count rate into flux for each point source assuming a {\tt powerlaw} spectrum with an index of 1.41 (De Luca \& Molendi 2004).

We extracted spectra from seven continuous annular regions from the central Chandra observation on the ACIS-S3 chip, 0--7$^{\prime\prime}$, 7-15$^{\prime\prime}$, 15--25$^{\prime\prime}$, 25--42$^{\prime\prime}$, 42--78$^{\prime\prime}$, 78--138$^{\prime\prime}$, 138--208$^{\prime\prime}$, and 208--300$^{\prime\prime}$.
The width of each annulus was chosen to contain sufficient photons for spectral analysis with approximately the same number ({\bf $\gtrsim2000$}) of background-subtracted counts.  Spectral response matrices were produced for each annulus with the CIAO tool {\tt mkwarf} and {\tt mkacisrmf}. Spectral fitting was performed with {\sl XSPEC} 12.7 using the C-statistic. We rebinned the spectra to ensure at least 20 photons per bin to aid in model selection and computational speed. The energy range for spectral fitting was restricted to 0.5--7.0 keV. 
All spectra were fitted simultaneously including components for the cluster emission and background.
We used a single thermal {\tt vapec} component to model the ICM emission. We included an additional 7.3 keV bremsstrauhlung component to account for the emission from unresolved low-mass X-ray binaries (LMXBs) in the central galaxy (Irwin et al.\ 2003). 
Since the number of LMXBs scales with the stellar light, the relative normalization of this component between each annulus was fixed to match the relative $K$-band luminosity in the associated regions. 
 
To account for the X-ray background, we employed a multi-component background model 
consisting of an {\tt apec} thermal emission model for the Local Bubble 
({\tt apec$_{\rm LB}$},  kT=0.08 keV, solar abundances), 
an additional {\tt apec} thermal emission model ({\tt apec$_{\rm MW}$}, kT=0.2 keV, solar abundances) for the Milky Way emission  
in the line of sight  (Smith et al.\ 2001), 
and a power law model {\tt pow$_{\rm CXB}$} (with index $\Gamma=1.41$) characterizing the unresolved CXB (De Luca \& Molendi 2004).
All these components but the Local Bubble were assumed to be absorbed
by foreground (Galactic) cooler gas, with the absorption characterized by 
the {\tt phabs} model for photoelectric absorption.  
Photoionization cross-sections were from Balucinska-Church \& McCammon (1992). 
We adopted a Galactic 
hydrogen column of $N_H=1.02\times10^{20}$ cm$^{-2}$ toward RXJ1159+5531, deduced 
from the Dickey and Lockman (1990) map incorporated in the {\sl HEASARC}
$N_H$ tool.
The ICM components were allowed to vary independently for each radial annulus. 
The normalization of each background component within each annulus was linked to scale with the extraction area, but the total normalizations were allowed to vary freely. Best-fit X-ray background components are listed in Table~2.
To accommodate the particle background. we included a number of Gaussian lines and a broken power-law model (see Humphrey et al.\ 2012 for detail), which were not folded through the ancillary response file (ARF). The normalization and shape of the instrumental components were allowed to vary freely. 

 \begin{table*}
\caption{X-ray Background Components$^{\ast}$}
  \begin{center}
    \leavevmode
 \begin{tabular}{lccc} \hline \hline 
 \colhead{}&\colhead{CXB$^{\S}$}&\colhead{Local Bubble$^{\dagger}$}&\colhead{Milky Way$^{\ddagger}$}\\ 
  \hline
Chandra-Center&$3.9^{+1.0}_{-0.4}$&$11.4^{+2.6}_{-1.9}$&$8.9^{+1.0}_{-2.7}$\\
Suzaku--North&${3.6^{+0.2}_{-0.3}}$&${16.7^{+2.5}_{-2.3}}$&${7.7^{+1.4}_{-1.4}}$\\
Suzaku--South&${8.4^{+0.3}_{-0.3}}$&${44.8^{+3.2}_{-2.1}}$&${10.5^{+1.6}_{-1.9}}$\\
Suzaku--East&${9.3^{+0.2}_{-0.3}}$&${62.2^{+3.7}_{-3.7}}$&${12.9^{+2.1}_{-1.9}}$\\
Suzaku--West&${7.9^{+0.3}_{-0.4}}$&${58.4^{+3.5}_{-2.9}}$&${12.1^{+1.8}_{-1.8}}$\\
\hline
    \end{tabular}
  \end{center}
$^{\ast}$: Results for the normalizations of the different components of the (non-instrumental) X-ray background for each observation.\\
$^{\S}$: Normalization of a power-law component with fixed slope ($\Gamma$=1.4) divided by the solid angle, in the unit of 
photons\,s$^{-1}$cm$^{-2}$keV$^{-1}$str$^{-1}$ at 1keV. Note that the result for the {\sl Suzaku} North direction is the remaining unresolved flux after accounting for the sources resolved by Chandra.\\
$^{\dagger}$: An unabsorbed {\tt apec} thermal component (kT=0.08 keV, solar abundances) with normalization expressed as an emission measure integrated over the line of sight, $\frac{1}{4\pi[D_{\rm A}(1+z)]^2}\int n_{\rm e}n_{\rm H}dV/d\Omega$ in the unit of $10^{-14}$ cm$^{-5}$str$^{-1}$, where $d\Omega$ is the solid angle.\\
$^{\ddagger}$: An absorbed {\tt apec} thermal component (kT=0.2 keV, solar abundances) with normalization expressed as an emission measure integrated over the line of sight, $\frac{1}{4\pi[D_{\rm A}(1+z)]^2}\int n_{\rm e}n_{\rm H}dV/d\Omega$ in the unit of $10^{-14}$ cm$^{-5}$str$^{-1}$, where $d\Omega$ is the solid angle.\\
\end{table*}

\subsection{\sl Suzaku}

{\sl Suzaku} data reduction and analysis were performed with the {\sl Heasoft} 6.15 software package 
using {\sl CalDB}20141001. 
Data were obtained in both $3\times3$ and $5\times5$ data readout modes;
$5\times5$  mode data were converted to $3\times3$ mode and merged with the $3\times3$ mode data. 
The events were filtered by retaining those with a geomagnetic cutoff rigidity $>6$ GeV/c, 
and an Earth elevation $>10^\circ$.
The calibration source regions and hot pixels were excluded. 
Light curves were filtered using {\sl CIAO}4.6 script {\tt lc\_clean}.
No anomalous event rates deviate more than 3$\sigma$ from the mean was found. 
The effective exposure time of each pointing is listed in Table~1.
Bright point sources were identified by eye and confirmed by {\sl Chandra} imaging whenever possible. We give the number of each point source as labeled in Figure~\ref{fig:suzaku}. We excluded a circular region of 2.5$^{\prime}$ radius centered on each source (we excluded a elliptical region of similar size for source $\#$1).

 \begin{figure*}
   \begin{center}
     \leavevmode 
\hspace{5mm}\subfloat{\includegraphics[width=0.48\textwidth]{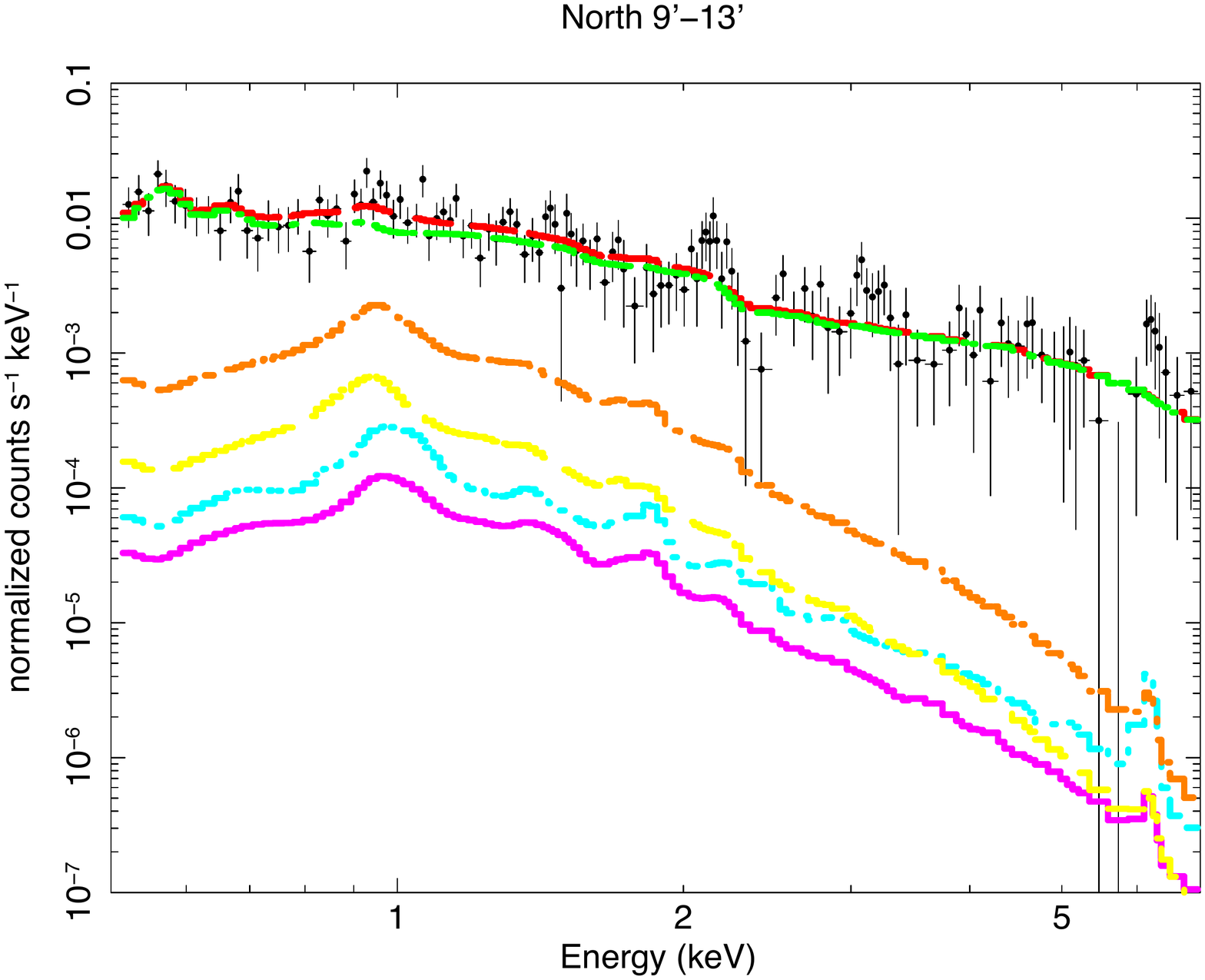}}\hspace{0mm}\subfloat{\includegraphics[width=0.48\textwidth]{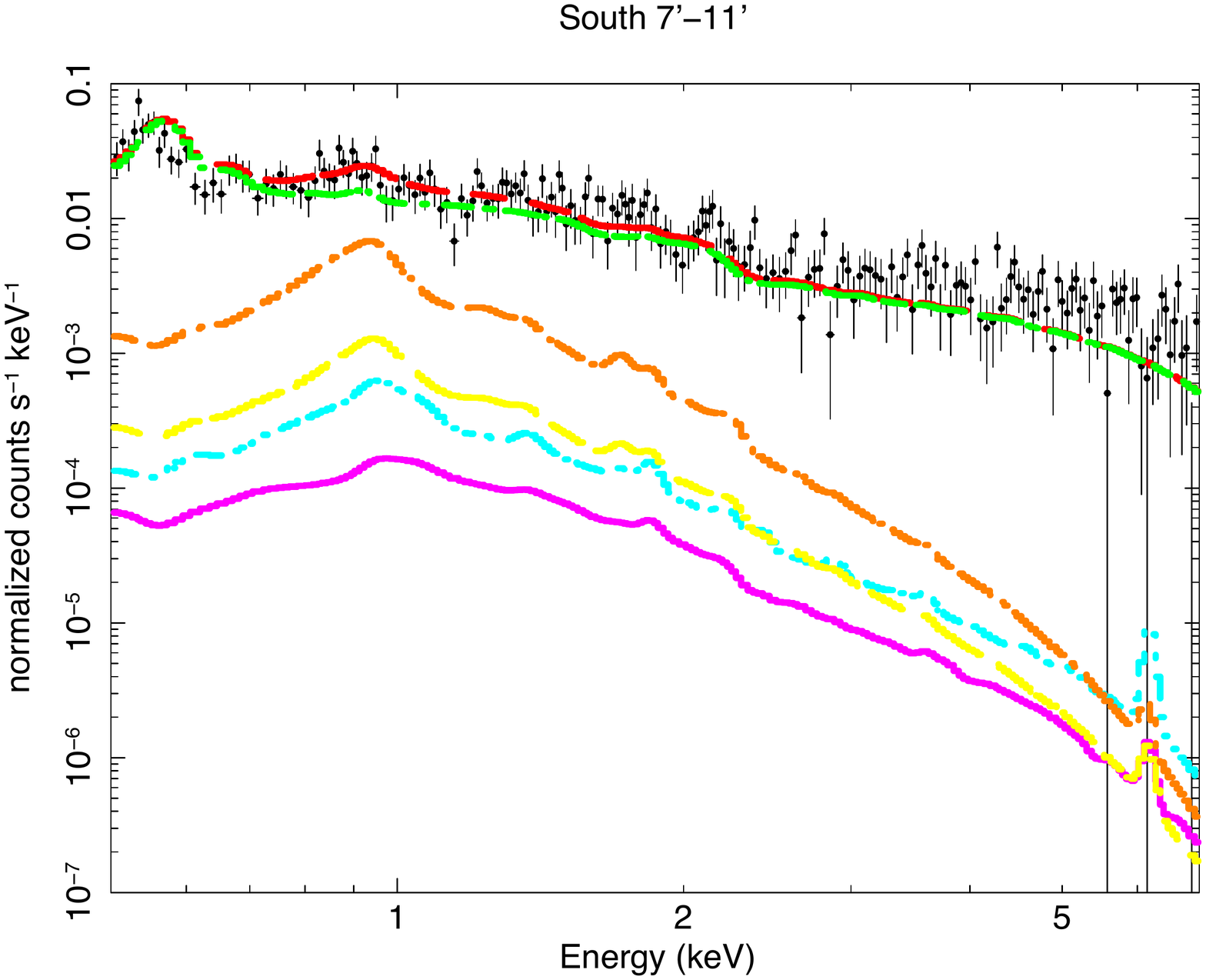}}\\
\vspace{-5mm}
\hspace{5mm}\subfloat{\includegraphics[width=0.48\textwidth]{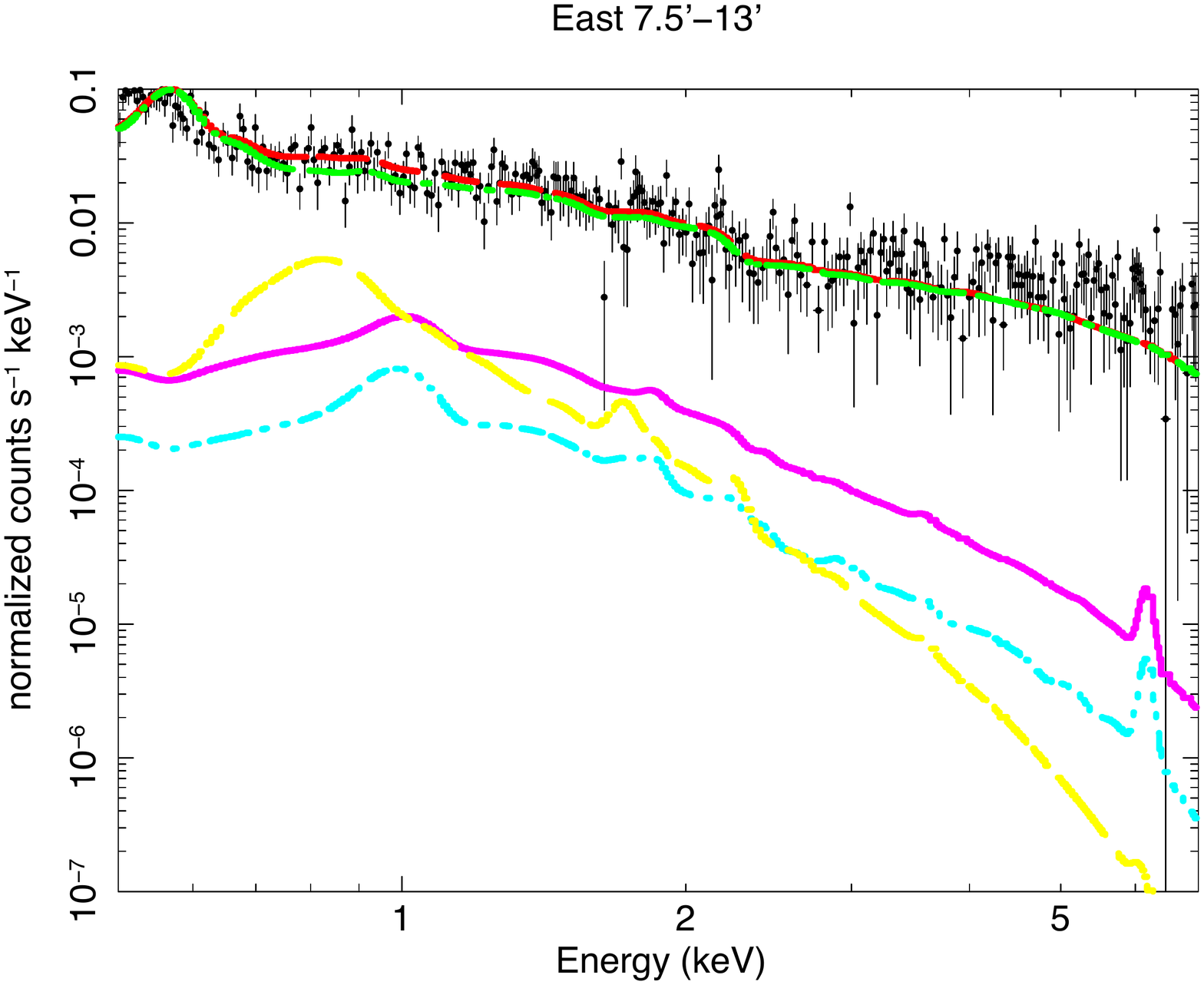}}\hspace{0mm}\subfloat{\includegraphics[width=0.48\textwidth]{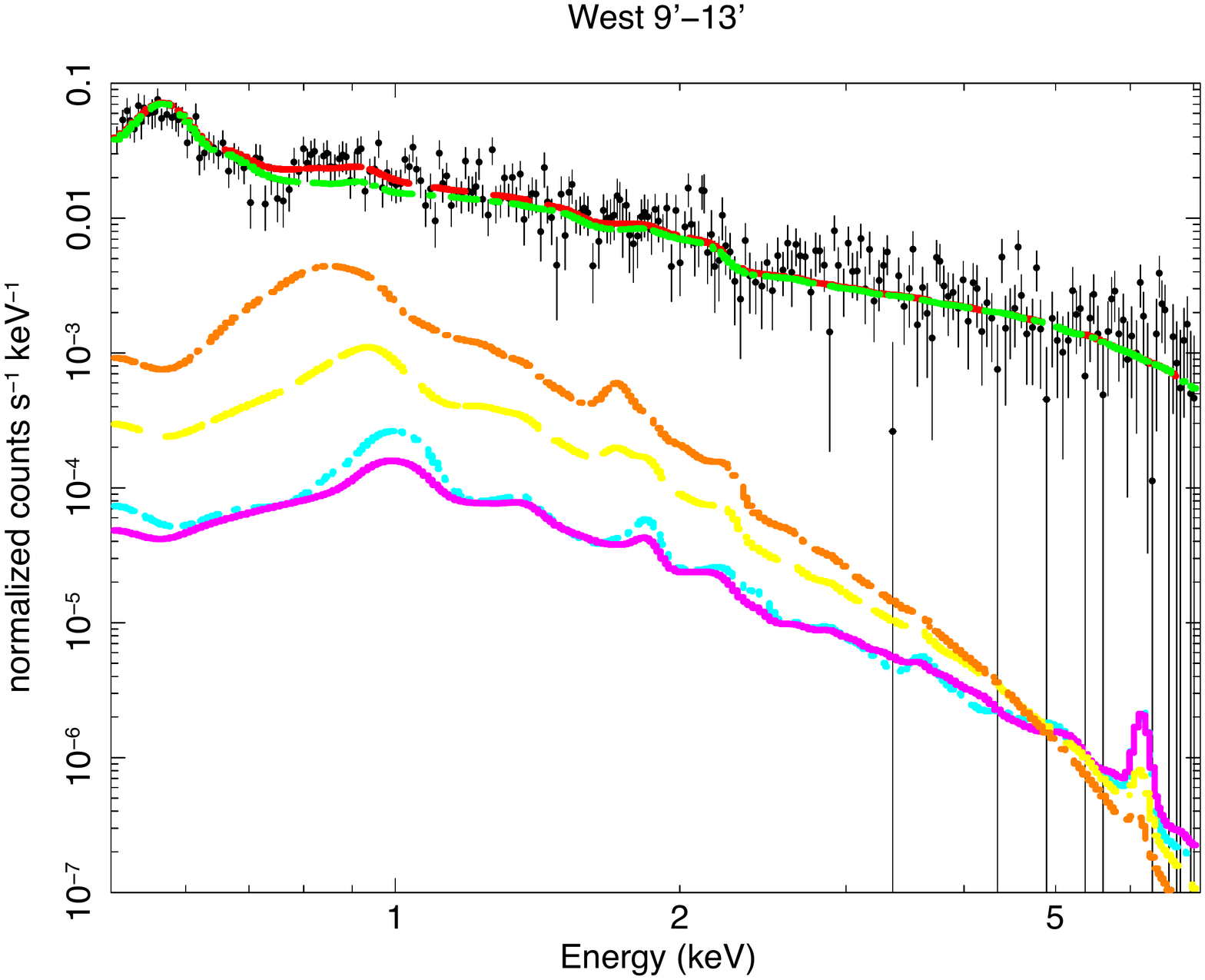}}\\
\caption{{\sl Suzaku} XIS1 spectra for RXJ1159+5531 from the outermost bin of the north, south, west, and east directions. Instrumental background has been subtracted. 
Black: observed spectra. 
Red: best-fit model. 
Green: the X-ray background. 
Magenta, Blue, Yellow, Orange: hot gas emission from 1st, 2nd, 3rd, and 4th radial annulus, respectively. 
[{\sl see the electronic edition of the journal for a color version of this figure.}]} 
\label{fig:spectra} 
\end{center}
\end{figure*}

We extracted spectra from four annular regions centered on RXJ1159+5531 and extending out to $R_{\rm vir}$:
0$^{\prime}$--2$^{\prime}$, 2$^{\prime}$--5$^{\prime}$, 5$^{\prime}$--9$^{\prime}$, and 9$^{\prime}$--13$^{\prime}$ from the {\sl Suzaku} observation in our previous analysis of the North pointing (Humphrey et al.\ 2012). In this work, we employed the same radii for the North and West pointings. However, we were unable to obtain reliable constraints on the ICM properties using the same radii for the South and East pointings due to their farther offset and point source contaminations. For the South pointing, we used spectra extracted from 0$^{\prime}$--2$^{\prime}$, 2$^{\prime}$--4$^{\prime}$, 4$^{\prime}$--7$^{\prime}$, and 7$^{\prime}$--11$^{\prime}$ annulus regions to model the ICM properties (and we used its 11$^{\prime}$--15$^{\prime}$ annulus spectrum to assist in constraining the background model for this pointing). For the East pointing, we find it is necessary to use a smaller number of radius sets of larger extraction regions:  0$^{\prime}$--2.5$^{\prime}$, 2.5$^{\prime}$--7$^{\prime}$, and 7$^{\prime}$--13$^{\prime}$ to model the ICM. 
The FTOOL {\tt xissimarfgen} was used to generate an ARF for each region and detector. To provide the appropriate photon weighting
for each ARF, we used a $\beta$-model surface brightness distribution
determined by fitting the surface brightness profile from the central {\sl Chandra} observations.
To model the X-ray background, we followed the standard procedure and generated
ARFs assuming uniform sky emission with a radius of $20^\prime$.
Redistribution matrix files 
(RMFs) were generated for each region and detector using the FTOOL {\tt xisrmfgen}. 
Non-X-ray background (NXB) spectra were generated with the FTOOL {\tt xisnxbgen}. 
Spectra from XIS0, 1, \& 3 were simultaneously fitted with {\sl XSPEC} v12.7.2 (Arnaud 1996). 
We adopted the solar abundance standard of Asplund at al.\ (2006) in thermal spectral models. 
Energy bands were restricted to 0.5-7.0 keV for the back-illuminated CCD (XIS1) and 0.6-7.0 keV for the front-illuminated CCDs (XIS0, XIS3), 
where the responses are best calibrated (Mitsuda et al.\ 2007). 
In order to account for spectral mixing between each annulus, 
we produced ``crossing" ARFs between each radial annulus using
the algorithm described in Humphrey et al.\ (2011).

We used {\sl XSPEC} to fit each spectrum with a multi-component background plus source model (as for the {\sl Chandra} data).
Since the {\sl Chandra} analysis reveals a significant cool core, we included a second {\tt apec} component to better model the ICM emission in the central bin. 
We were unable to constrain the ICM metal abundance of the outer most bin in each direction separately; we linked the metal abundance of the outer most bin to the adjacent annulus interior to it for each direction.
To account for the discrepancies in the responses between the front-illuminated CCDs and back-illuminated CCDs, 
the normalizations of the model for BI-chip (XIS1) and FI-chips (XIS0, XIS3) were allowed to vary independently; the discrepancy of their normalizations is $\sim$10\%.  
In Figure~\ref{fig:spectra}, we show the XIS1 spectra of the outmost annulus in each direction and their individual components. 


\subsection{\sl Refine Suzaku analysis with point sources resolved by Chandra}

 \begin{figure*}
   \begin{center}
     \leavevmode 
  \plottwo{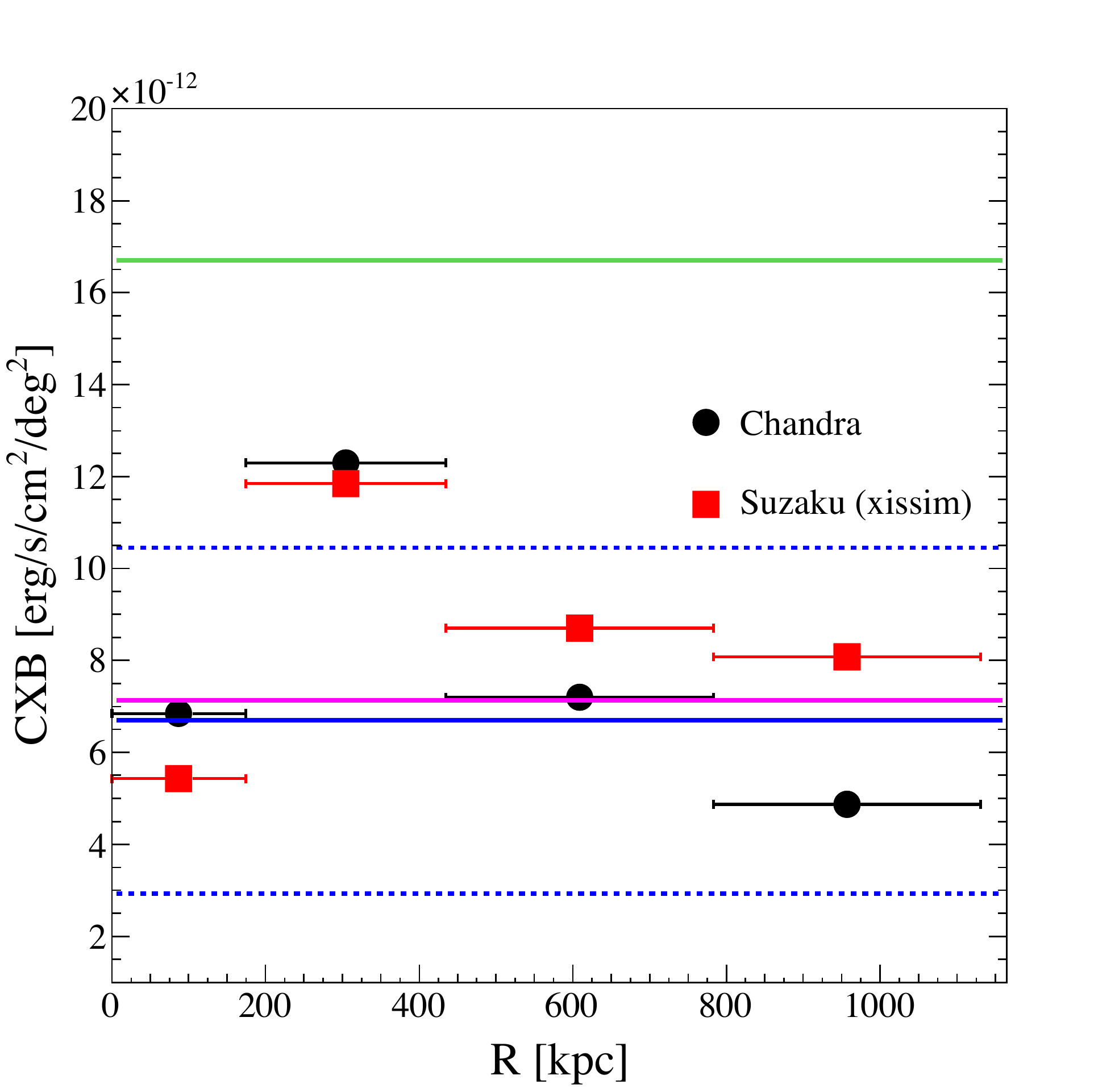}{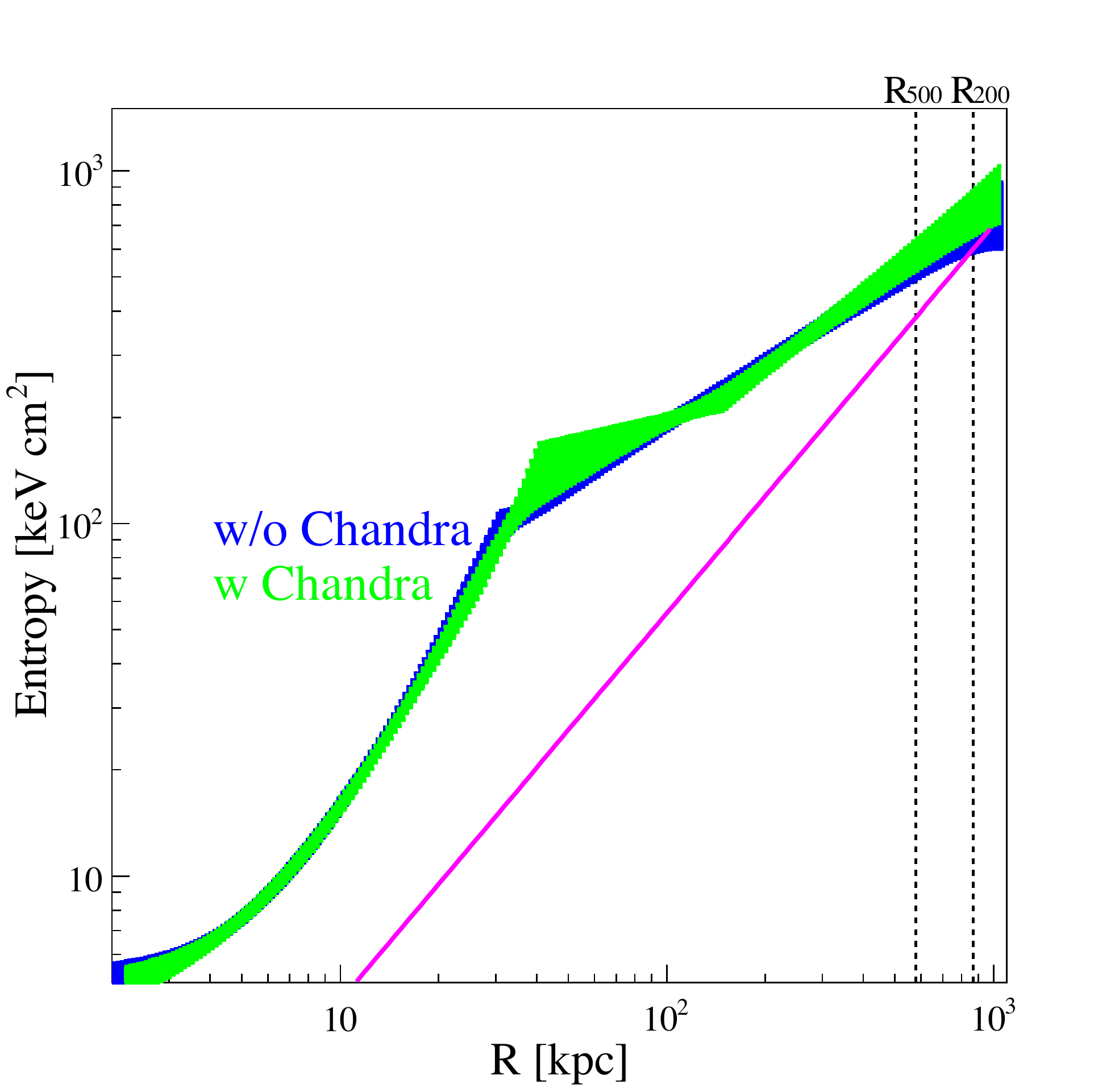}
\caption{{\it left}: black: surface brightness profile of point sources resolved by {\sl Chandra}. red: surface brightness profile of simulated point sources for {\sl Suzaku}. Green: best-fit surface brightness profile of the CXB component determined with {\sl Suzaku}. Magenta: best-fit surface brightness profile of the remaining CXB component determined by spectral fitting with {\sl Suzaku} after excluding point sources resolved by {\sl Chandra}. Blue: the expected surface brightness of the remaining CXB component (dashed blue: 1$\sigma$ error).
{\it right}: Entropy profile of the north direction. Blue: results measured with {\sl Suzaku} observation only. Green: results measured with {\sl Suzaku} and the point source refinement of {\sl Chandra}.}
\label{fig:cxb} 
\end{center}
\end{figure*}

The unresolved CXB component is the dominant and most uncertain component of the X-ray background in the {\sl Suzaku} spectra of cluster outskirts for energies above 1 keV. In order to constrain this component, we followed the approach described in Walker et al.\ (2013).  
We obtained the coordinates and  2.0--10.0 keV count rates of the point sources resolved by the three {\sl Chandra} pointings (Figure~\ref{fig:chandra}). We convert their count rates to fluxes in 2.0-10.0 keV assuming a absorbed {\tt powerlaw} spectrum with an index of 1.41. 
We detected 148 point sources in the FOV of {\sl Chandra} and 85 of them are within the FOV of {\sl Suzaku}. The faintest resolved point source has a flux of 4.01$\times10^{-17}$\,erg\,s$^{-1}$\,cm$^{-2}$. 

The surface brightness profile of point sources in {\sl Suzaku} FOV is shown in Figure~\ref{fig:cxb}.   
Using the {\sl Heasoft} tool {\tt xissim}, we simulated a {\sl Suzaku} observation of point sources resolved with {\sl Chandra} in the field-of-view of the {\sl Suzaku} observation. The simulated exposure time is set to a very large number to guarantee good statistics.
We extracted spectra from each annular section used in the {\sl Suzaku} analysis from this simulated observation and fit the spectra to a absorbed {\tt powerlaw} with an index of 1.41, 
we obtained the total flux of these simulated point sources within each annulus. Their corresponding surface brightness profiles are shown in Figure~\ref{fig:cxb}-left.

In our {\sl Suzaku} spectral fitting of the north direction, we include a {\tt powerlaw} component with an index of 1.41 to account for the emission of the {\sl Chandra} resolved point sources and with their normalizations fixed at their associated fluxes.
We also included another {\tt powerlaw} component with an index of 1.41 to model the remaining unresolved point sources; its normalization is set to vary freely but all regions were linked together since the fluctuations of the remaining unresolved point sources should be much smaller. 
We obtained a best-fit of 7.14$\times10^{-12}$\,erg\,s$^{-1}$\,cm$^{-2}$\,deg$^{-2}$ for the remaining unresolved point sources in 2.0--10.0 keV. 

Following Moretti et al.\ (2003), the expected level of the remaining unresolved point sources can be estimated through the integral,
\begin{equation}
F_{\rm CXB}=2.18\pm0.13\times 10^{-11} -\int_{S_{\rm excl}}^{S_{\rm max}}\left(\frac{dN}{dS}\right)\times S dS,
\end{equation}
in units of erg\,s$^{-1}$\,cm$^{-2}$\,deg$^{-2}$. 
We take an analytical form of the source flux distribution in the 2.0--10.0 keV band given by Moretti et al.\ (2003),
\begin{equation}
N (> S) = \nonumber \\
\end{equation}
\begin{equation}
N_{S(H)} \left[\frac{(2\times10^{-15})^{\alpha_{1,S(H)}}}{S^{\alpha_{1,S(H)}}+S_{0,H}^{\alpha_{1,S(H)}-\alpha_{2,S(H)}}S^{\alpha_{2,S(h)}}}\right]
\end{equation}
where $\alpha_{1,H}=1.57^{+0.10}_{-0.08}$, $\alpha_{2,H}=0.44^{+0.12}_{-0.13}$, $S_{0,H}=(4.5^{+3.7}_{-1.7})\times10^{-15}$ erg\,s$^{-1}$\,cm$^{-2}$, and $N_H=5300^{+2800}_{-1400}$.
Using the best-fit values of these parameters, we estimate the level of remaining unresolved point sources to be $6.69\pm3.76\times10^{-12}$\,erg\,s$^{-1}$\,cm$^{-2}$\,deg$^{-2}$.
The best-fit surface brightness of unresolved point sources are in good agreement with the expected emission of unresolved point sources as compared in Figure~\ref{fig:cxb}-left.  

For most annuli, about 50\% of the point sources were resolved (see Figure~\ref{fig:cxb}-left).  
The total flux of the resolved point sources plus the flux of the remaining unresolved point sources are comparable to the flux for the originally determined unresolved point sources with {\sl Suzaku} data only.
In Figure~\ref{fig:cxb}-right we compare the {\sl Suzaku} measurements of the entropy profile of RXJ1159+5531 (see below in \S4.3) in the north direction, obtained {\it with} and {\it without} the refinement of {\sl Chandra}.
The values are consistent while the statistical uncertainty of the virial entropy has been reduced by 18\% with the incorporation of {\sl Chandra} data. 
\medskip

\section {\sl Hydrostatic Models}

The forward fitting approach allows us to fit the projected density and temperature profiles directly and also allows the cluster emission outside the largest spectral extraction annulus to be treated self-consistently. These properties are advantageous for background-dominated X-ray emission in the outskirts of a cluster where spectral deprojection (e.g., using the standard Òonion-peelingÓ approach) further degrades the data quality and traditionally assumes there is no cluster emission outside the largest annulus used. We focus on solutions of the hydrostatic equilibrium equation in terms of the entropy because the additional constraint of convective stability is easily enforced by requiring the entropy to increase with radius. In addition, unlike the temperature and density, the much smoother entropy profile is easily parameterized with a simple broken power law model.

Rewriting the hydrostatic equation in terms of the entropy and a variable depending on the pressure yields
\begin{equation}
\frac{{\rm d}\xi}{{\rm d}r}=-\frac{2}{5}\frac{GM(<r)}{r^2}K^{-3/5}.  ~~~ \xi\equiv P^{2/5}
\end{equation}
where $P$ is the gas pressure, and $M(<r)$ is the total enclosed mass within radius r, which includes the contributions from stars, gas, and dark matter. Equation (4) can be solved directly given models for $K(r)$ and $M(r)$ provided the gas mass can be neglected, since the pressure (and therefore $\xi$) depends on the gas density. For a self-consistent solution Equation (4) is rearranged and differentiated with respect to $r$, taking advantage of the mass continuity equation to give,

\begin{equation}
\frac{{\rm d}}{{\rm d}r}\left(r^2K^{3/5}\frac{{\rm d}\xi}{{\rm d}r}\right)+\frac{8\pi r^2G}{5}K^{-3/5}\xi^{3/2}
 \nonumber \\
 \end{equation}
 \begin{equation}
=-\frac{8\pi r^2G}{5}(\rho_{\rm stars}+\rho_{\rm DM}).
\end{equation}

To parameterize the dark matter component we used the NFW profile (Navarro et al. 1997),
\begin{equation}
\rho_{\rm DM}(r)=\frac{\rho_0}{\frac{r}{R_{\rm s}}\left(1+\frac{r}{R_{\rm s}}\right)^2}
\end{equation}
where the free parameters are the scale radius ($R_{\rm s}$) and a characteristic density ($\rho_0$). The stellar component has no free parameters and is described in \S4.4. For $K(r)$ we adopt a simple broken power-law model plus a constant, which takes the form of $K = {\rm k_0} + f(r)$, where
\begin{equation}
f(r) = \left \{
\begin{array}{l}
{\rm k_1}\,r^{\beta_1}  \quad r \leq r_1\\
{\rm k_2}\,r^{\beta_2}  \quad {r_1 <r \leq r_2}\\
{\rm k_3}\,r^{\beta_3}  \quad {r > r_2}
\end{array}  \right .\
\end{equation}
There are two boundary conditions that need to be specified Ñfor $\xi(r)$ and $d\xi(r)/dr$. Since the gas mass is negligible at small radius, Equation (4) can be solved directly at some small radius to give the boundary condition for $d\xi(r)/dr$. The boundary condition for $\xi(r)$ amounts to specifying the pressure or entropy at some radius which we treat as a free parameter (see Humphrey et al.\ 2008 and Buote \& Humphrey 2012). For a given set of parameters for $K(r)$, $M(r)$, and the $\xi(r)$ boundary condition, we solve Equation (7) for $\xi(r)$ from which we compute the density (emission measure) and emission-weighted temperature profiles in projection that are then compared to the measurements via the $\chi^2$ statistic. We explored the parameter space using a Bayesian Monte Carlo Method $-$ version 2.7 of the MultiNest code (Feroz \& Hobson 2008; Feroz et al.\ 2009). Flat priors were assumed for all free parameters.

 \begin{figure*}
   \begin{center}
     \leavevmode 
 \epsscale{1.2}
        \centering
\hspace{5mm}\subfloat{\includegraphics[width=0.45\textwidth]{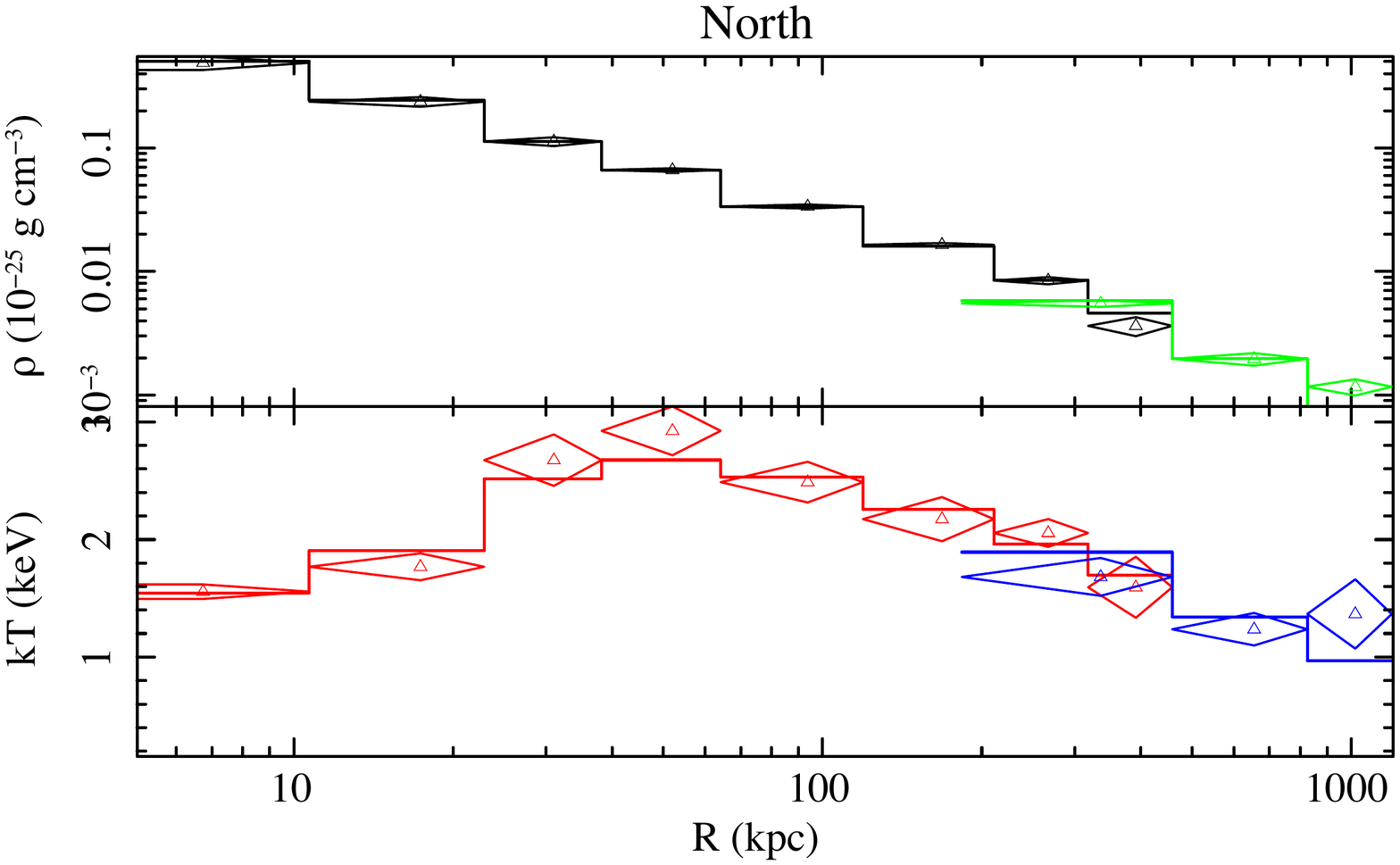}}\hspace{0mm}\subfloat{\includegraphics[width=0.45\textwidth]{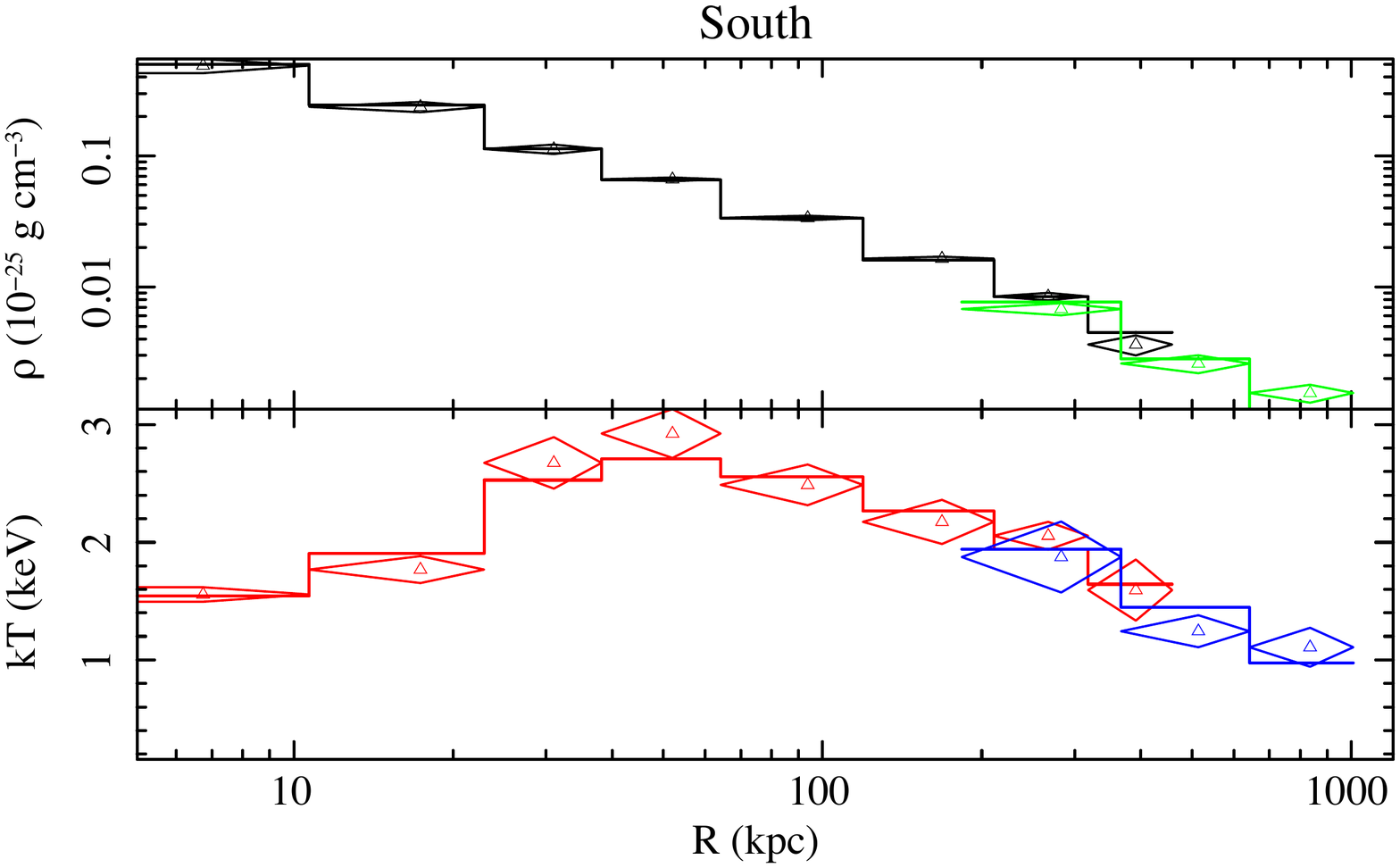}}\\
\vspace{-15mm}
\hspace{5mm}\subfloat{\includegraphics[width=0.45\textwidth]{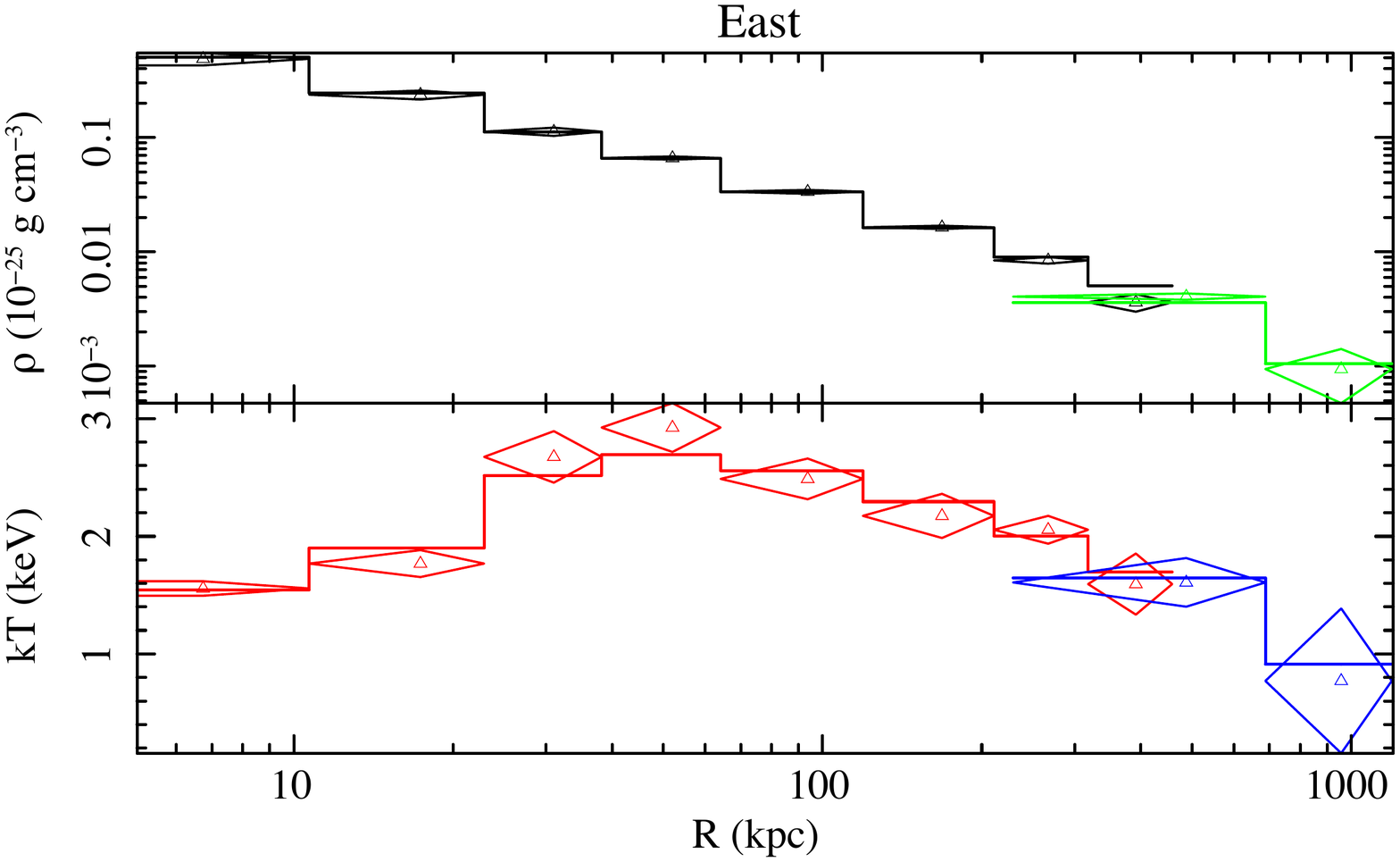}}\hspace{0mm}\subfloat{\includegraphics[width=0.45\textwidth]{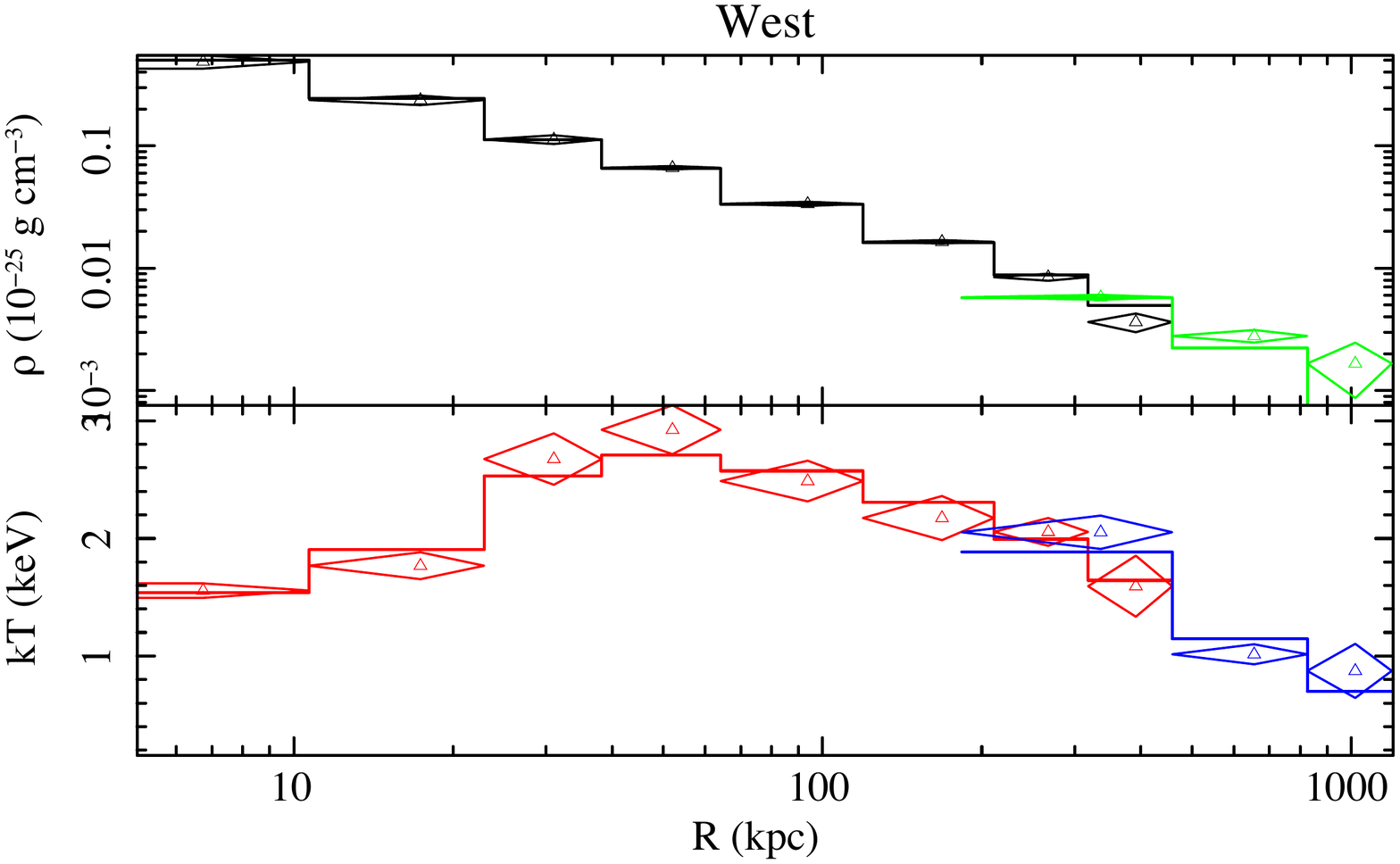}}\\
\vspace{-15mm}
\caption{Top: Projected density profiles of central regions measured with {\sl Chandra} (black diamonds) and of each direction measured with {\sl Suzaku} (green diamonds). Black and green solid lines indicate the values of best-fits. Bottom: Projected temperature profiles of central regions measured with {\sl Chandra} (red) and of each direction measured with {\sl Suzaku} (blue). red and blue solid lines indicate the values of best-fits. }
\label{fig:2d} 
\end{center}
\end{figure*}

In this work, we applied a pair of parameterized dark matter density and entropy profiles that can uniquely determine the pressure profile by solving the hydrostatic equilibrium equation (there are also free parameters associated with the boundary conditions). We can obtain three dimensional temperature and density profiles through entropy and pressure profiles. We project the three dimensional temperature and density profiles on to the sky assuming spherical symmetry of the ICM and compare them to the observed projected temperature and density profiles. We repeat this process through monte-carlo simulations until we can reproduce the observed results. 
The ability to fit the projected data directly is very valuable for low quality, noisy data such as cluster outskirts whose integrity would be greatly degraded by deprojection noise.
In this work, we applied this technique to data in each direction separately to obtain their gas properties. 
We display the projected temperature and density profiles along with their best-fitting models in 
Figure~\ref{fig:2d} (top and bottom panels). We obtained a best-fit $\chi^2/$d.o.f of $13.5/11$, $12.6/11$, $10.8/9$, and $12.8/11$ for the North, South, East, and West directions respectively.
As shown in \S 4, very similar results for the entropy and other quantities are obtained for each direction. Thus, for convenience and clarity of presentation we adopt fiducial values of $R_{500}$, $R_{200}$, $R_{\rm 108}$, and the total mass within $R_{\rm 108}$\,($M_{\rm tot}$=9.7$\times10^{13}$\,M$_{\odot}$) obtained from the North pointing as the parameters of RXJ1159+5531. 
We note that in our joint fit of the {\sl Chandra} and {\sl Suzaku} data the high S/N Chandra data at small radii are crucial for constraining the model components near the center (e.g., M/L of the stellar component, NFW scale radius). However, the {\sl Suzaku} data provide the crucial constraints on the model properties obtained at large radius; the constraints on the mass profile derived using only the {\sl Chandra} data (e.g., Gastaldello et al.\ 2007) were not as good as our joint fit.

\section {\sl Results}

\subsection{\sl Temperature and density profiles}

 \begin{figure*}
   \begin{center}
     \leavevmode 
\hspace{5mm}\subfloat{\includegraphics[width=0.48\textwidth]{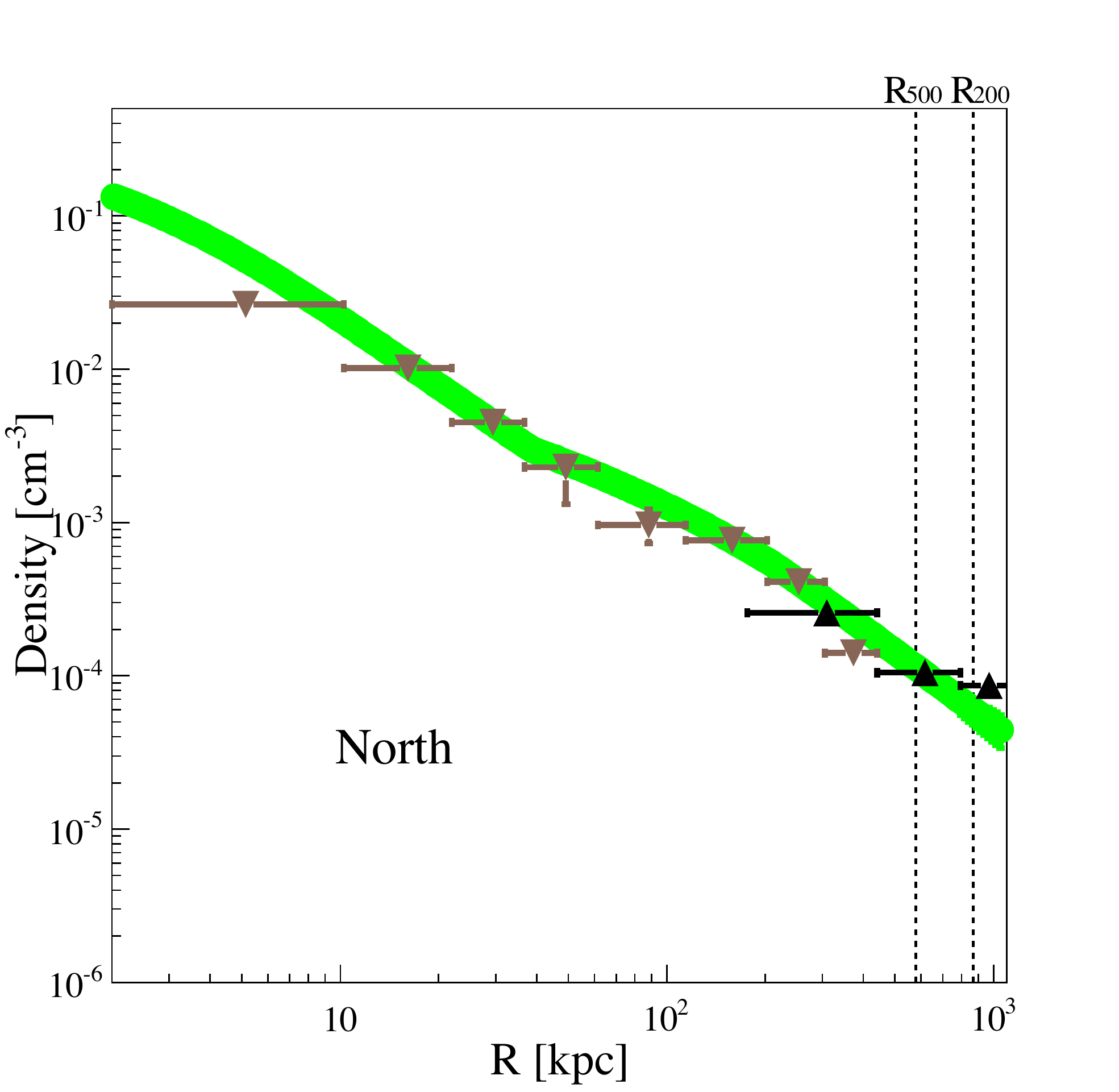}}\hspace{0mm}\subfloat{\includegraphics[width=0.48\textwidth]{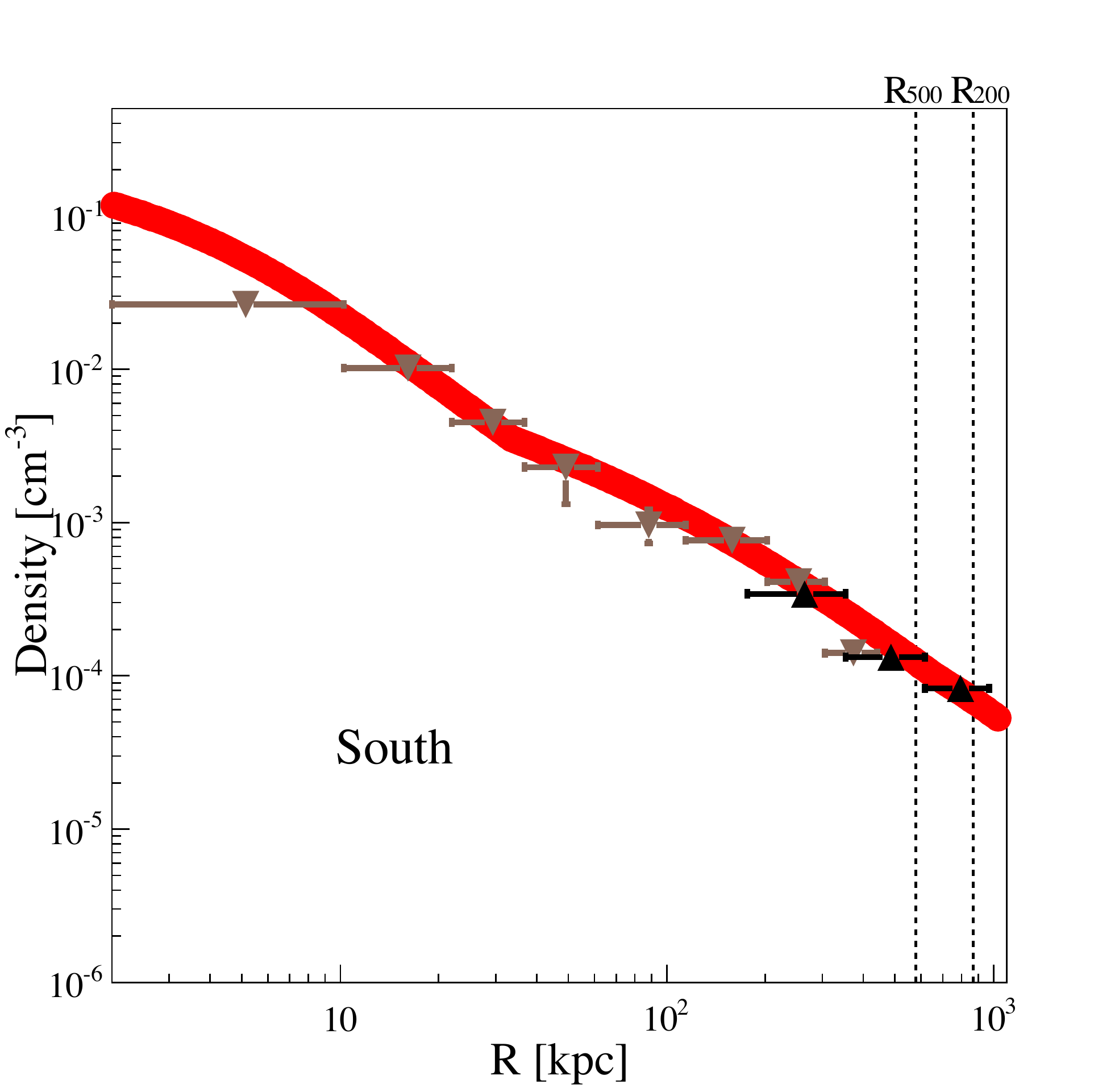}}\\
\vspace{-2.5mm}
\hspace{5mm}\subfloat{\includegraphics[width=0.48\textwidth]{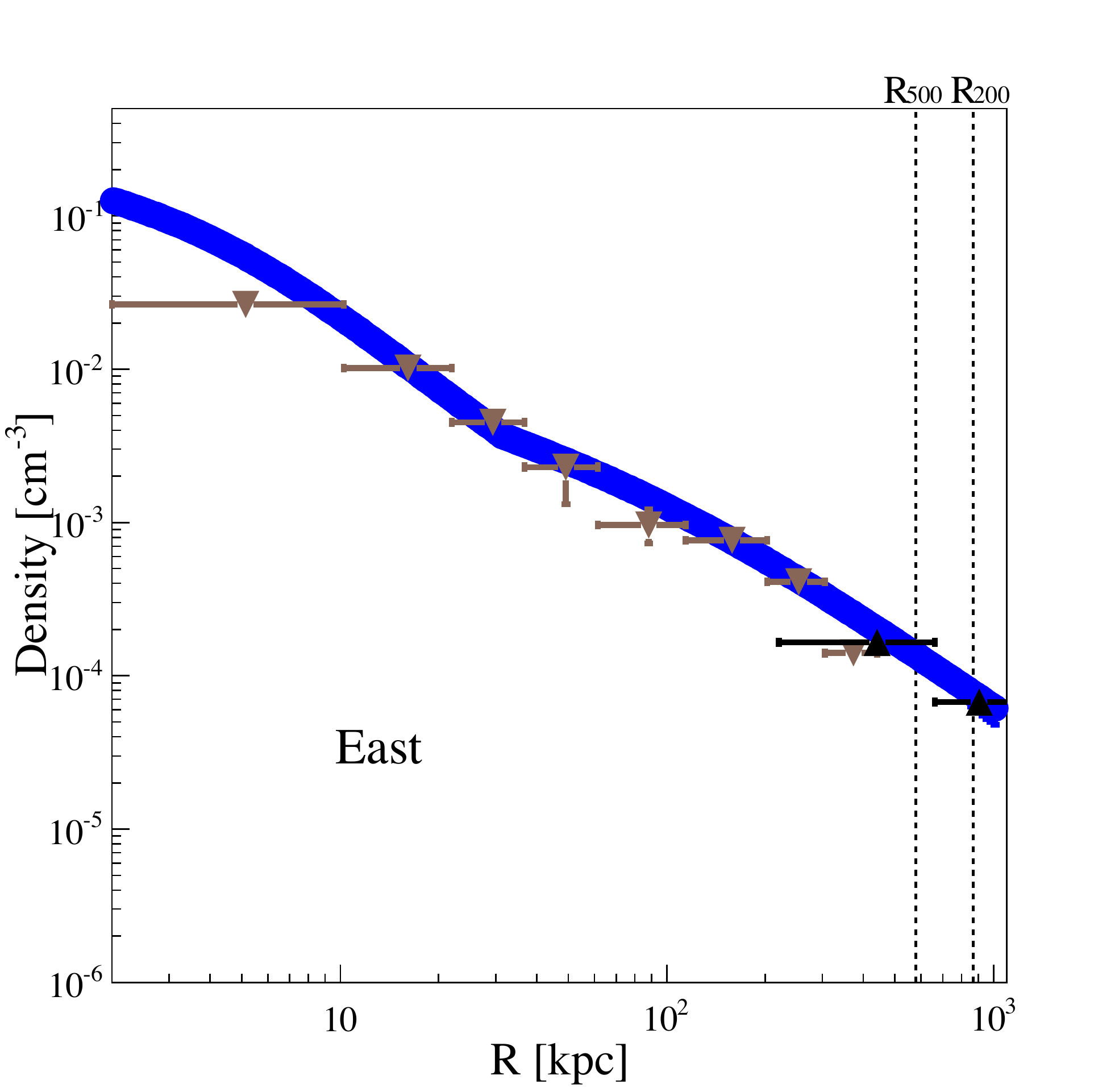}}\hspace{0mm}\subfloat{\includegraphics[width=0.48\textwidth]{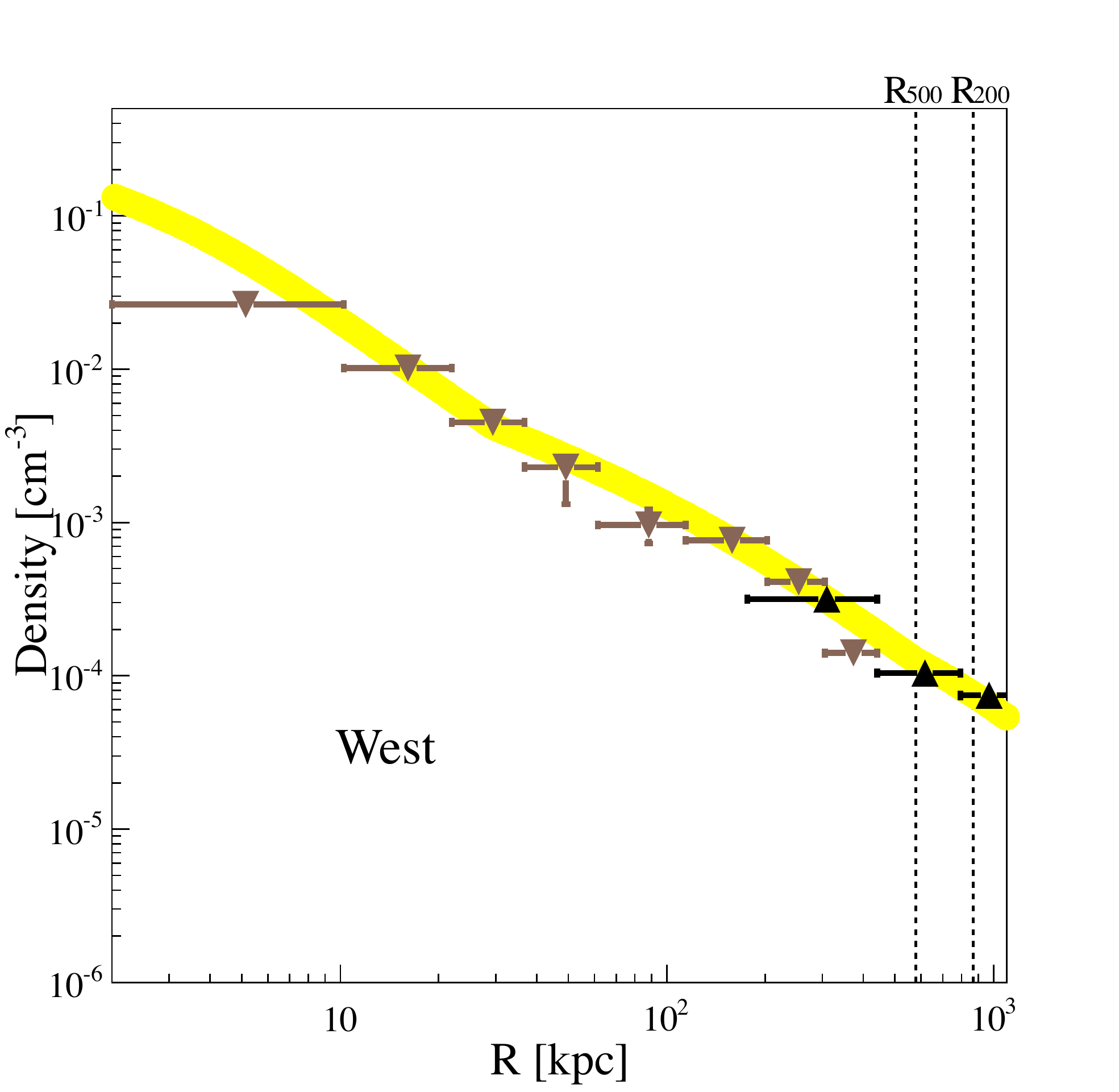}}\\
\caption{Solid lines: 3D density profiles derived with our primary method (entropy-based forward fitting) for each direction. 
Shaded regions indicate 1$\sigma$ uncertainties.  
Triangles: 3D density profiles obtained  via spectral deprojection {\tt projct} (Brown measured with {\sl Chandra}; Black measured with {\sl Suzaku}).}
\label{fig:density} 
\end{center}
\end{figure*}

 \begin{figure*}
   \begin{center}
     \leavevmode 
\hspace{5mm}\subfloat{\includegraphics[width=0.48\textwidth]{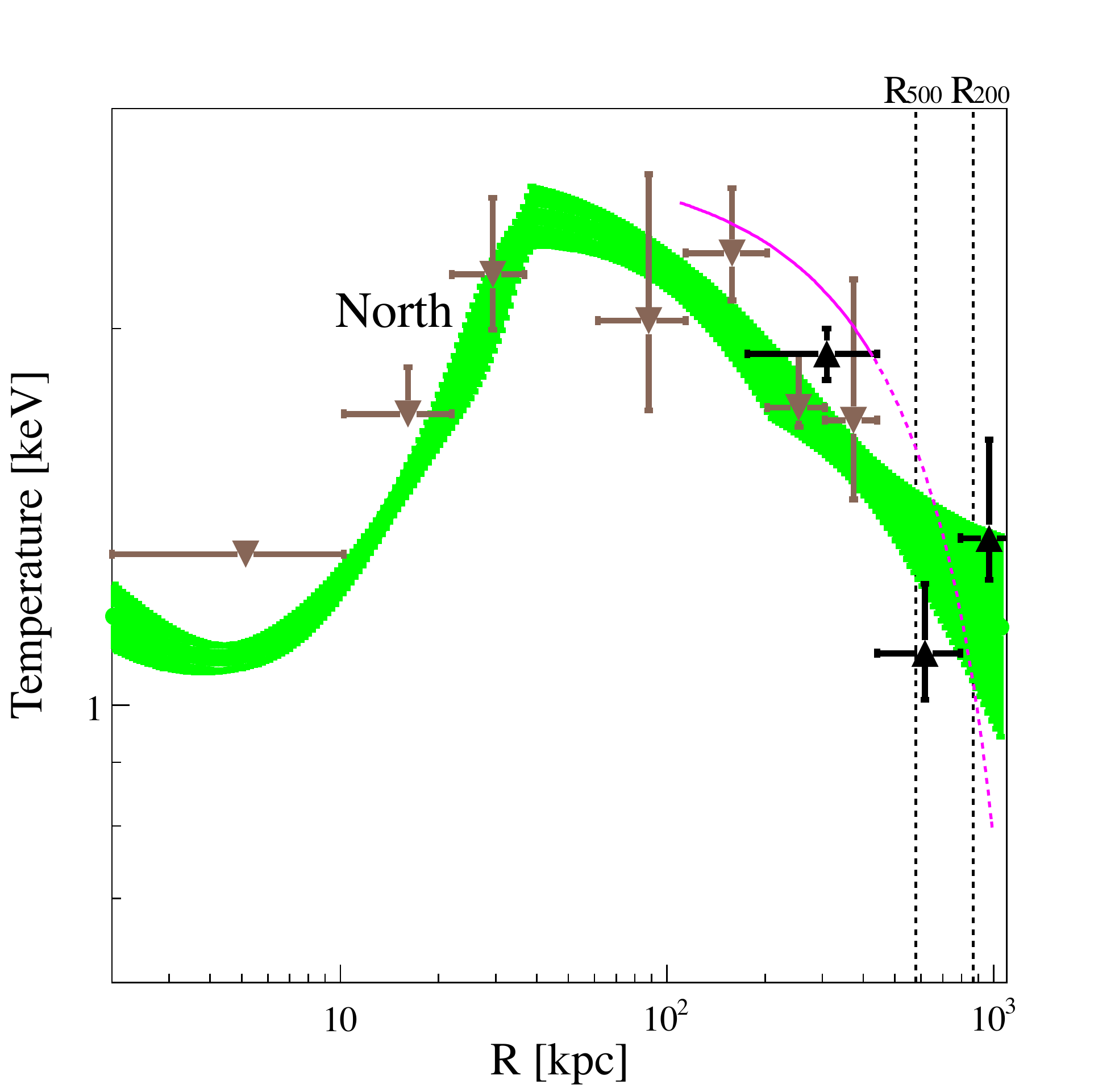}}\hspace{0mm}\subfloat{\includegraphics[width=0.48\textwidth]{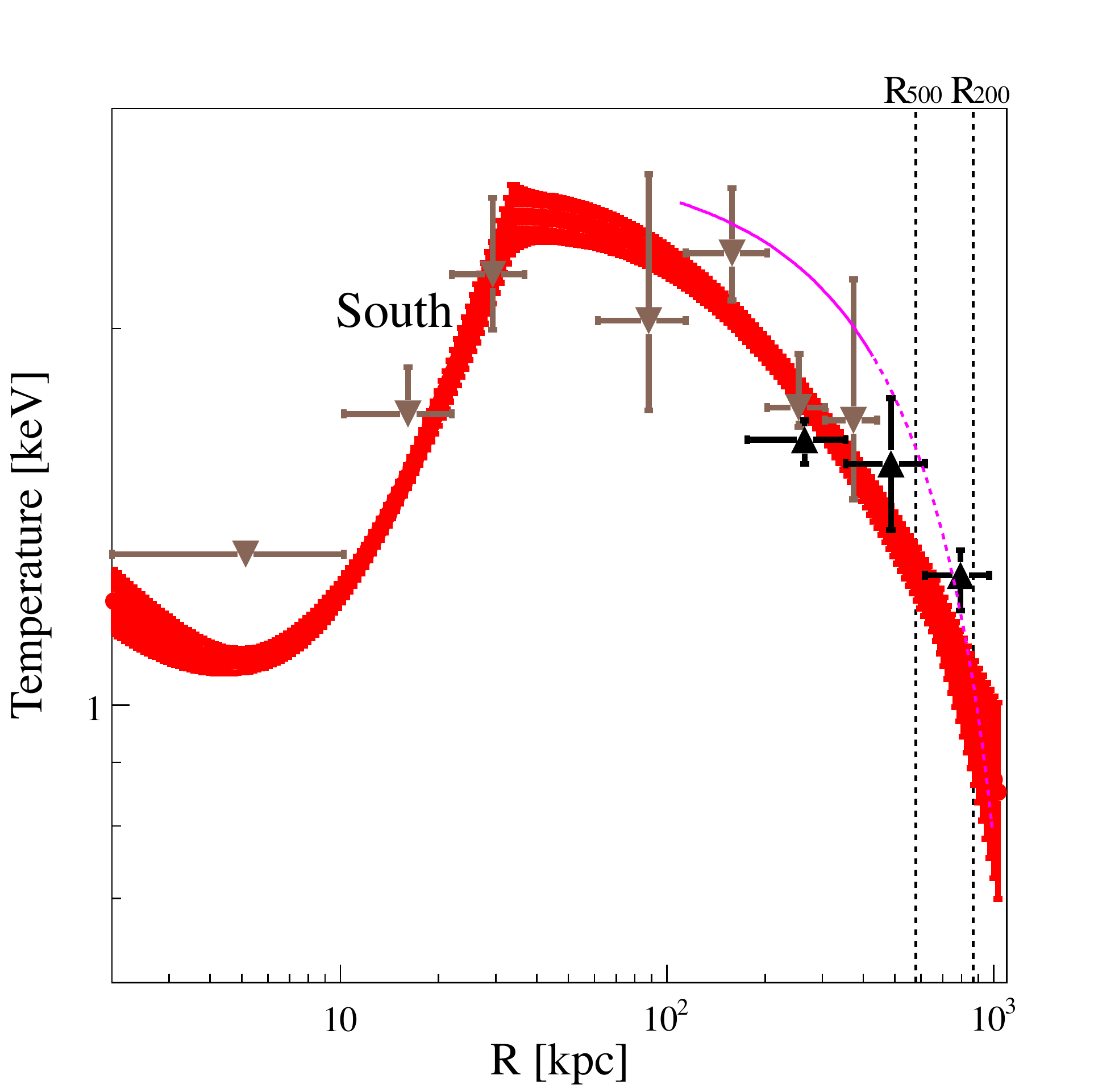}}\\
\vspace{-2.5mm}
\hspace{5mm}\subfloat{\includegraphics[width=0.48\textwidth]{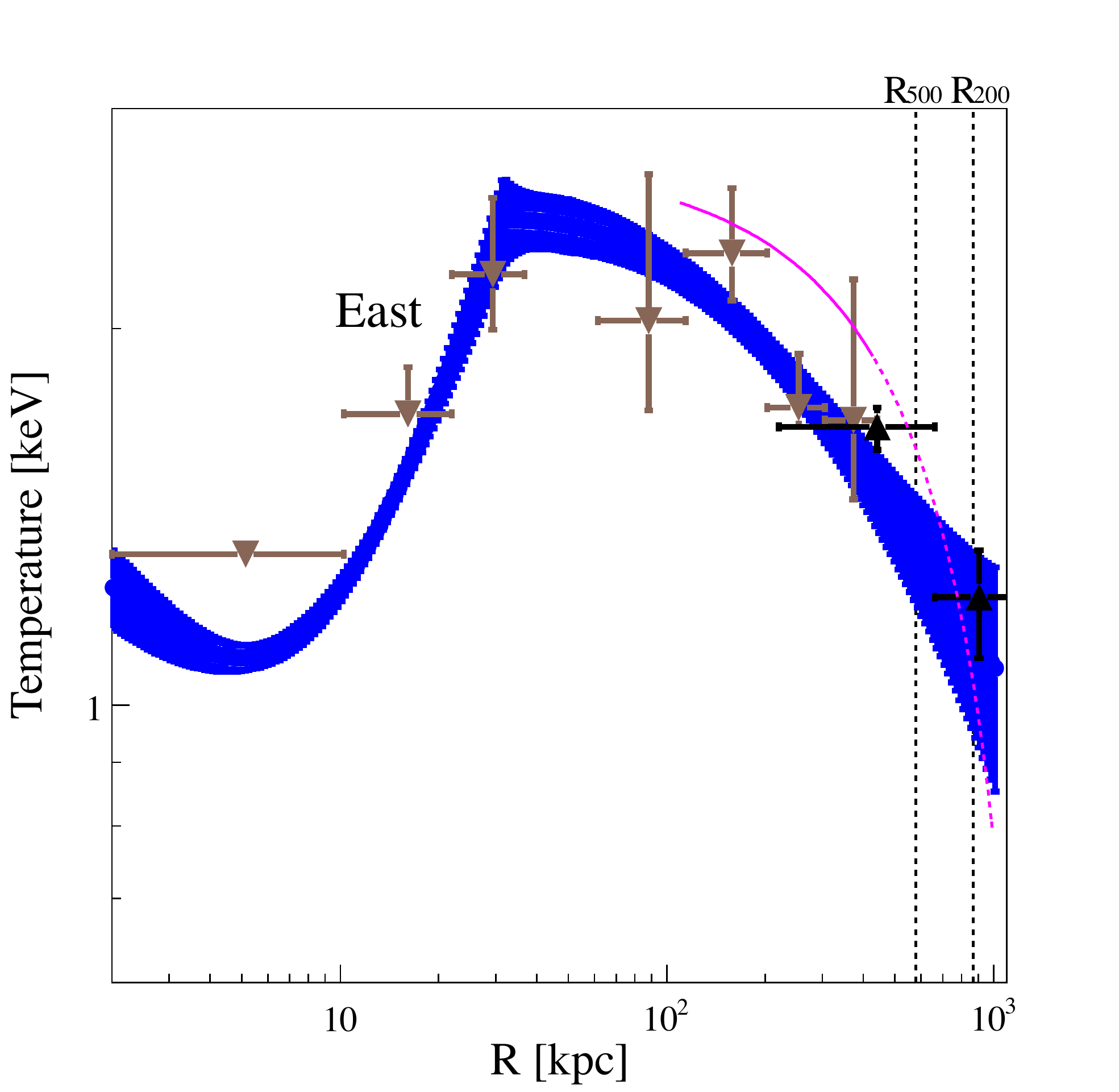}}\hspace{0mm}\subfloat{\includegraphics[width=0.48\textwidth]{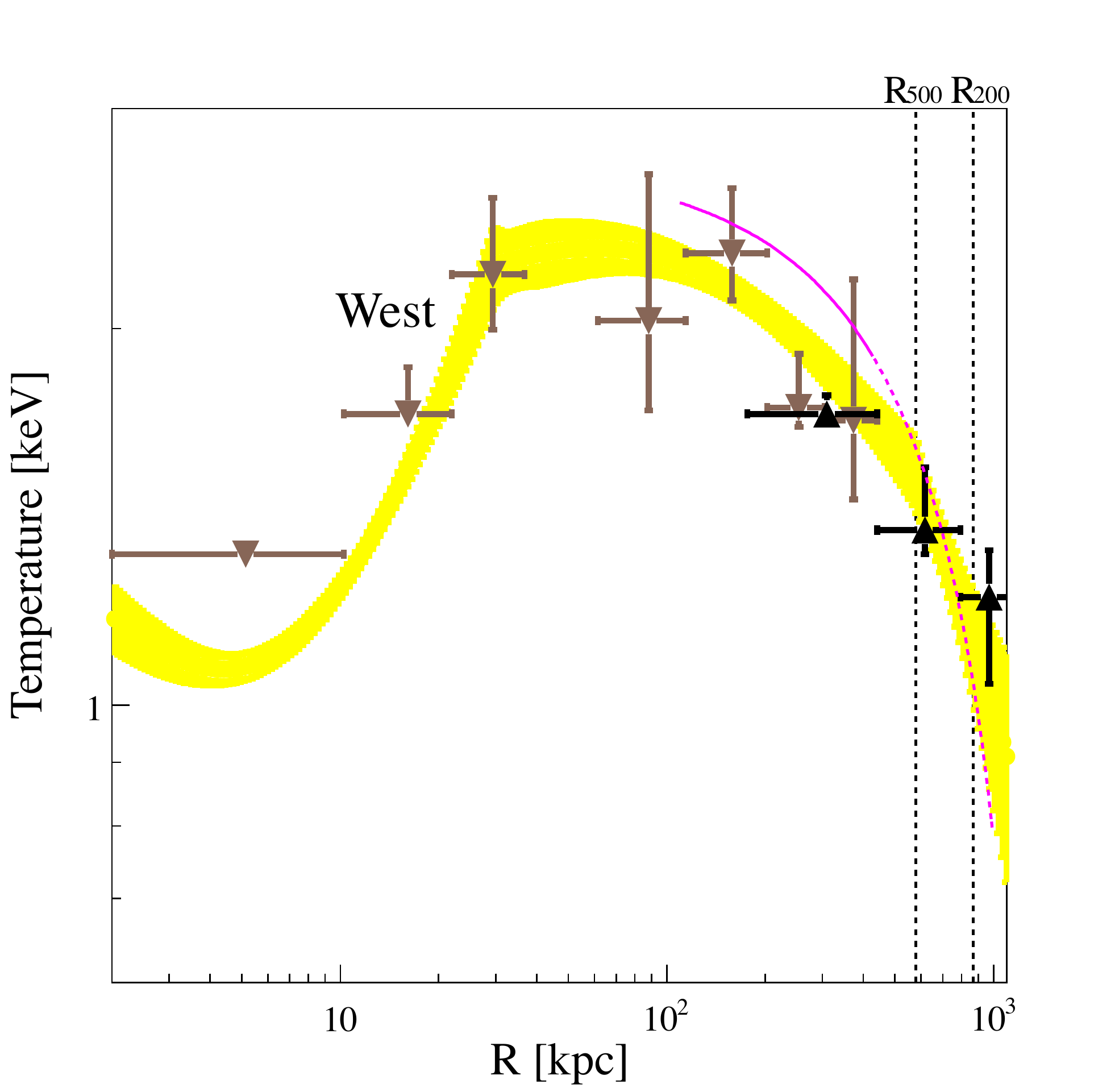}}\\
\caption{Solid lines: 3D temperature profiles derived with our primary method (entropy-based forward fitting) for each direction.  
Shaded regions indicate 1$\sigma$ uncertainties.
Magenta: the
heuristic formula calibrated by Pratt et al.\ (2007) for clusters. Triangles: 3D temperature profiles obtained  via spectral deprojection {\tt projct} (Brown measured with {\sl Chandra}; Black measured with {\sl Suzaku}).}
\label{fig:temperature} 
\end{center}
\end{figure*}

The best-fit two-dimensional (2D) gas temperature and density profiles of RXJ1159+5531 in all directions are shown in Figure~\ref{fig:2d} (top and middle panels). 
For each direction, {\sl Suzaku} results are obtained with $\sim90^{\circ}$ sectional  $\sim$$3.0^{\prime}$ radial bins beyond 0.3 $R_{\rm vir}$. We did not plot the results of the innermost bin of {\sl Suzaku} 
since our two-temperature fit is only a crude representation of the spatially resolved temperature profile measured by {\sl Chandra}. 
We also plotted the results obtained with {\sl Chandra} in several $\sim1^{\prime}$, $\sim360^{\circ}$ annuli for the central regions. 
The combined profiles span from the center out to $R_{\rm vir}$. 
The width of our extracted annuli vary from 4 kpc in the center to 350 kpc out to $R_{\rm vir}$. In the overlapping regions, the results from {\sl Chandra} and {\sl Suzaku} show good agreement.

\begin{figure*}
   \begin{center}
     \leavevmode 
\hspace{5mm}\subfloat{\includegraphics[width=0.48\textwidth]{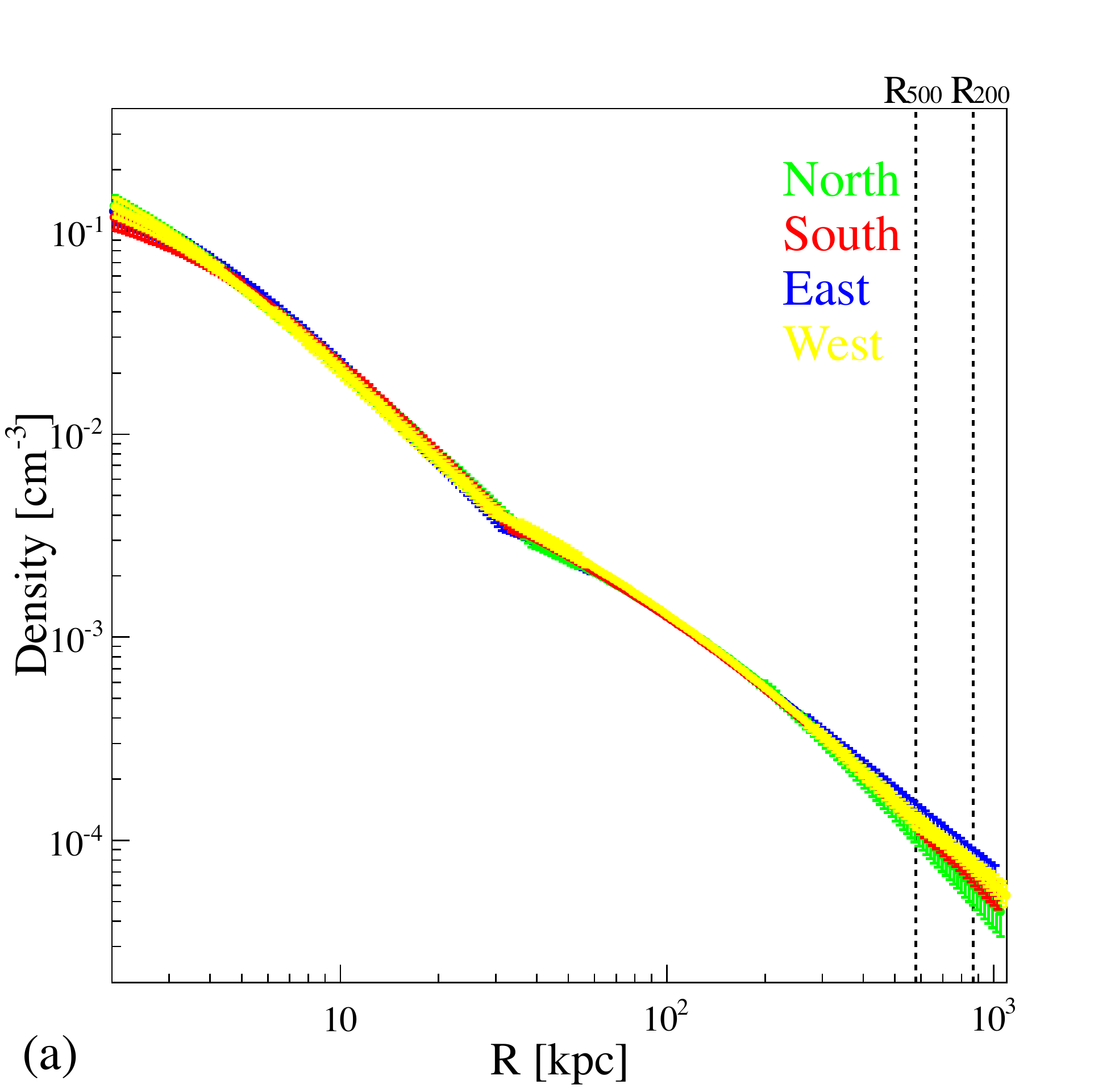}}\hspace{0mm}\subfloat{\includegraphics[width=0.48\textwidth]{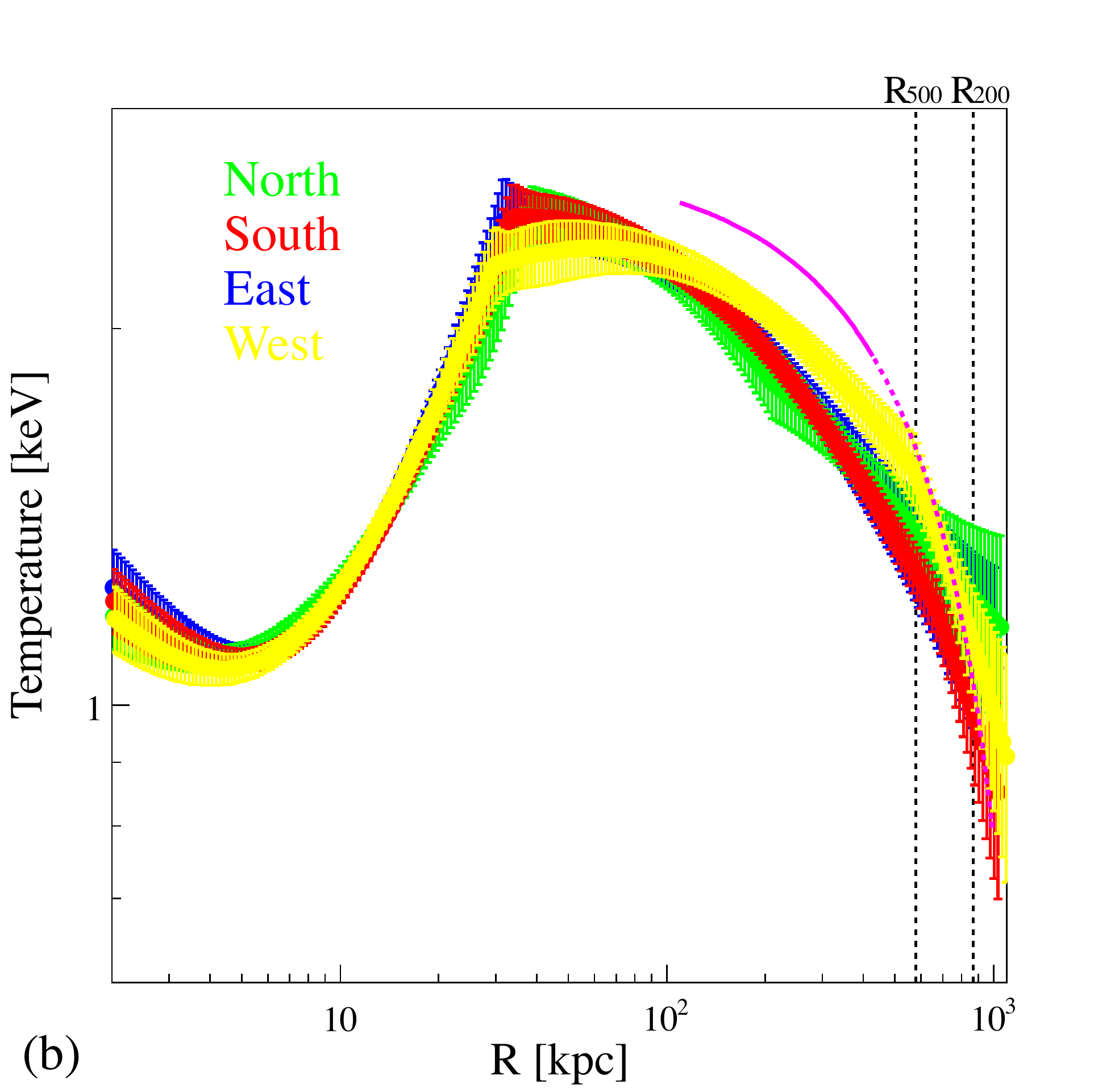}}\\
\vspace{-2.5mm}
\hspace{5mm}\subfloat{\includegraphics[width=0.48\textwidth]{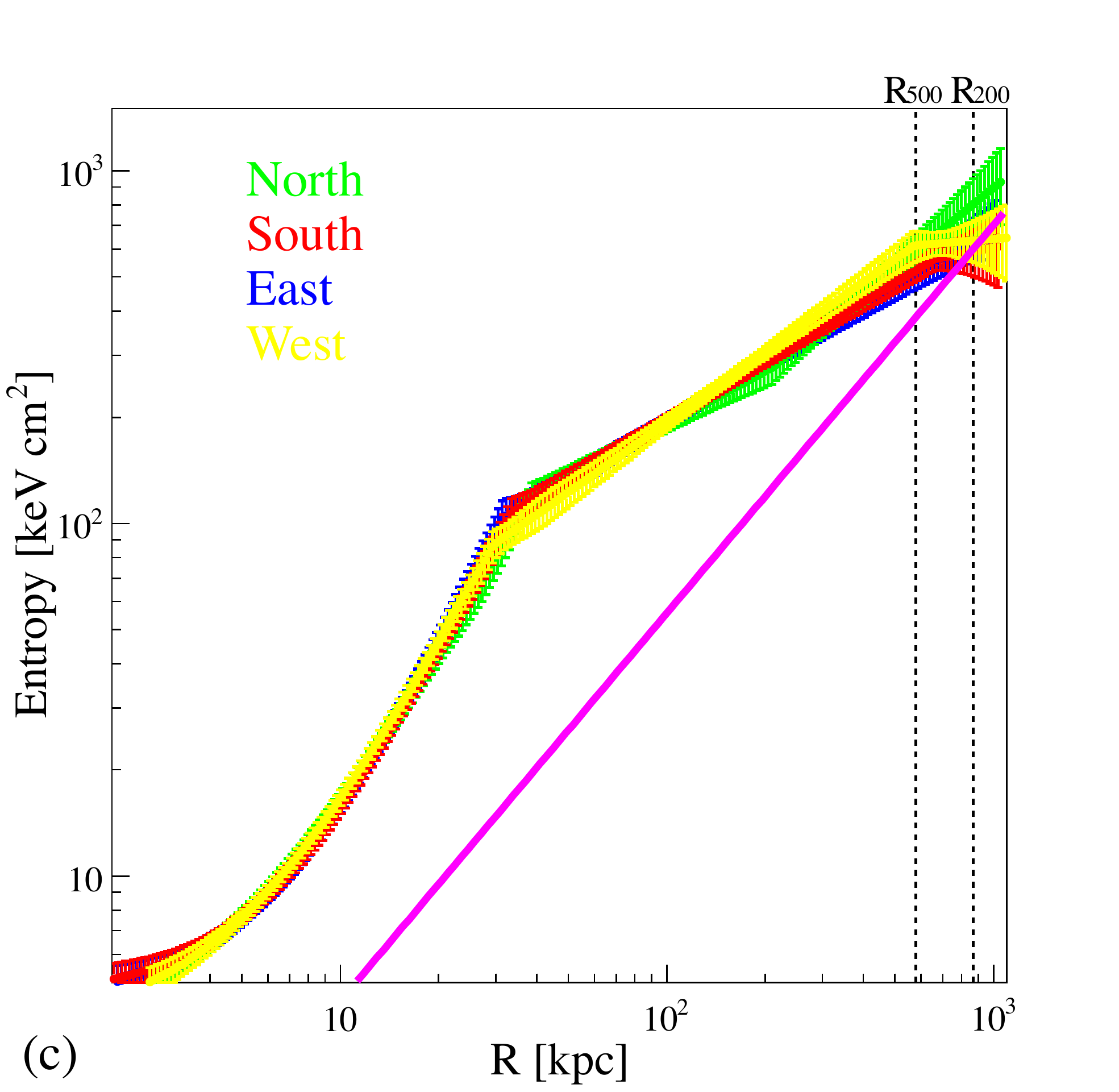}}\hspace{0mm}\subfloat{\includegraphics[width=0.48\textwidth]{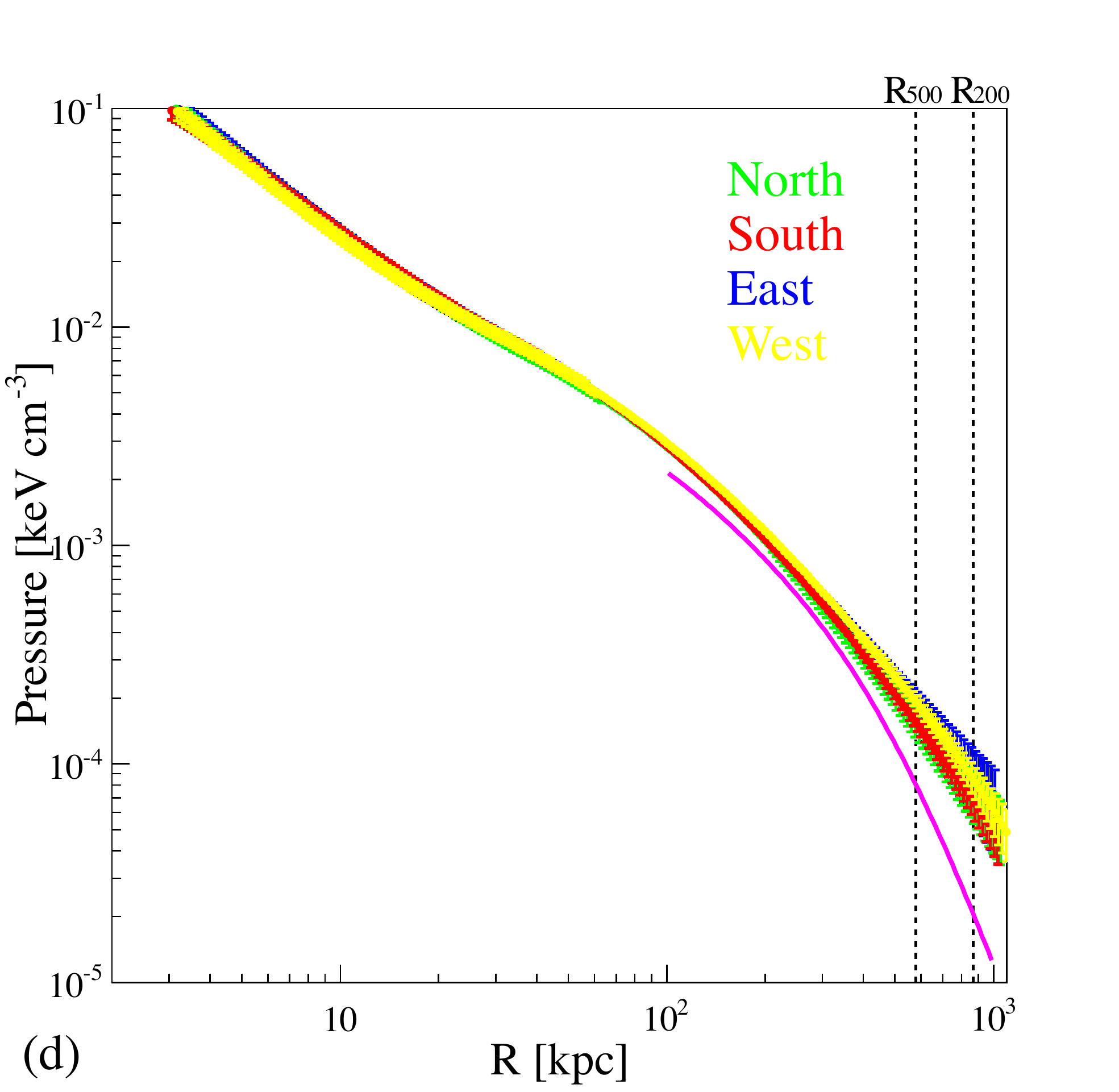}}\\
\caption{Gas properties of all directions obtained with our primary method (entropy-based forward fitting). Shaded regions indicate 1$\sigma$ uncertainties.  
(a). 3D density profile. (b). 3D temperature profile. (c). 3D entropy profile. Magenta: entropy profile from gravity-only cosmology simulations, 
$K\propto r^{1.1}$; normalization is derived from Voit et al.\ (2005). (d). 3D pressure profile. Magenta: universal pressure profile derived by Arnaud et al.\ (2010).}
\label{fig:rxj} 
\end{center}
\end{figure*}

We plot the three-dimensional (3D) density and temperature profiles for each direction in Figure~\ref{fig:density} and Figure~\ref{fig:temperature} respectively (marked in solid lines) obtained with the forward-fitting approach described in \S3.   
We compare the density and temperature profiles of all directions together in Figure~\ref{fig:rxj} (a) and (b). Each direction shows remarkably similar density profiles.
The virial temperatures of all directions are consistent within 1$\sigma$ uncertainties. 
The 3D temperature profiles decline by more than a factor of 2 from a peak temperature of $\sim$3\,keV at 0.3 $R_{\rm vir}$ to $\sim$1.0\,keV at $R_{\rm vir}$.
We compared the 3D temperature profiles to the
heuristic formula 
calibrated by Pratt et al.\ (2007) for clusters over a radial range of $0.125 < R/R_{200} < 0.5$:
\begin{equation}
T/T_X=1.19-0.74~R/R_{200}, 
\end{equation} 
where $T_X$ is the peak temperature. We observe a similar decline of the temperature profiles out to large radii in this system as observed in other galaxy clusters (Akamatsu et al.\ 2011).

\subsection {\sl Entropy and pressure profiles}

 \begin{figure*}
   \begin{center}
     \leavevmode 
\hspace{5mm}\subfloat{\includegraphics[width=0.48\textwidth]{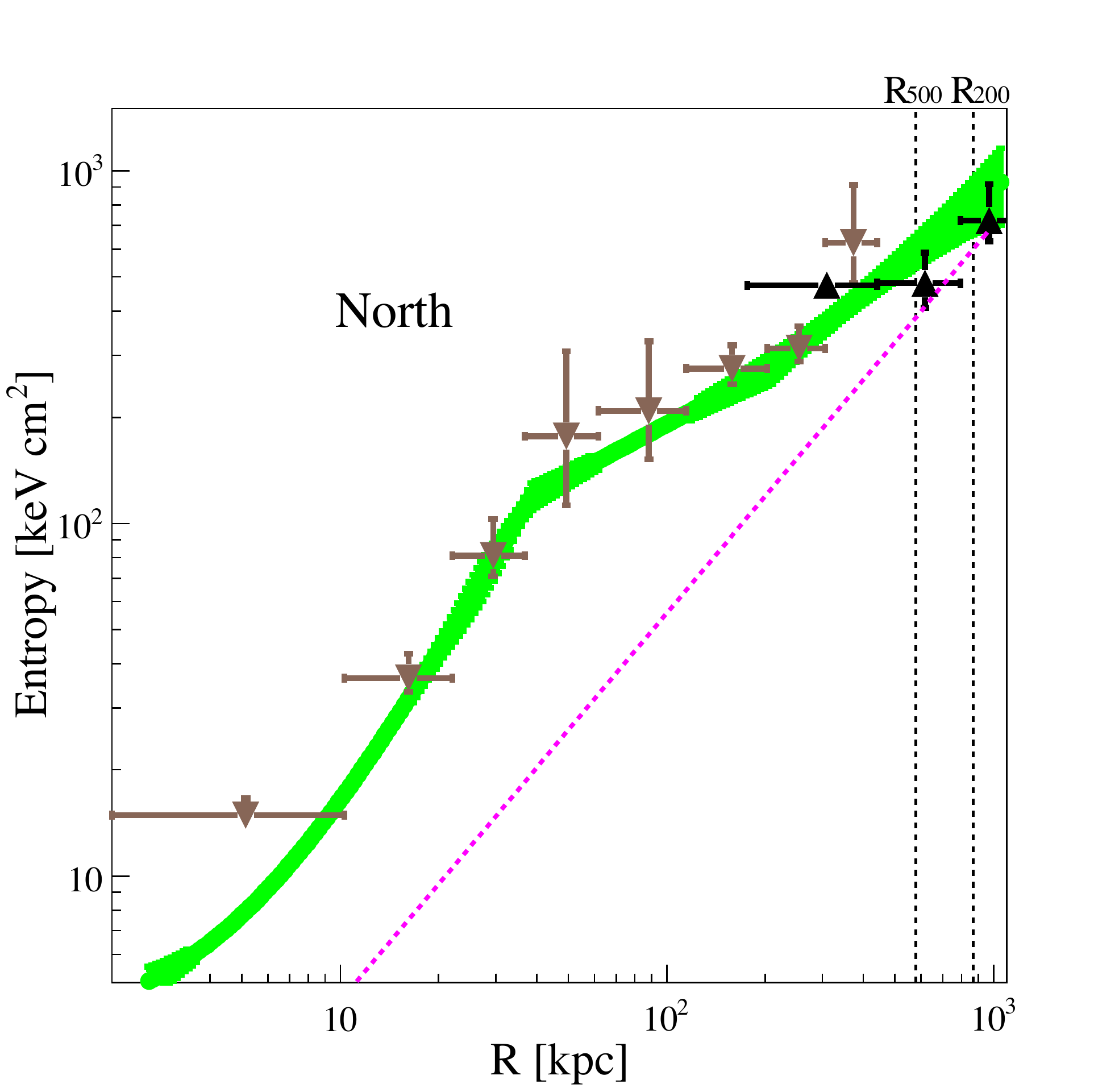}}\hspace{0mm}\subfloat{\includegraphics[width=0.48\textwidth]{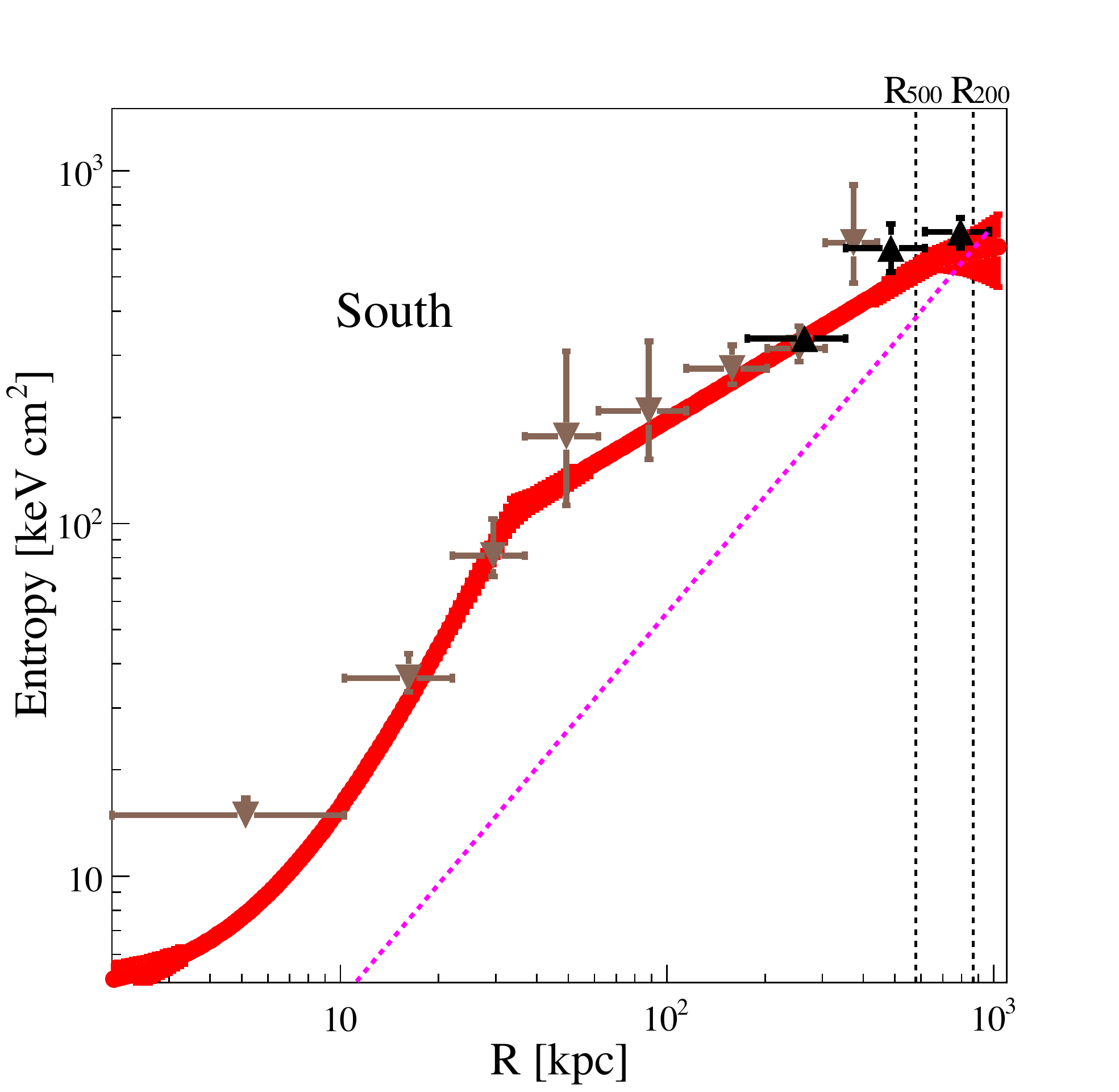}}\\
\vspace{-2.5mm}
\hspace{5mm}\subfloat{\includegraphics[width=0.48\textwidth]{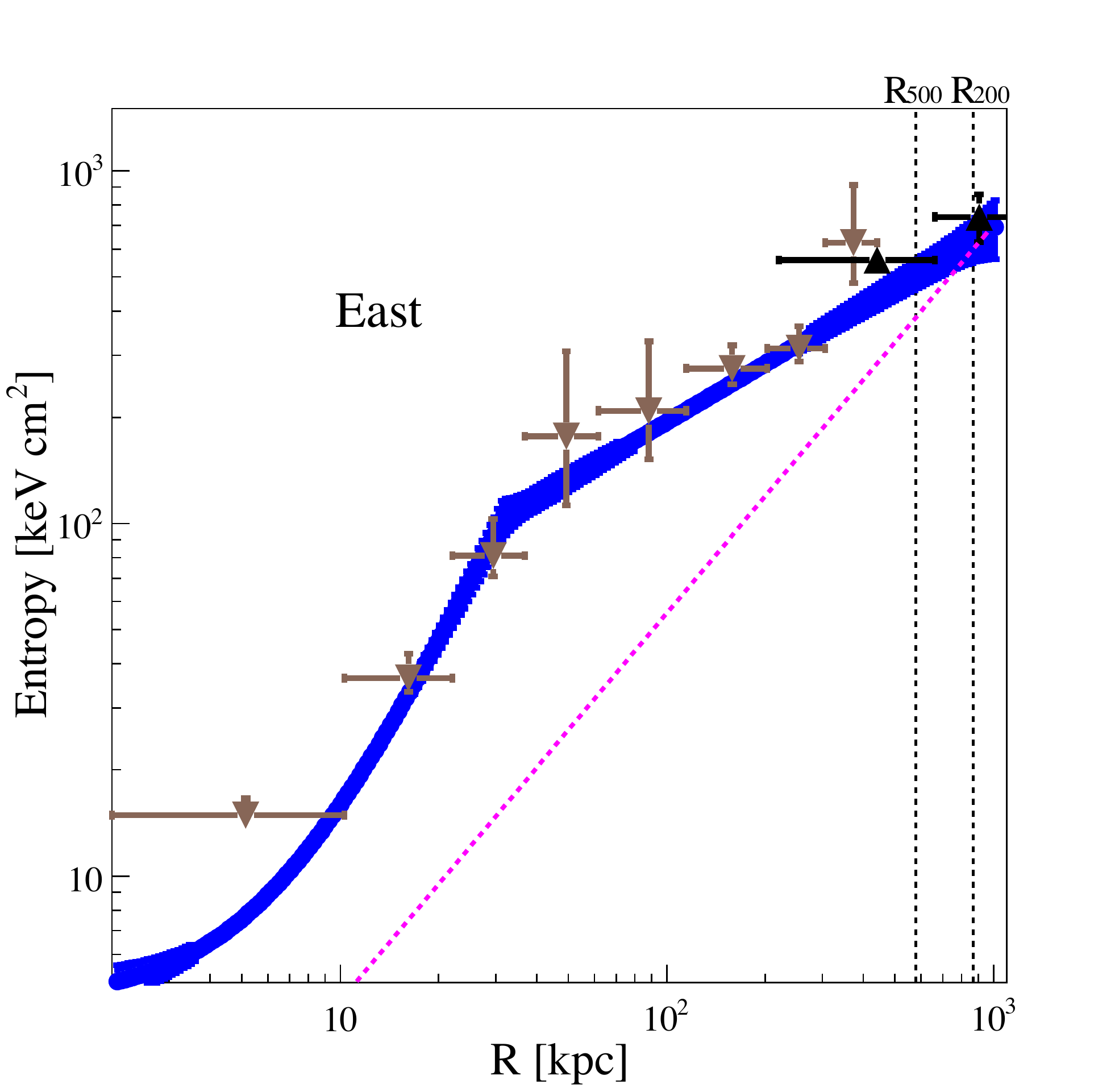}}\hspace{0mm}\subfloat{\includegraphics[width=0.48\textwidth]{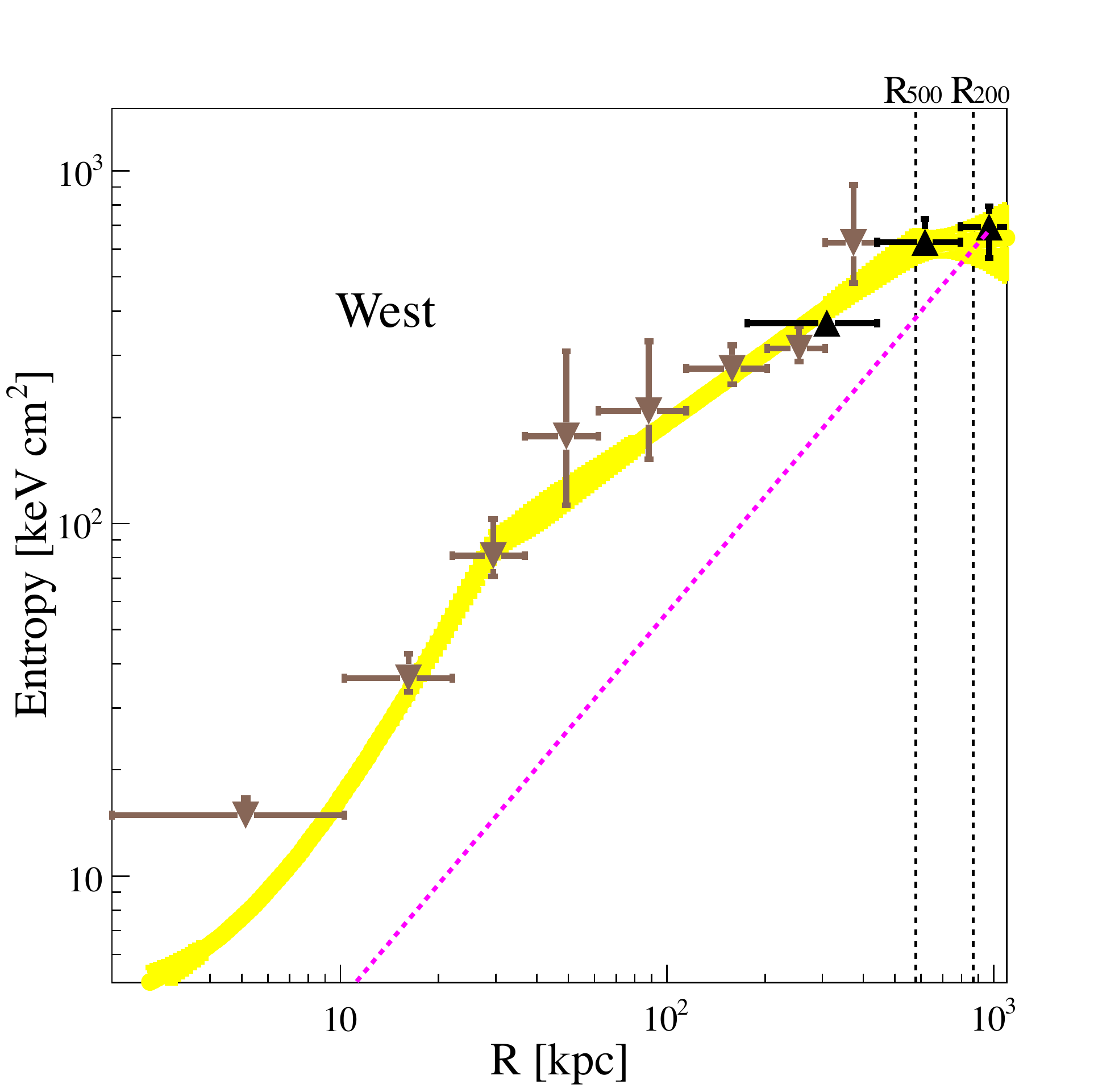}}\\
\caption{Solid lines: 3D entropy profiles derived with our primary method (entropy-based forward fitting) for each direction. Shaded regions indicate 1$\sigma$ uncertainties. 
Magenta line: entropy profile from simulations, 
$S\propto r^{1.1}$; normalization is derived from Voit et al.\ (2005). 
Triangles: 3D density profiles obtained via spectral deprojection {\tt projct} (Brown measured with {\sl Chandra}; Black measured with {\sl Suzaku}).}
\label{fig:entropy} 
\end{center}
\end{figure*}

 \begin{figure*}
   \begin{center}
     \leavevmode 
\hspace{5mm}\subfloat{\includegraphics[width=0.48\textwidth]{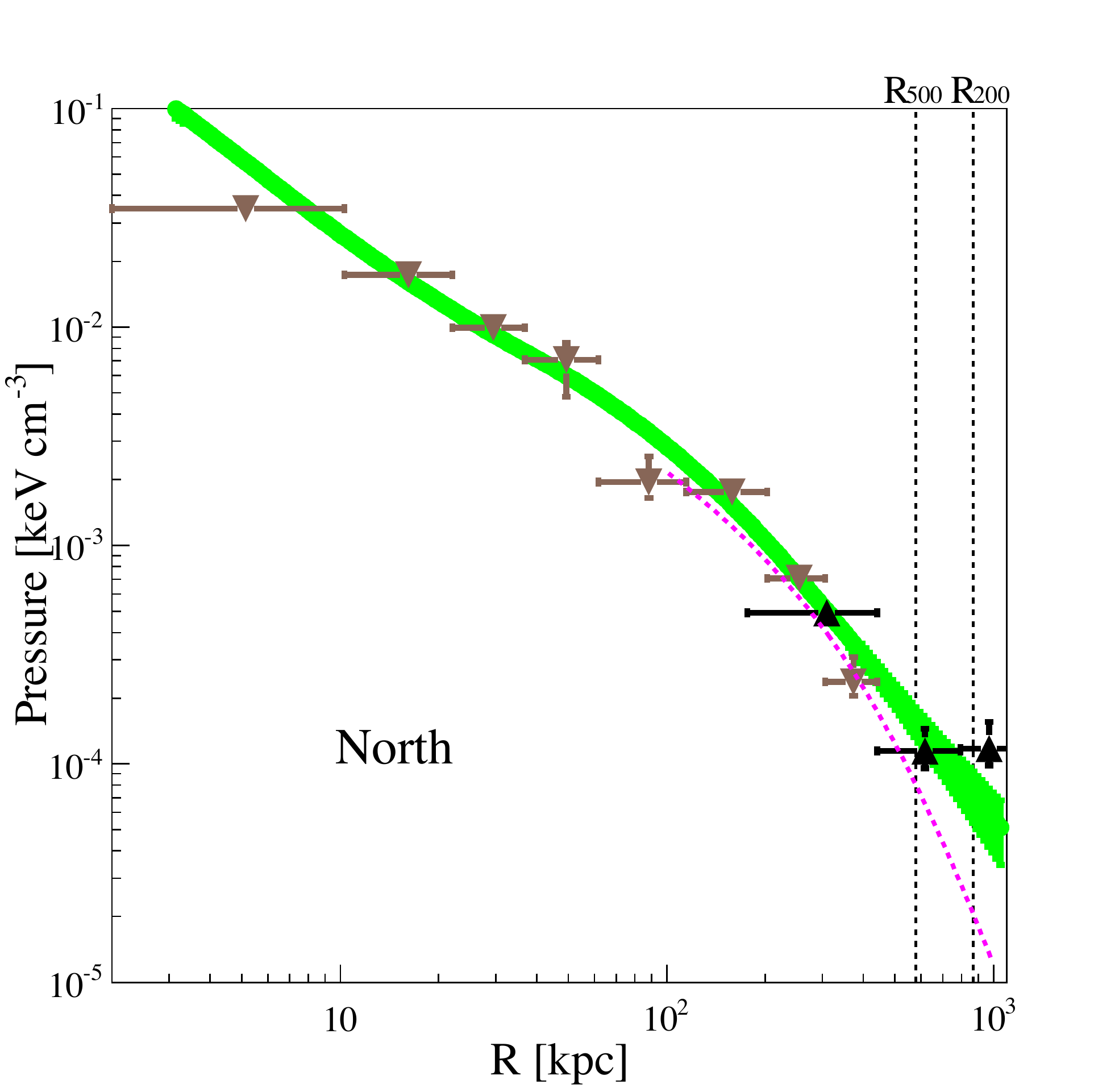}}\hspace{0mm}\subfloat{\includegraphics[width=0.48\textwidth]{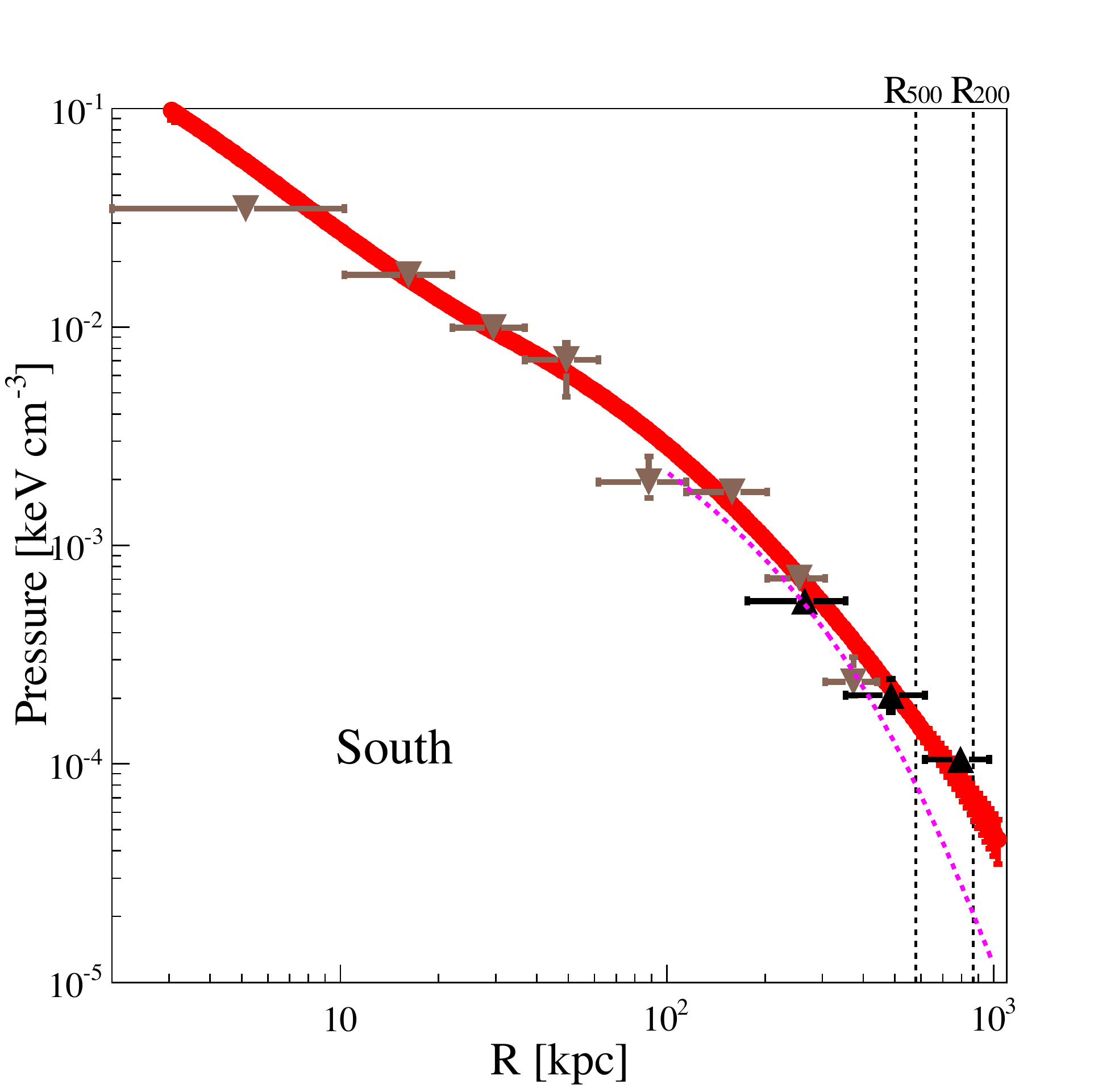}}\\
\vspace{-2.5mm}
\hspace{5mm}\subfloat{\includegraphics[width=0.48\textwidth]{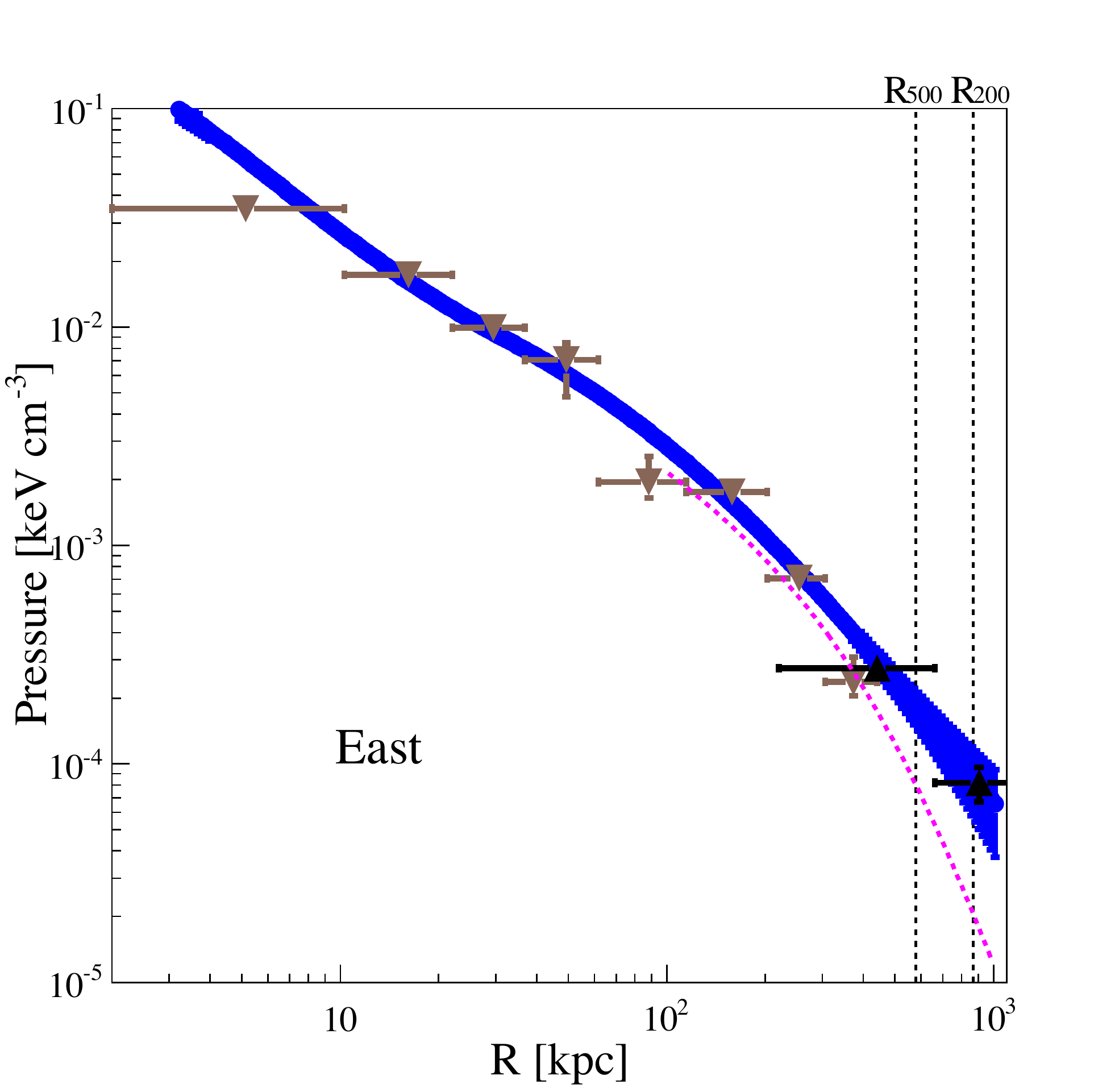}}\hspace{0mm}\subfloat{\includegraphics[width=0.48\textwidth]{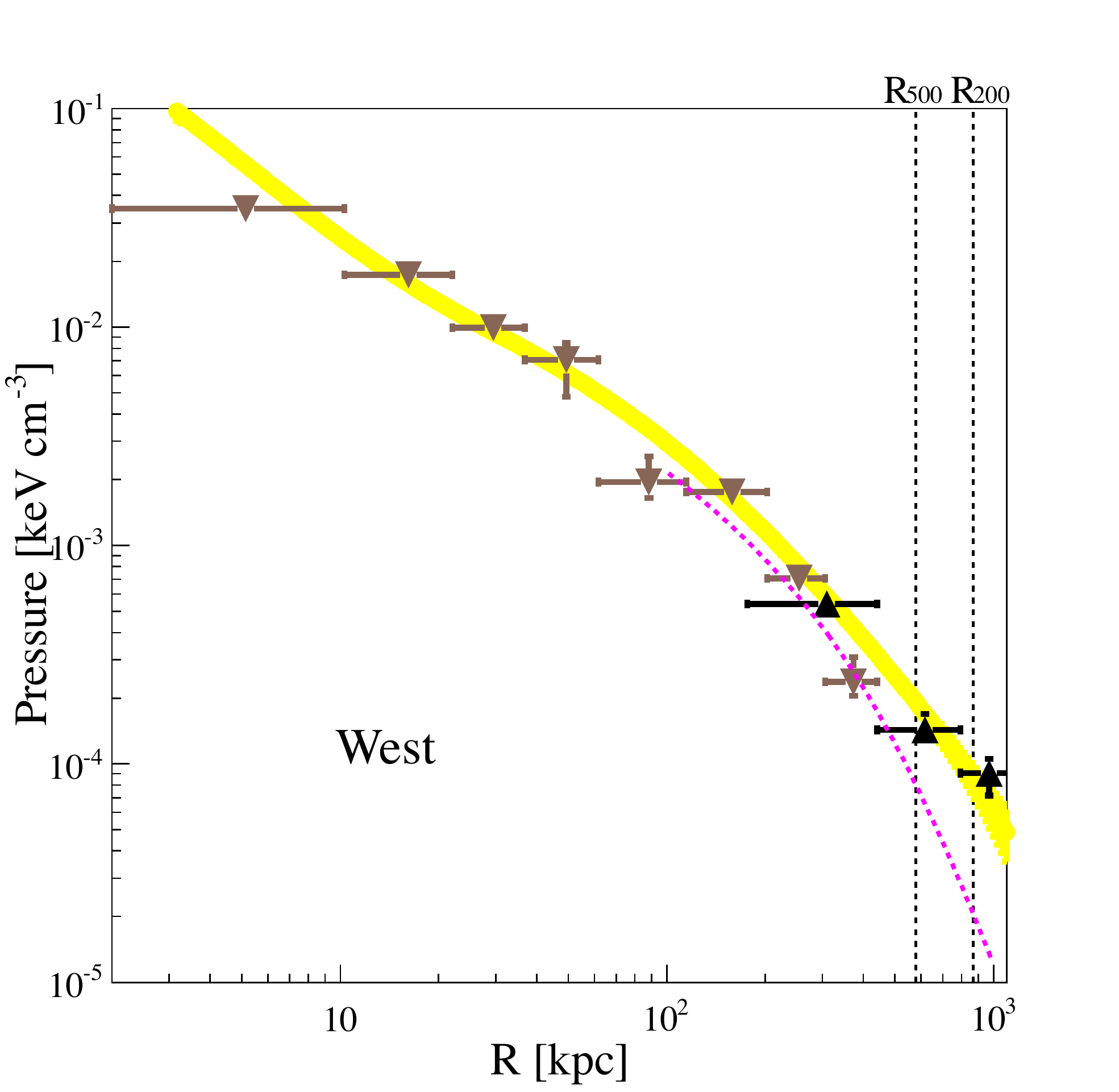}}\\
\caption{Solid lines: 3D pressure profiles derived with our primary method (entropy-based forward fitting) for each direction. Shaded regions indicate 1$\sigma$ uncertainties.
Magenta: universal pressure profile derived by Arnaud et al.\ (2010).
Triangles: 3D density profiles obtained  via spectral deprojection {\tt projct} (Brown measured with {\sl Chandra}; Black measured with {\sl Suzaku}).
}
\label{fig:pressure} 
\end{center}
\end{figure*}

We fitted the broken power-law model (equation 6) for the entropy to the data and enforced a monotonically rising profile (for convective stability) by requiring the slope parameters to be positive or zero. 
This model employed
a power-law component with two breaks and a constant component. The
normalizations of the power-law and constant components, the
radii of the breaks  and their slopes
were all free parameters. 
We plot the best-fit model of the 3D entropy profile for each direction in Figure~\ref{fig:entropy}.
Shaded regions indicate 1$\sigma$ confidence regions. We marked in magenta the ÒbaselineÓ predictions from gravitational structure formation (Voit et al. 2005), given by:
\begin{equation}
K_{\rm gra}(R)=1.32~K_{200}(R/R_{200})^{1.1},
\end{equation}
where the normalization, $K_{200}$, is given by 
$$K_{200}=362~\frac{GM_{200}\mu m_p}{2R_{200}} \nonumber \\
$$
$$
\left(\frac{1}{\rm keV}\right)\times\left[\frac{H(z)}{H_0}\right]^{-4/3}\left(\frac{\Omega_m}{0.3}\right)^{-4/3} ~{\rm keV ~cm^{-2}}.$$ 
The central entropy profile is more elevated and extended than this baseline $K_{\rm gra}$. 
The entropy profiles of the north and east directions beyond 0.5 $R_{\rm vir}$ are very well behaved, rising linearly all the way out to the $R_{\rm vir}$, following the $K\propto r^{1.1}$ expectation. 
The entropy profiles of the south and west directions have a flatter slope but stay consistent with the theoretical expectation near $R_{\rm vir}$.

Figure~\ref{fig:pressure} shows the three dimensional pressure profile of each direction.
We compared the observed pressure profile of each direction to 
 a semi-analytic universal pressure profile derived by Arnaud et al.\ (2010)
from comparison of their numerical simulations 
to {\sl XMM-Newton} observations of clusters within $R_{500}$.
This pressure profile is characterized as
\begin{equation}
P(r)=P_{500}\left[\frac{M_{500}}{1.3\times10^{14} h_{70}^{-1}M_{\odot}}\right]^{a_p+a^{\prime}_p
(x)} \nonumber \\
\end{equation}
\begin{equation}
\times\frac{P_0}{(c_{500}x)^{\gamma}[1+(c_{500}x)^{\alpha}]^{(\beta-\gamma)/\alpha}},
\end{equation}
where 
$$x=\frac{r}{R_{500}}; ~ ~ ~ ~ ~ a^{\prime}_p(x)=
0.10-(a_p+0.10)\frac{(x/0.5)^3}{[1+(x/0.5)^3]};$$ 
and $P_{500}$ and $M_{500}$ are respectively the pressure and total mass at $R_{500}$. 
Arnaud et al.\ (2010) adopted parameter values of 
$$[P_0, ~c_{500}, ~\gamma, ~\alpha, ~\beta] ~=
 \nonumber \\$$
$$
~ [8.403h_{70}^{-3/2}, ~1.177, ~0.3081, ~1.0510, ~5.4905].$$

The pressure profiles of the all direction of RXJ1159+5531 from 100 kpc out to about half $R_{500}$ are in good agreement with this universal profile. 
However, its pressure profiles exceed this universal profile at $R_{500}$ by 60\% to 200\%. Note that this universal profile is an average result of a large number of different clusters which have a 300\% internal discrepancy at $R_{500}$ itself. 
We do not expect the observed pressure profiles of RXJ1159+5531 to be in perfect agreement with this average profile.
At least, this result suggests that the pressure profiles of RXJ1159+5531 are not  significantly smaller than other systems and its outskirts are unlikely to be dominated by non-thermal pressure support. 
We compared the pressure profile of all directions in Figure~\ref{fig:rxj} (d). They have remarkably similar behaviors out to the large radii. 


\subsection{\sl Gas and baryon fractions}

To compute the gas and baryon mass, 
we included the stellar mass of the central galaxy, the 
gas mass, and the two additional baryon reservoirs: intracluster light and other member
galaxies. Vikhlinin et al. (1999) found that $\sim$25\%
of the $V$-band stellar light is associated with other member galaxies. We assumed a similar ratio in the $K$-band and adopted a $M_*/L_{\rm K}$ ratio of 1 for these galaxies.
Furthermore, we assume that the intracluster light within $R_{\rm vir}$ contains  
twice as much as the stellar mass of the central galaxy
(Purcell et al. 2007).

 \begin{figure*}
   \begin{center}
     \leavevmode 
\hspace{5mm}\subfloat{\includegraphics[width=0.48\textwidth]{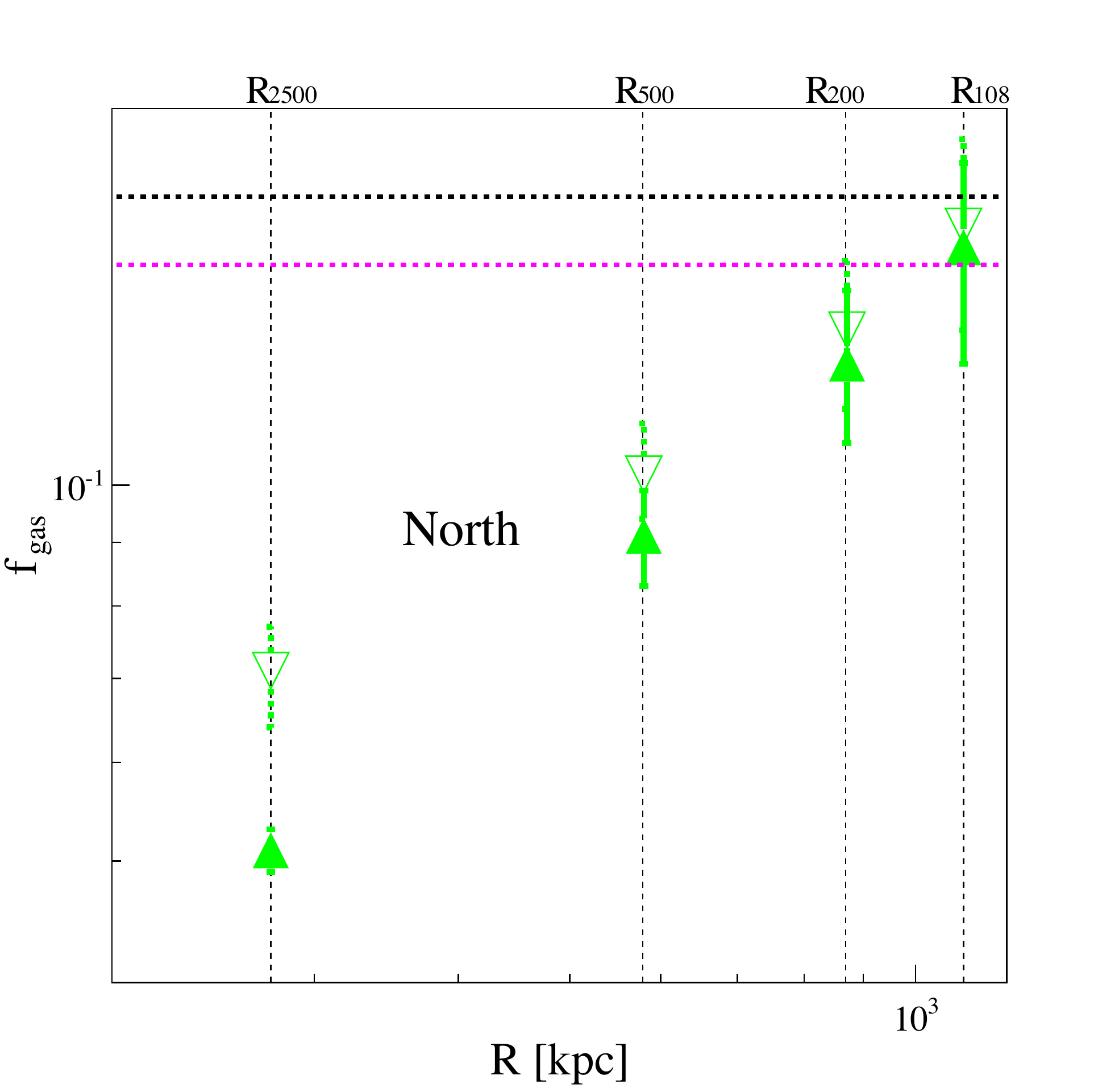}}\hspace{0mm}\subfloat{\includegraphics[width=0.48\textwidth]{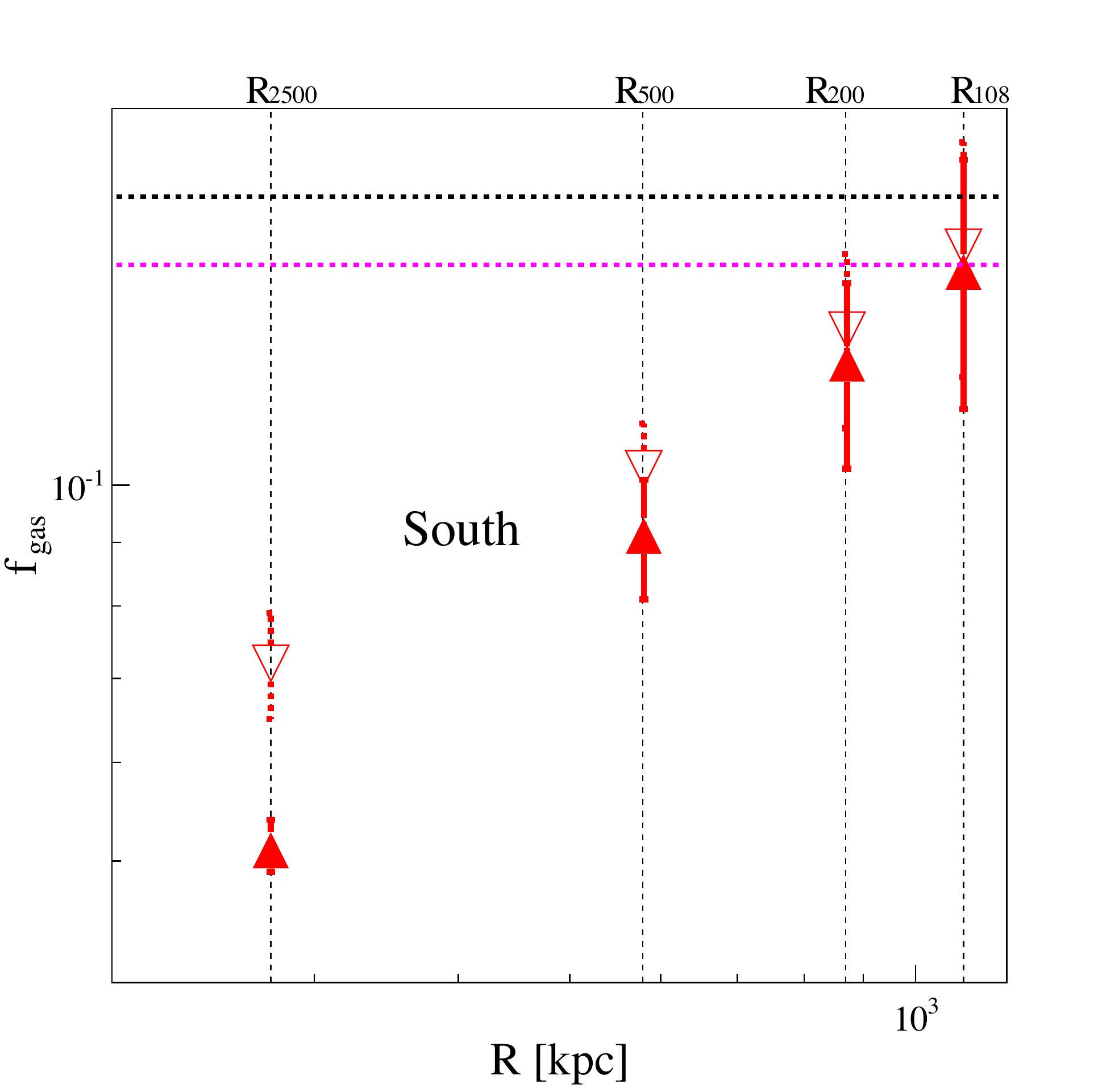}}\\
\vspace{-2.5mm}
\hspace{5mm}\subfloat{\includegraphics[width=0.48\textwidth]{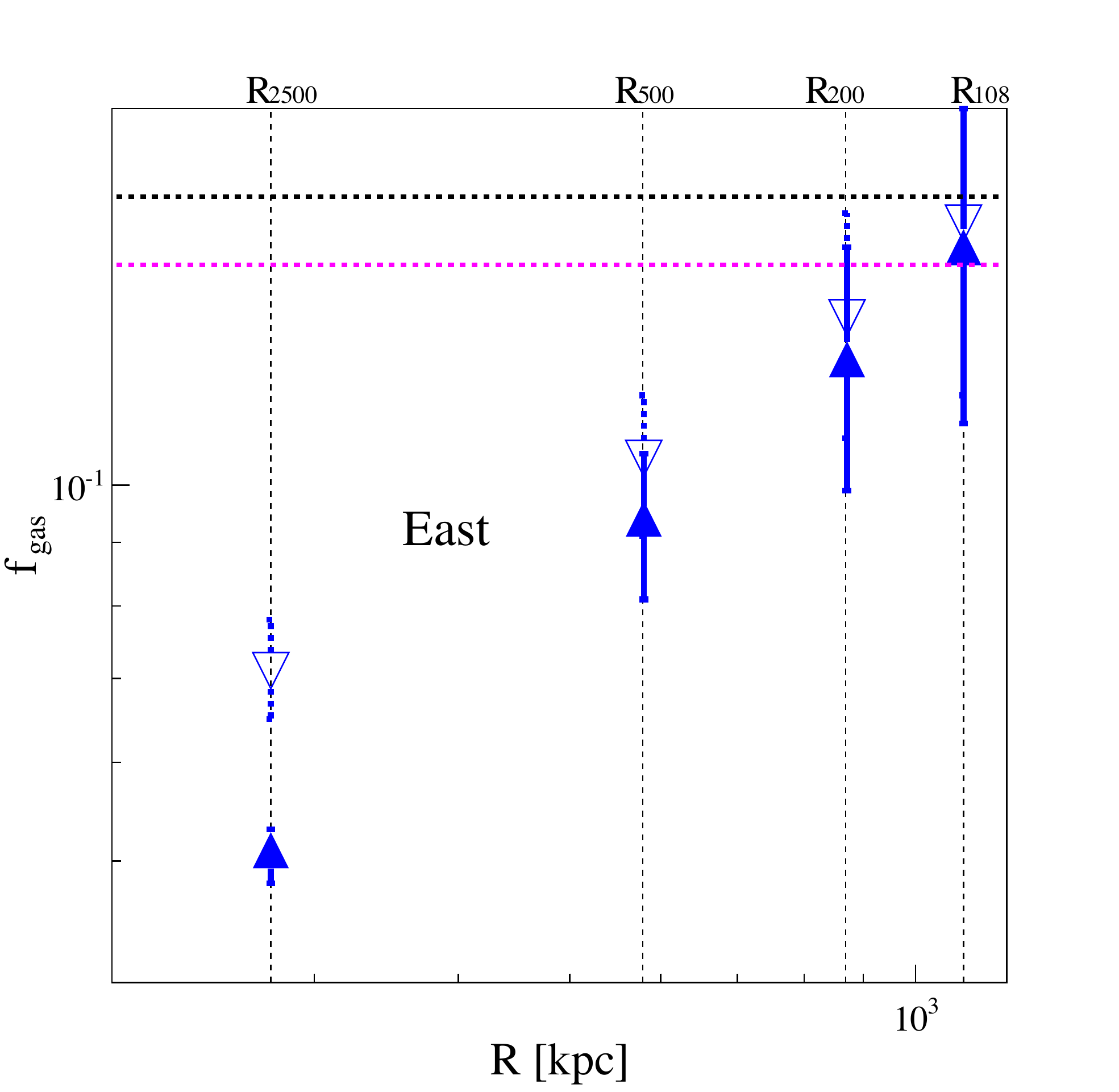}}\hspace{0mm}\subfloat{\includegraphics[width=0.48\textwidth]{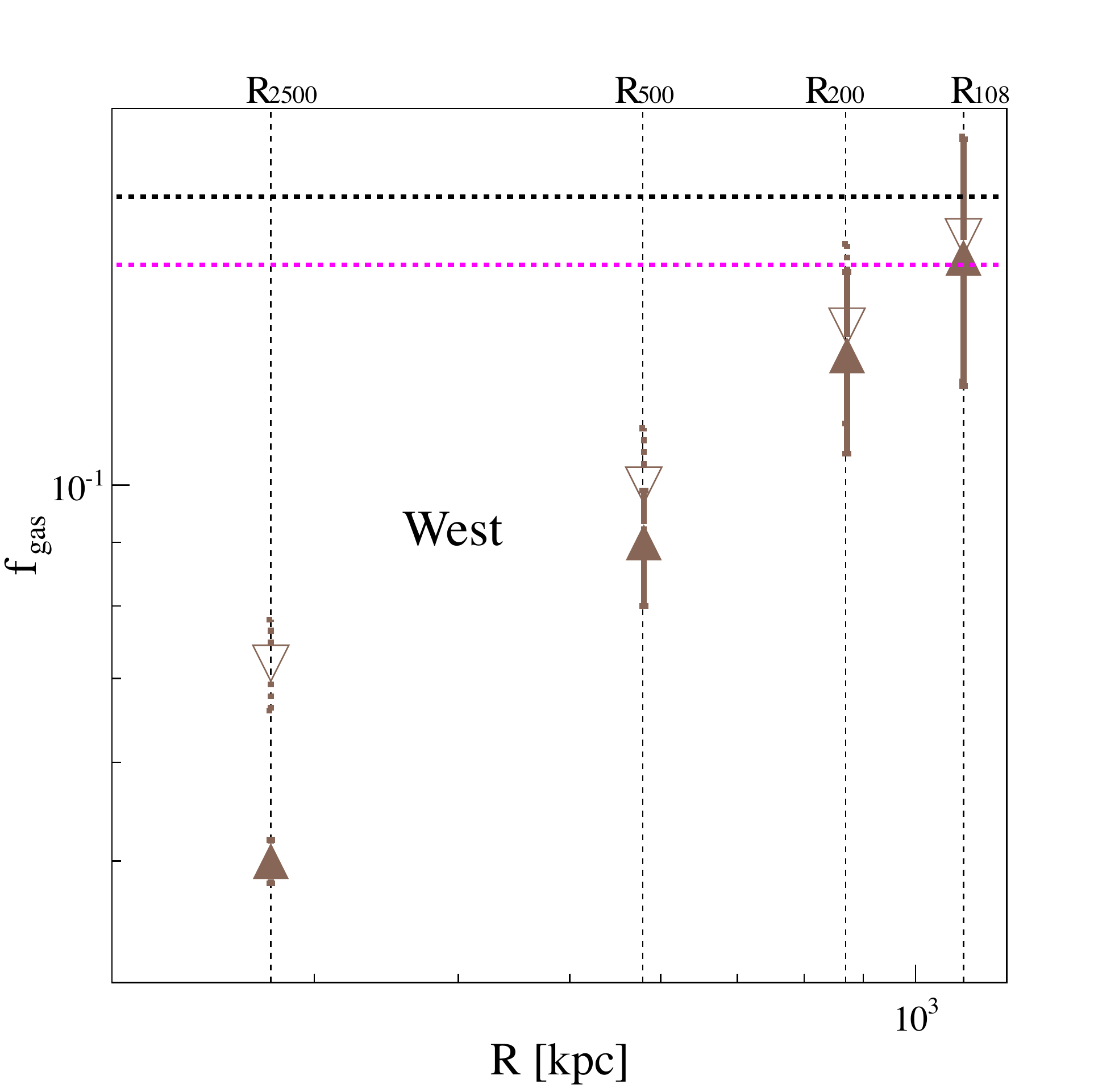}}\\
\caption{Solid triangles: enclosed gas mass fraction for each direction.  Open triangles: enclosed baryon fraction for each direction.
Horizontal magenta line: the cosmic value determined by {\sl Planck}. Horizontal black line: the cosmic value determined by WMAP.}
\label{fig:fgas} 
\end{center}
\end{figure*}

$R_{200}$ (871 kpc) is near the effective center of the outermost radial bin in all directions.
We confirm that we are able to measure the gas properties at $R_{200}$ with these data as listed in Table~3. The baryon fractions of each direction are consistent within uncertainties. 
$R_{\rm 108}$ is within the outermost bins of north, east, and west direction, and just outside that of the south direction. Based on our entropy-based Òforward-fittingÓ method, we can put constrain on its gas properties around $R_{\rm 108}$. 
The radial profiles of enclosed gas mass fraction and baryon fraction of each direction are shown in Figure~\ref{fig:fgas}.
We obtained total masses within $R_{\rm 108}$ of 9.72$\pm$1.44 (N), 9.91$\pm$1.74 (S), 9.66$\pm$2.10 (E), and 10.9$\pm$2.0 (W)$\times10^{13}$M$_{\odot}$ for each direction as described in \S3. 
The enclosed baryon fractions within $R_{\rm 108}$ are 
$0.161^{+0.028}_{-0.028}$ (N), $0.155^{+0.033}_{-0.033}$ (S), $0.162^{+0.045}_{-0.044}$ (E), and $0.158^{+0.032}_{-0.037}$ (W) for each direction. 
All of them are very consistent with the cosmic baryon fraction (0.15 Plank Collaboration 2013; 0.17 Komatsu et al.\ 2011). 


\section {\sl Systematic Error Budget}

We have constructed a detailed error budget considering a variety of possible systematic effects.
In Table~3, we list error budgets for the densities, temperatures, entropies and pressures measured at $R_{\rm 200}$, as well as enclosed baryon and gas fractions within $R_{\rm 200}$, for the north, south, west, and east directions.
We consider any systematic error to be significant once it equals or exceeds the statistical error of the same variable.
Our discussion pays special attention to the impact of each effect on the measurement of the entropy at $R_{\rm 200}$.

\subsection{Background}
In the outermost annulus, the 
background dominates the X-ray emission so that the ICM contributes only 8 (1)\%, 11 (1)\%, 35 (0.2)\%, and 6 (0.4)\% of the total emission (including the NXB) in 0.5--2.0 keV (2.0--8.0 keV) for north, south, east, and west directions, respectively. Consequently, our results depend on the determination of the various background components. To examine our sensitivity to the background, we first increased and decreased the level of particle background by 5\% since the variation of the XIS NXB is expected to be $\leq$3\% (Tawa et al.\ 2008). We list the impact of these variations in Table~3 (as $\Delta$NXB). The systematic uncertainty associated with variations in the particle background are smaller than the statistical uncertainty on the parameters of all directions. 

For {\sl Suzaku}, the most important and also the most uncertain component of the X-ray background is the CXB. 
To assess the systematic uncertainties associated with the CXB for each direction, we alternatively fixed the normalization of the CXB power law $\Gamma=1.41$ component in the model at the expected value according to Equation\,(1) in \S2.3 for each direction (see Table~3 $\Delta$CXB). These changes do not have significant impact on our results of entropy. 
Despite that the slope of the CXB power law component is well determined (e.g., De Luca et al.\ (2004): $\Gamma=1.41\pm0.06$--{\sl XMM-Newton};  Moretti et al.\ (2009): $\Gamma=1.47\pm0.07$--{\sl Swift}; Tozzi et al.\ (2001): $\Gamma=1.44\pm0.03$--{\sl Chandra}), we examined this uncertainty by fixing its slope at $\Gamma=1.5$ and $\Gamma=1.3$, respectively. 
The impact of using $\Gamma=1.3$ appears to be a small effect on our results, while using $\Gamma=1.5$ would increase the entropy at $R_{200}$ by $\sim30\%$ for the south and east pointings (see $\Delta$CXB-$\Gamma$ in Table~3).
The unresolved CXB component may have harder spectrum. 
Using {\sl Chandra} and {\sl Swift}, Moretti et al.\ (2012) determined that the slope of the unresolved CXB power law component can be as small as 0.1. We tested this effect by applying power law $\Gamma=0.1$ to model the remaining CXB component of the north pointing; its results are listed in Table~3 (as $\Delta$CXB-$\Gamma=0.1$). This variation causes a systematic error on the entropy at $R_{200}$ slightly larger than its statistical uncertainty. However, we consider $\Gamma=0.1$ for the unresolved CXB to be an extreme case and its justification requires more work in the future.

\subsection{SWCX}
\begin{figure*}
   \begin{center}
     \leavevmode 
\hspace{5mm}\subfloat{\includegraphics[width=0.48\textwidth]{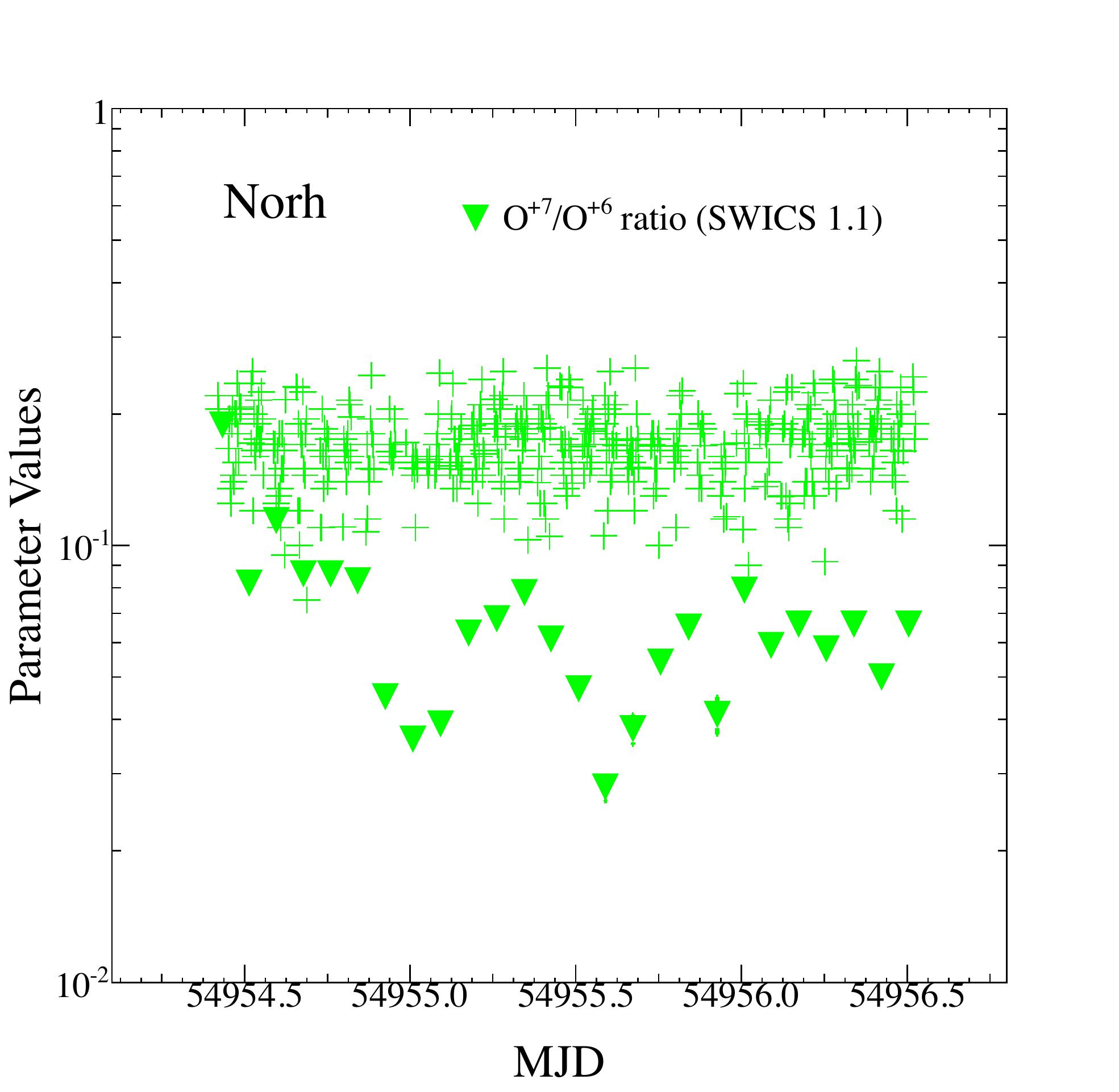}}\hspace{0mm}\subfloat{\includegraphics[width=0.48\textwidth]{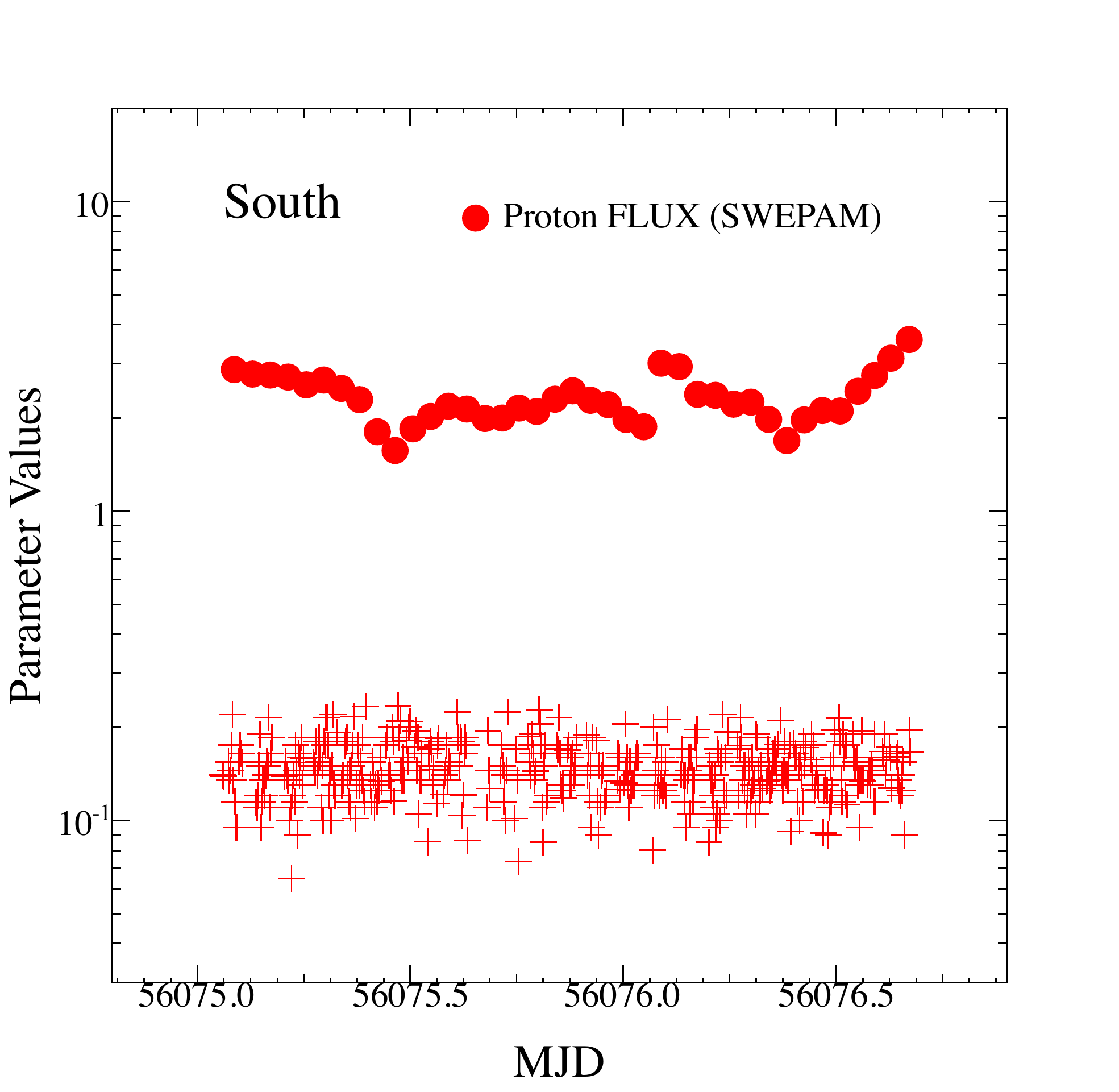}}\\
\vspace{-5mm}
\hspace{5mm}\subfloat{\includegraphics[width=0.48\textwidth]{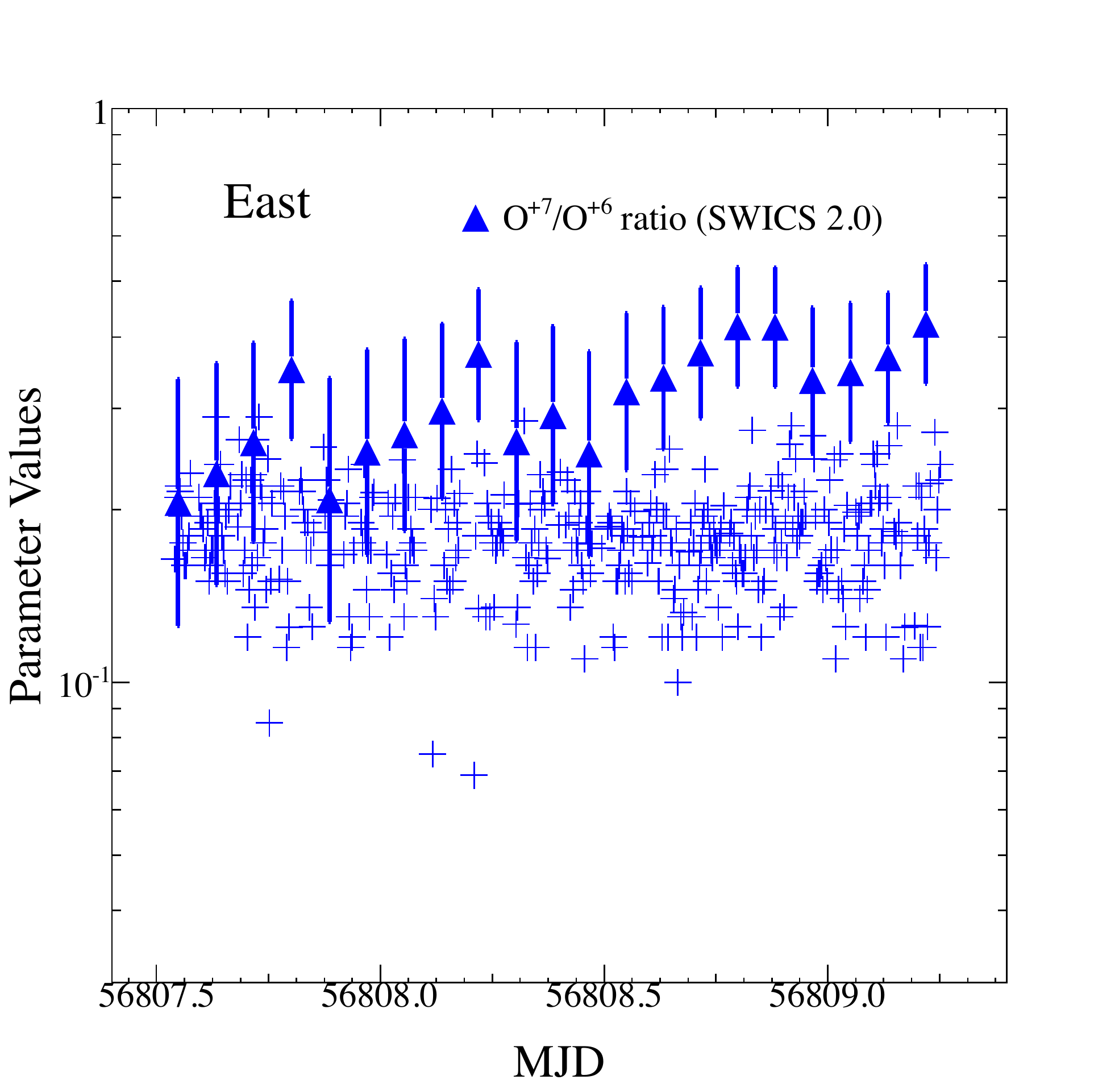}}\hspace{0mm}\subfloat{\includegraphics[width=0.48\textwidth]{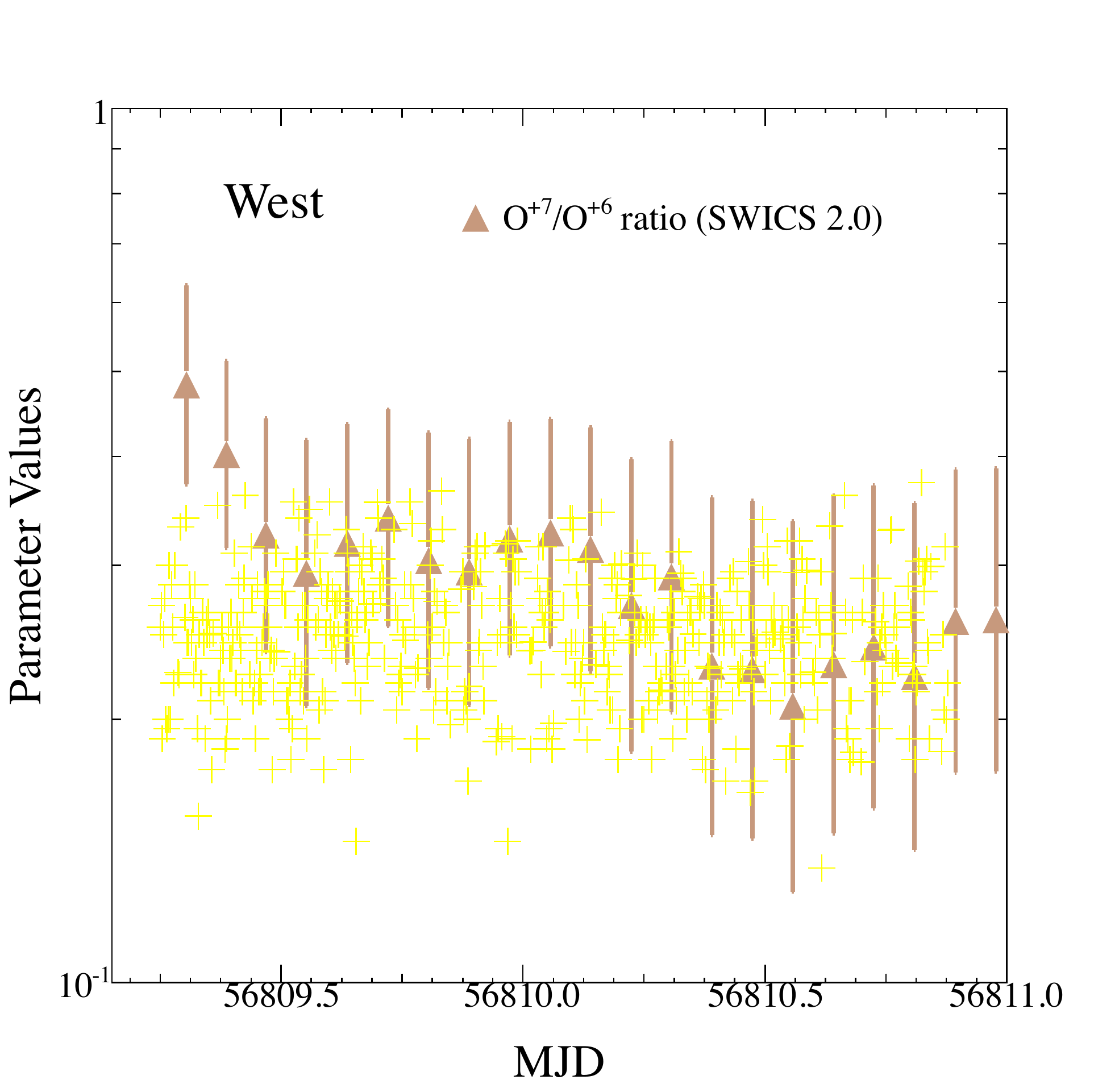}}\\
\caption{Crosses: {\sl Suzaku} XIS1 light curve in 0.5-2.0 keV energy band in each direction, in the unit of counts/s. {\it top-left}: triangles represent the light curve of O$^{+7}$$/$O$^{+6}$ ratio taken from ACE-SWICS (1.1) during {\sl Suzaku} observation of the north direction. {\it top-right}: circles represent the light curve of proton flux taken from ACE-SWEPAM during {\sl Suzaku} observation of the south direction, in the unit of 10$^{8}$ cm$^{-2}$\,s$^{-1}$. {\it bottom-left}: triangles represent the light curve of the O$^{+7}$$/$O$^{+6}$ ratio taken from ACE-SWICS (2.0) during {\sl Suzaku} observation of the east direction. {\it bottom-right}: same as {\it bottom-left} but for the west pointing. ACE data corrected for travel time to Earth.
[{\sl see the electronic edition of the journal for a color version of this figure.}]} 
\label{fig:solar} 
\end{center}
\end{figure*}

In order to probe the effect of the activity of the Solar Wind Charge eXchange (SWCX), we compared the light curve of {\sl Suzaku} observations in the soft band (0.5-2.0 keV) to the data taken from the {\sl Advanced Composition Explorer} (ACE) during these observations as shown in Figure~\ref{fig:solar}. For the south direction, we managed to obtain the light curve of proton fluxes taken from the {\sl Solar Wind Electron Proton Alpha Monitor} (SWEPAM) onboard ACE. Its variation ranges within 1.5--3.6$\times 10^{8}$\,cm$^{-2}$\,s$^{-1}$.  According to Snowden et al.\ (2004), a proton flux of $\lesssim$\,4$\times 10^{8}$\,cm$^{-2}$\,s$^{-1}$ is considered to be a quiescent level. Unfortunately, the SWEPAM data during the {\sl Suzaku} observations of the other three pointings were indicated to be bad or missing. Instead, we inspected the O$^{+7}$$/$O$^{+6}$ ratio taken from the {\sl Solar Wind Ion Composition Spectrometer} (SWICS) onboard ACE. Their light curves lack obvious variations and their levels ($\lesssim$\,0.3) are consistent with the expectations of quiescent emission interval (Snowden et al.\ 2004).  
Over all, the activity of SWCX
was inferred to be low during these {\sl Suzaku} observations. 
Still, we added a few gaussian lines to model the SWCX component as part of the X-ray background component, which contains 6 gaussian lines at energies of 0.46 keV, 0.56 keV, 0.65 keV, 0.81 keV, 0.91 keV, and 1.34 keV (Snowden et al.\ 2004). We compared our best-fit when 
including and excluding this component in the model in Table~3.
The differences are typically much smaller than the statistical uncertainty on those parameters.

\subsection{PSF}

The large, energy dependent PSF of  {\sl Suzaku} causes some photons from the central part of clusters to be scattered by the optics out to radii of $\sim$\,10$^{\prime}$. In our analysis, we accounted for this effect by including spectral mixing between each annulus in the model. 
To explore how sensitive our results are to the mixing level,
we experimented with adjusting the amount of light that
is scattered into each annulus by $\pm$5\%. This did not appreciably
affect the results of the north and west directions, but it did have significant impact on the east and south directions  (change the entropy at $R_{200}$ by $\pm$35\% and $\pm$38\% respectively), as shown in Table~3.

\subsection{Entropy model}

The entropy profiles we calculated assume a broken power-law model consisting of two breaks and a constant. This model has more freedom than the one used in Humphrey et al.\ (2012) which consists of only one break. In order to test how sensitive our result is to the chosen entropy model, we added one more break in the model of entropy profile. 
We listed the corresponding results in Table~3 ($\Delta$model). This effect causes little impact on our results of the east, south, and west directions but it gives a higher entropy value for the north direction.

\subsection{Solar abundance table}
Plasma emission is a strong function of metallicity for low temperature systems (particularly for kT $\lesssim$ 1 keV). Adopting an accurate solar abundance table is crucial for determining both the thermal properties and the metal abundance of hot gas (e.g., Su \& Irwin 2013). 
The discrepancy in measurements, especially in thermal properties, caused by using different solar abundance tables have been largely underappreciated. 
Comparing the systematic uncertainty caused by using different solar abundance tables allow us to compare our results to previous studies.    
In our analysis, we adopt Asplund et al.\ (2006) as the abundance table for the plasma emission model.  
We performed our fit again using one of the most outdated but widely used solar abundance table Anders \& Grevesse (1989). We compared these two sets of results in Table~3 ($\Delta$solar).  The discrepancy in the entropy values is quite large for all pointings. 
RXJ1159+5531 is a low mass cluster for which the contribution of metal line emissions to the X-ray emission is more significant than more massive clusters. Thus its gas properties are more sensitive to the choice of abundance table.  
Our test shows that it is important to use updated solar abundance tables and caution needs to be taken when comparing recent studies with previous studies, at least for the outermost virial region of low mass clusters where kT $\sim$ 1keV. 
We did the same exercise by using another commonly used solar abundance table Lodders (2003); the results are also listed in Table~3. Its deviation from our best-fit is similar compared with using Anders \& Grevesse (1989).

 \begin{figure}
 \epsscale{1.2}
        \centering
        \plotone{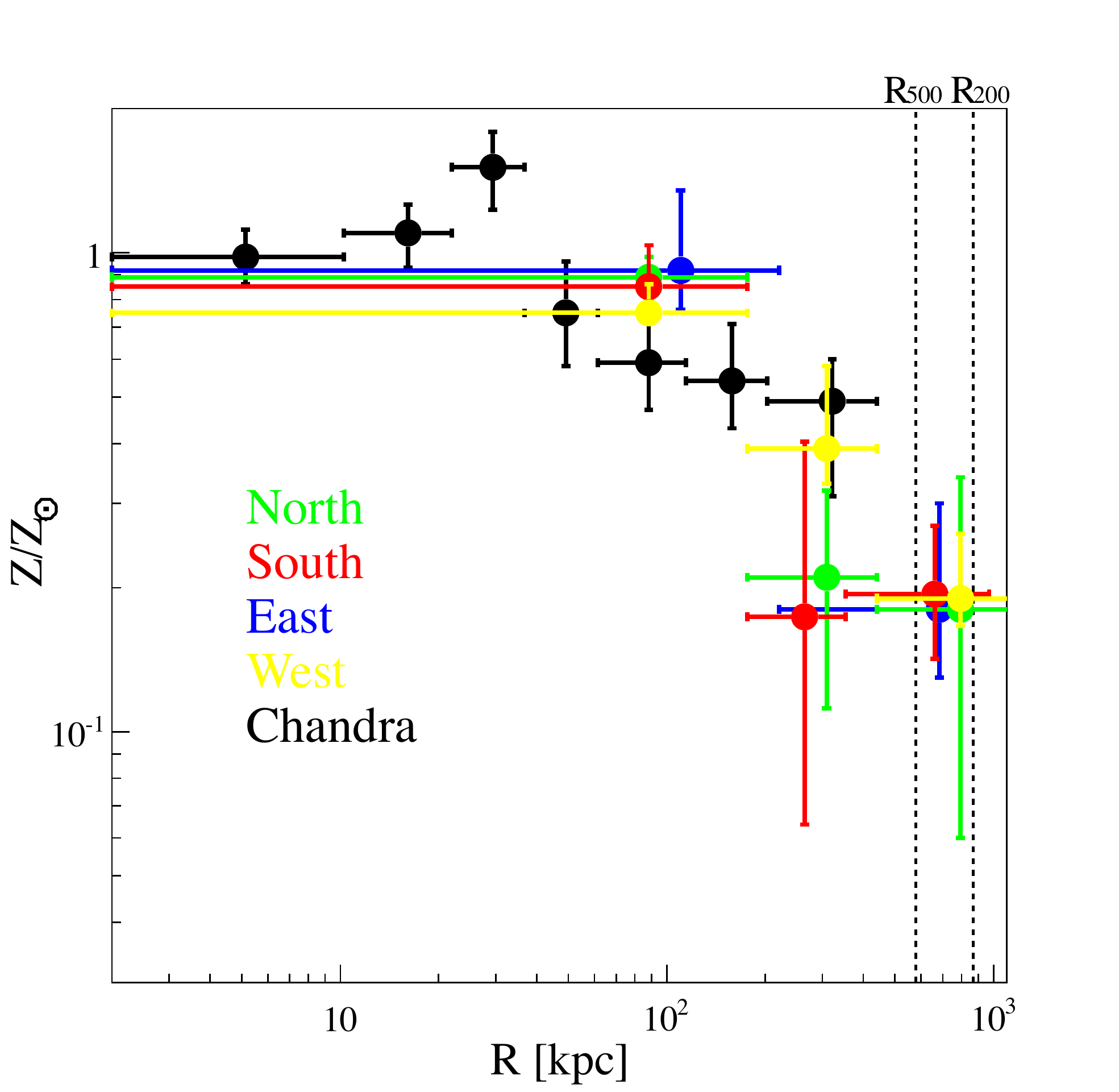}
\figcaption{\label{fig:a} The Fe abundance profile of each direction.}
\end{figure}

The Fe abundance profile of RXJ1159+5531 of each direction are shown in Figure~\ref{fig:a}.
The central dip of the metallicity profile observed by {\sl Chandra} is due to the ``Fe-bias" (Buote 2000) in the central region, as has also been noted and discussed in Humphrey et al.\ (2012).
We obtained a similar best-fit hot gas metallicity of $\sim 0.2$\,Z$_{\odot}$ in the spectral analysis for the outermost two annuli in each direction (the ICM metallicity of the outermost annulus was tied to their adjacent inner annulus). 
We examined the impact of the metallicity determination on our results by fixing the 
hot gas metallicity of the outermost two annuli at 0.1\,Z$_{\odot}$ and 0.3\,Z$_{\odot}$ respectively. Using 0.3\,Z$_{\odot}$ typically has little impact on our results while using 0.1\,Z$_{\odot}$ would reduce the entropy of the north and south direction by 30\%, as shown in Table~3 ($\Delta$abun).

\subsection{nH and distance}

To test the sensitivity of our result to the absorbed Galactic $N_{\rm H}$ value.
We varied nH from the Dickey and Lockman (1990) map by 20\%. This did not appreciably
affect our conclusions except for the east direction as shown in Table~3. 
We also examined the error associated with distance uncertainty. 
We varied the value of redshift in the model by 5\%. The impact of this latter factor is always very small with respect to the statistical errors. 

\subsection{FI-BI}

Although there have not been detailed calibration works on the level of consistency between different XIS chips that we are aware of, 
we allow the normalizations of the XIS1 spectra and that of XIS0 and XIS3 spectra to vary independently in the fit in order to account for the discrepancies between the response of FI-chips and BI-chips (we assume the same normalization for the XIS0 and XIS3). 
We found a $\sim$10\% difference in the normalizations determined by {\sl Suzaku} FI-chips and BI-chips for RXJ1159+5531.
In order to test the significance of this discrepancy, we also performed our fit by linking their normalizations. 
The difference it made is smaller than the statistical uncertainty in the value of entropy. Details are listed in Table~3 ($\Delta$FI-BI).

\subsection{Techniques}

In order to assess the sensitivity of our results to the forward-fitting analysis of the projected temperature and density profiles, we compared our results to those obtained using the well-known spectral deprojection model {\tt projct} in {\sl Xspec}.  
The 3D density and temperature profiles obtained with both methods are compared in Figures~\ref{fig:density} and ~\ref{fig:temperature}. 
Using the 3D temperature and density profiles derived with the {\tt projct} method, we also calculated their corresponding entropy and pressure profiles which we also show in Figure~\ref{fig:entropy} and Figure~\ref{fig:pressure}. This procedure (similar to the classic Òonion peelÓ) assumes constant ICM emission in each three-dimensional shell and no emission outside the bounding shell/annulus.

Both methods have their pros and cons (for a more detailed comparison see Buote \& Humphrey 2012). 
A key advantage of spectral deprojection with {\tt projct} is that no parameterized model for the radial ICM properties (e.g., entropy) is required nor is hydrostatic equilibrium assumed.
On the other hand, 
the Òforward-fittingÓ method allows us to self-consistently account for ICM emission beyond $R_{\rm vir}$ (we evaluate the hydrostatic model, and hence the entropy profile, out to 2\,$R_{\rm vir}$) while the method using {\tt projct} may over subtract the background component by assuming there is no emission outside of the outmost extracted region. 
Moreover, 
by projecting our 3D model onto the sky, the Òforward-fittingÓ method allows us to correctly account for the radial variation of the spectral properties within each bin. Using {\tt projct} or the standard onion-peeling method has to assume the spectral properties are uniform in the entire bin, which can be a substantial error for very large apertures such as used for {\sl Suzaku} analysis. 
We find that the three-dimensional radial profiles of the ICM temperature, density, entropy, and pressure in each direction obtained by both methods are consistent with each other as demonstrated in Figures~\ref{fig:density} -- \ref{fig:pressure}.

\begin{table*}
\caption{Gas properties at $R_{200}$ with their systematic budgets}
  \begin{center}
    \leavevmode
 \begin{tabular}{llcccccc} \hline \hline  
\colhead{}&\colhead{}&\colhead{Temperature}&\colhead{Density}&\colhead{Entropy}&\colhead{Pressure}&\colhead{$f_{\rm gas}$}&\colhead{$f_{\rm b}$}\\
\colhead{}&\colhead{}&\colhead{(keV)}&\colhead{(10$^{-5}$\,cm$^{-3}$)}&\colhead{(keV\,cm$^2$)}&\colhead{(10$^{-5}$\,keV\,cm$^{-3}$)}&\colhead{}&\colhead{}  \\ \hline 
\multirow{14}{*}{North}&best-fit&$1.21\pm0.17$&5.98$\pm1.20$&800$\pm145$&7.29$\pm1.90$&0.125$^{+0.018}_{-0.017}$&0.133$^{+0.018}_{-0.018}$\\
\hline
&$\Delta$NXB&$\pm0.06$&$\pm0.60$&$\pm94$&$\pm0.46$&$\pm0.002$&$\pm0.003$\\
&$\Delta$CXB&$-0.05$&$+0.20$&$-55$&$+0.12$&$+0.010$&$+0.011$\\
&$\Delta$CXB-$\Gamma$&$+0.05,-0.10$&$-1.18,+0.62$&$+135,-115$&$-0.94,+0.10$&$-0.017,+0.003$&$-0.019,+0.003$\\
&$\Delta$CXB-$\Gamma$\,(Extreme)&$+0.13$&$+4.02$&$-180$&$+3.01$&$+0.023$&$+0.024$\\
&$\Delta$SWCX&$+0.08$&$-0.02$&$+48$&$+0.39$&$+0.003$&$+0.002$\\
&$\Delta$PSF&$\pm0.05$&$\pm1.03$&$\pm63$&$\pm1.55$&$\pm0.008$&$\pm0.006$\\
&$\Delta$model&$-0.015$&$-0.79$&$+62$&$-1.09$&$-0.001$&$-0.003$\\
&$\Delta$solar&$-0.012, +0.11$&$-0.26, -1.12$&$+9.6, +194$&$-0.44, -0.87$&$-0.005, -0.009$&$-0.004, -0.009$\\
&$\Delta$abun&$-0.40, -0.01$&$-0.20, -0.88$&$-256, +75$&$-2.60, -1.17$&$-0.002, -0.016$&$-0.001, -0.017$\\
&$\Delta$nH&$\pm0.09$&$\pm0.26$&$\pm86$&$\pm0.30$&$\pm0.04$&$\pm0.03$\\
&$\Delta$Distance&$\pm0.01$&$\pm0.15$&$\pm1$&$\pm0.29$&$\pm0.001$&$\pm0$\\
&$\Delta$FI-BI&$-0.02$&$-0.44$&$+22$&$-0.67$&$+0.002$&$+0.001$\\
\hline
\multirow{13}{*}{South}&best-fit&$0.98\pm0.11$&6.77$\pm0.78$&591$\pm90$&6.59$\pm1.1$&0.125$^{+0.020}_{-0.022}$&0.133$^{+0.020}_{-0.022}$\\
\hline
&$\Delta$NXB&$\pm0.09$&$\pm0.87$&$\pm2.37$&$\pm1.44$&$\pm0.002$&$\pm0.003$\\
&$\Delta$CXB&$-0.21$&-1.5&$-43$&-2.62&$-0.006$&$-0.008$\\
&$\Delta$CXB-$\Gamma$&$+0.28,+0.12$&$-1.07,+0.57$&$+244,+41$&$+0.40,+1.48$&$-0.017,+0.008$&$-0.017,+0.008$\\
&$\Delta$SWCX&$+0.09$&$+0.19$&$+67$&$+0.34$&$+0.005$&$+0.004$\\
&$\Delta$PSF&$\pm0.12$&$\pm1.89$&$\pm227$&$\pm1.2$&$\pm0.015$&$\pm0.015$\\
&$\Delta$model&$-0.01$&$+0.66$&$-49$&$+0.64$&$+0.001$&$+0.001$\\
&$\Delta$solar&$+0.27, -0.01$&$-1.38, +0.79$&$+287, -50$&$+0.02, +0.61$&$-0.007, -0.009$&$-0.007, -0.009$\\
&$\Delta$abun&$-0.27, +0.06$&$+0.97, +0.13$&$-202, +26$&$-1.2, +0.45$&$-0.002, -0.017$&$-0.002, -0.018$\\
&$\Delta$nH&$\pm0.065$&$\pm1.08$&$\pm114$&$\pm0.78$&$\pm0.003$&$\pm0.004$\\
&$\Delta$Distance&$\pm0.125$&$\pm1.12$&$\pm159$&$\pm0.49$&$\pm0.003$&$\pm0.003$\\
&$\Delta$FI-BI&$+0.005$&$-0.02$&$+2$&$-0.07$&$+0.002$&$+0.002$\\
\hline
\multirow{13}{*}{East}&best-fit&$1.13\pm0.19$&7.45$\pm1.44$&643$\pm129$&8.43$\pm3.00$&0.126$^{+0.029}_{-0.027}$&0.136$^{+0.029}_{-0.027}$\\
\hline
&$\Delta$NXB&$\pm0.11$&$\pm1.39$&$\pm127$&$\pm0.58$&$\pm{0.018}$&$\pm{0.018}$\\
&$\Delta$CXB&$+0.06$&$+0.83$&$-15$&$+1.4$&$+0.007$&$+0.004$\\
&$\Delta$CXB-$\Gamma$&$+0.55,-0.01$&$+1.51,+2.41$&$+187,-110$&$+6.35,+2.72$&$+0.005,+0.019$&$+0.006,+0.015$\\
&$\Delta$SWCX&$+0.07$&$-0.42$&$+13$&$+1.00$&$0$&$-0.002$\\
&$\Delta$PSF&$\pm0.50$&$\pm0.75$&$\pm223$&$\pm4.94$&$\pm0.012$&$\pm0.013$\\
&$\Delta$model&$0$&$-0.64$&$-37$&$+0.71$&$-0.004$&$-0.002$\\
&$\Delta$solar&$-0.15, +0.08$&$+2.14, -1.28$&$-174, +134$&$+0.97, -0.96$&$+0.020,-0.011$&$+0.021,-0.012$\\
&$\Delta$abun&$-0.06, -0.11$&$-0.83, -1.99$&$+12, +68$&$-1.34, -2.86$&$-0.004, -0.023$&$-0.005, -0.023$\\
&$\Delta$nH&$\pm0.44$&$\pm1.0$&$\pm175$&$\pm4.83$&$\pm0.007$&$\pm0.004$\\
&$\Delta$Distance&$\pm0.04$&$\pm0.5$&$\pm52$&$\pm0.24$&$\pm0.005$&$\pm0.004$\\
&$\Delta$FI-BI&+0.23&+1.43&+42&+3.65&+0.020&+0.020\\
\hline
\multirow{13}{*}{West}&best-fit&$1.11\pm0.12$&7.41$\pm0.74$&634$\pm88$&8.20$\pm1.30$&0.127$^{+0.021}_{-0.021}$&0.134$^{+0.022}_{-0.022}$\\
\hline
&$\Delta$NXB&$\pm0.005$&$\pm1.51$&$\pm78$&$\pm1.6$&$\pm0.006$&$\pm0.007$\\
&$\Delta$CXB&$-0.11$&$-0.64$&$-30$&$-1.43$&$-0.008$&$-0.008$\\
&$\Delta$CXB-$\Gamma$&$-0.16,+0.03$&$-0.65,+1.48$&$-56,-51$&$-1.81,+1.80$&$-0.004,+0.011$&$-0.003,+0.012$\\
&$\Delta$SWCX&$+0.06$&$+0.49$&$-62$&$+0.10$&$+0.010$&$+0.013$\\
&$\Delta$PSF&$\pm0.07$&$\pm0.06$&$\pm45$&$\pm0.40$&$\pm0.005$&$\pm0.004$\\
&$\Delta$model&$-0.07$&$+1.66$&$-117$&$+1.23$&$-0.003$&$-0.003$\\
&$\Delta$solar&$+0.16, +0.04$&$+0.04, +1.07$&$+85, -36$&$-1.26, +1.55$&$+0.005, +0.017$&$+0.009, +0.018$\\
&$\Delta$abun&$+0.18, +0.22$&$+3.37, +0.37$&$-63, +98$&$+5.71, +2.14$&$+0.016, -0.004$&$+0.017, -0.002$\\
&$\Delta$nH&$\pm0.11$&$\pm0.02$&$\pm66$&$\pm0.77$&$\pm0.002$&$\pm0$\\
&$\Delta$Distance&$\pm0.11$&$\pm0.04$&$\pm67$&$\pm0.75$&$\pm0.002$&$\pm0.002$\\
&$\Delta$FI-BI&$-0.07$&$+0.69$&$-77$&$+0.22$&$+0.011$&$+0.013$\\ \hline
    \end{tabular}
  \end{center}
($\Delta$NXB): vary the particle background component by 5\%.\\
($\Delta$CXB): fix the normalization of power law $\Gamma=1.4$ at the expected value instead of letting it free to vary in the spectral fitting.\\
($\Delta$CXB-$\Gamma$): fix the slope of power law at $\Gamma=1.5$ and $\Gamma=1.3$, respectively, for the CXB component.\\
($\Delta$CXB-$\Gamma$ (Extreme)): fix the slope of power law at $\Gamma=0.1$ for the unresolved CXB component for the north pointing.\\
($\Delta$SWCX): include solar wind charge exchange components in the model.\\
($\Delta$PSF): vary the mixing photons between each annulus by 5\%.\\
($\Delta$model): using a power law with three breaks (instead of two breaks) to model entropy profile.\\
($\Delta$solar): using solar abundance table Anders \& Grevesse (1989) and Lodders (2003) respectively.\\
($\Delta$abun): fix the metal abundance of the outermost bin at 0.1$Z_{\odot}$ and 0.3$Z_{\odot}$ respectively.\\
($\Delta$nH): vary the hydrogen column density by 20\%.\\
($\Delta$Distance): vary the redshift parameter by 5\%.\\
($\Delta$FI-BI): do not consider the difference in the responses between FI-CCD (XIS0 and XIS3) and BI-CCD (XIS1). 
\end{table*}

\section {\sl Discussion}

 \begin{figure}
 \epsscale{1.2}
        \centering
        \plotone{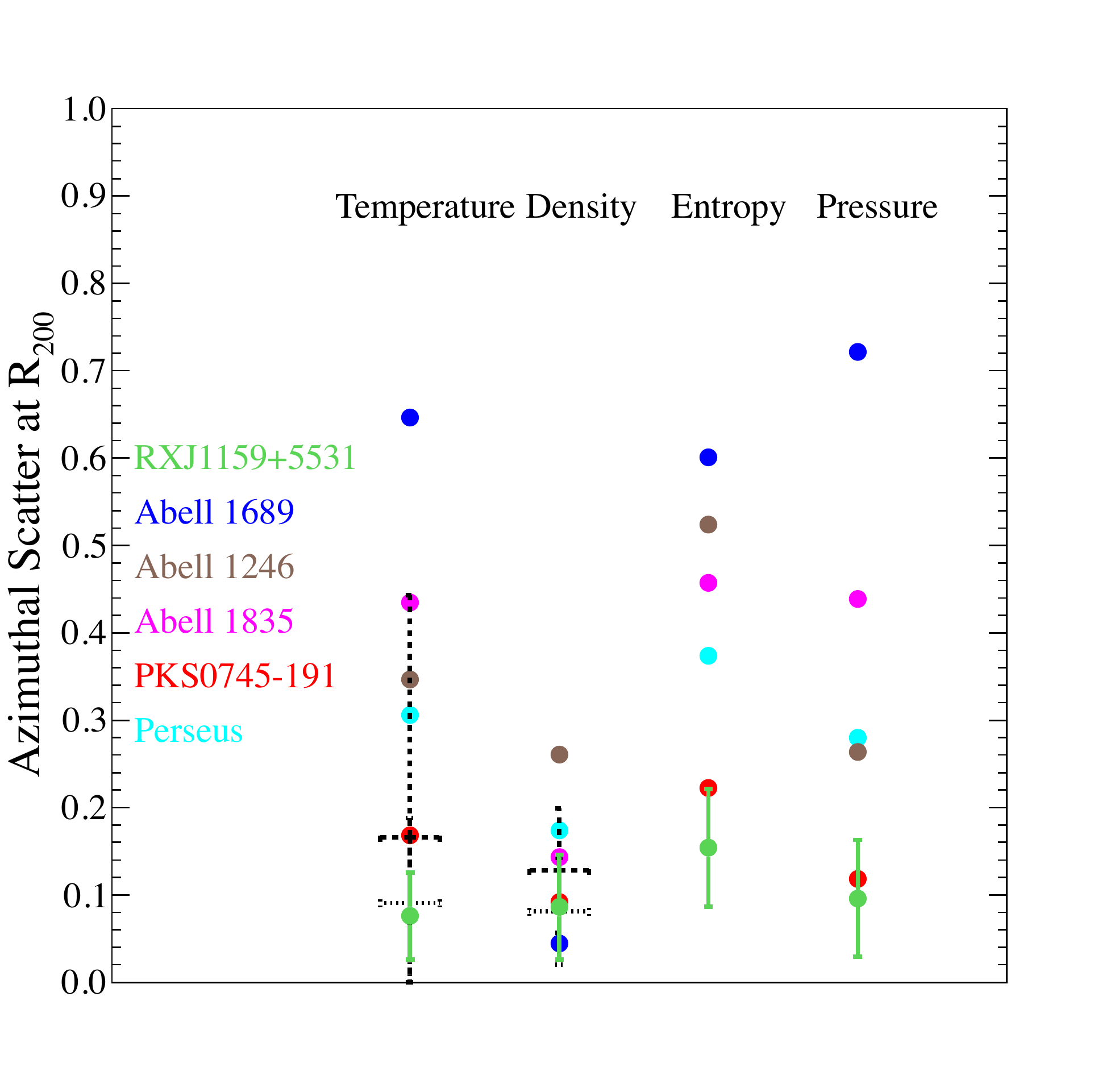}
\figcaption{\label{fig:scatter} Azimuthal scatters in temperature, density, entropy, and pressure calculated from eq. (10) for clusters studied in at least four directions with {\sl Suzaku}. The results of other clusters are taken from the literature: Abell~1689 (Kawaharada et al.\ 2010), Abell~1246 (Sato et al.\ 2014), Abell~1835 (Ichikawa et al.\ 2013), PKS0745-191 (George et al.\ 2009), and the Perseus Cluster (Urban et al.\ 2014). Solid black cross: azimuthal scatter of simulated relaxed clusters; Dashed black cross: azimuthal scatter of simulated disturbed clusters (Vazza et al.\ 2011).}
\end{figure}

An elliptical cluster in hydrostatic equilibrium, where the ICM is everywhere single-phase, should exhibit no substructure in its X-ray emission. While smooth radial profiles in the ICM properties are expected from such a system, any azimuthal spatial fluctuations present in the ICM reflect deviations from hydrostatic equilibrium and/or a single-phase ICM. Numerical simulations demonstrate that the azimuthal variation of the ICM properties increases with radius beyond 0.5\,$R_{200}$ (e.g., Burns et al.\ 2010; Roncarelli et al.\ 2013). For typical cool-core clusters, which are known to have mostly regular X-ray images and appear to be approximately relaxed, recent X-ray observations reveal that approximate azimuthal asymmetry is commonly found in the outskirts of such clusters (e.g. Eckert et al.\ 2013; Walker et al.\ 2012a; Urban et al.\ 2014).

\subsection{\sl Azimuthal variations}

To quantitively describe the level of azimuthal variation in RXJ1159+5531, we adopt the azimuthal scatter following Vazza et al.\ (2011):
\begin{equation}
S_{\rm c}(r)=\sqrt{\frac{1}{N}\sum_{i=1}\frac{[y_{i}(r)-Y(r)]^2}{[Y(r)]^2}},
\end{equation}
where $y_{i}(r)$ is the radial profile of a given quantity in a section $i$, $Y(r)$ is the azimuthal average of this quantity, and $N$ is the number of azimuthal sections. {\sl Suzaku} observed RXJ1159+5531 in four directions, $N=4$ and $Y(r)=$$\frac{\sum y_{i}(r)}{4}$. We obtained $S_{\rm c}(R_{200})$ of 0.076$\pm{0.050}$, 0.086$\pm{0.060}$, 0.154$\pm{0.060}$, and 0.096$\pm{0.067}$ for the density, temperature, entropy, and pressure, respectively, as shown in Figure~\ref{fig:scatter}. Its associated systematic uncertainties are listed in Table~4. These measured values are comparable to the predictions for relaxed systems (Vazza et al.\ 2011). 
Not many galaxy clusters have been observed with good azimuthal coverage near $R_{\rm vir}$ since the moderate spatial resolution of {\sl Suzaku} restricts the observations to nearby systems and their azimuthal studies can be observationally expensive. 
We found in the literature that four other galaxy clusters---Abell~1689 (Kawaharada et al.\ 2010), Abell~1246 (Sato et al.\ 2014), Abell~1835 (Ichikawa et al.\ 2013), and PKS0745-191 (George et al.\ 2009)\footnote{We note that the PKS0745-191 results given by George et al.\ (2009) may be compromised by inadequate background subtraction, and, as argued by Walker et al. (2012a), its radial entropy profiles should start to deviate from the baseline at larger radii. However, it is unclear how this affects its azimuthal variation that we are interested in.}---also have been studied in four different directions with {\sl Suzaku} out to $R_{\rm vir}$. We compared the azimuthal scatters in their gas properties at $R_{200}$ to that of RXJ1159+5531 in Figure~\ref{fig:scatter}.
The nearby Perseus Cluster has been observed extensively with {\sl Suzaku} in eight directions (still only a small fraction of its volume is covered). 
We also included the results of the {\sl Suzaku} analysis of the Perseus Cluster (Urban et al.\ 2014). Azimuthal scatters obtained with $N=8$ are expected to be $\sim$20\% larger than that obtained with $N=4$ (Vazza et al.\ 2011); thus we scaled the azimuthal scatters of the Perseus Cluster to its corresponding values for $N=4$ in Figure~\ref{fig:scatter}.
The range of the azimuthal scatter of temperatures is very large among these clusters. It is remarkable that the temperature scatter of RXJ1159+5531 is smaller than all the comparison clusters. In contrast, the range of the azimuthal density scatter is quite small and these clusters show comparable (and low level) density scatters (We note that the typical non-cool-core merging cluster Abell 1689, with the largest temperature scatter, has the smallest density scatter; thus the association of the density scatter with the cluster relaxation state is unclear).
Consequently, the azimuthal variation in its entropy and pressure of RXJ1159+5531 at $R_{200}$ is the smallest among these systems. 
In Figure~\ref{fig:scatter}, we also overplotted the predicted levels of azimuthal scatters for relaxed and disturbed clusters, respectively, based on simulations (Vazza et al.\ 2011). 
The lack of azimuthal scatter out to the $R_{\rm vir}$ in the gas properties of RXJ1159+5531 is a strong indication of it being a highly relaxed system (Roncarelli et al.\ 2013).

In order to better understand the origin of the relaxed ICM, we inspected the Sloan Digital Sky Survey (SDSS) galaxy distribution on a scale of 20 Mpc within a $\lvert z \rvert<0.005$ dispersion using spectroscopic redshift centered on RXJ1159+5531 as shown in Figure~\ref{fig:env}. 
We marked the boundary of 3 Mpc centered on RXJ1159+5531. 
Some simulations show that galaxies start to be affected by ram pressure stripping within $\sim$ 3 Mpc from the center of the cluster and the accretion flows of sub-structures also start around a few times $R_{\rm vir}$ (Cen et al.\ 2014). 
The surrounding large-scale galaxy environment of RXJ1159+5531 indicates that the southwest directions seem connected to high density filamentary regions while the east direction is adjacent to ``void" regions. However, 
there is no obvious connection between its ICM properties around $R_{\rm vir}$ and its large scale structures.

Many other factors have been associated with large azimuthal scatter. These include unresolved substructures (e.g., galaxies or subgroups, or even point sources for {\sl Suzaku}), the presence of clumping gas or relativistic plasma, and gas displacement associated with bulk motions.  The strikingly symmetrical gas properties found in RXJ1159+5531 place constraints on the importance of the above processes. 
The lack of significant azimuthal scatter in the gas properties out to $R_{\rm vir}$ of RXJ1159+5531 is strong evidence of it being a highly evolved system and hydrostatic equilibrium to be a very good approximation.

\subsection{\sl The origin of the unexpected entropy profile}

As discussed in the Introduction, cluster entropy  ($K \equiv kT/n_e^{2/3}$)
profiles at large radii have been frequently observed to deviate from the power-law profile ($K\propto
r^{1.1}$) predicted by ``adiabatic'' cosmological simulations that
account only for gravity (Voit et al.\ 2005), and several explanations have been proposed to reconcile this discrepancy.

 \begin{figure}
 \epsscale{1.2}
        \centering
        \plotone{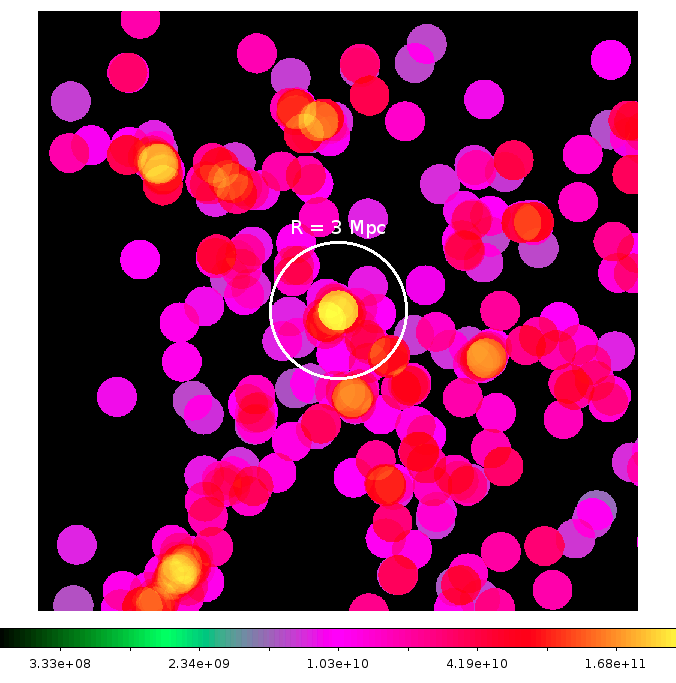}
\figcaption{\label{fig:env}Galaxy number density map coded with their r-band luminosities from SDSS catalogue centered on RXJ1159+5531. The spectroscopic redshifts of these galaxies are used for selection and only those within a 1500 km\,s$^{-1}$ dispersion are included. The white circle indicates the physical radius of 3 Mpc.}
\end{figure}

The proposed mechanisms all depend on the dynamical state or mass scale of the cluster, which provides us an opportunity to place observational constraints on them by comparing the entropy profiles of galaxy clusters of different masses and at different evolutionary stages. For example, gas clumpiness has been predicted to be more prominent among more massive galaxy clusters. This is because more massive clusters formed more recently and are more perturbed (Vazza et al.\ 2013; Battaglia et al.\ 2010), although Nagai \& Lau (2011) argue this is because lower mass systems have a larger fraction of lower-temperature gas, and cooling removes high-density gas out of the detectable energy range. The presence of non-thermal pressure support (associated with departures from hydrostatic equilibrium) is also expected to be more pronounced in dynamically younger (and usually more massive) clusters (Shi \& Komatsu 2014). Avestruz et al.\ (2014) show that the non-thermal equilibrium between ions and electrons is more significant in hotter clusters due to their longer equilibration time. Fujita et al.\ (2013) also found that cosmic ray acceleration is more effective in hotter clusters. 
On the contrary, the scenario proposed by Cavaliere et al.\ (2011) predicts that the entropy drop should be minimal in dynamically young systems since merger shocks are more active among them and highly evolved clusters should
deviate more from the adiabatic $\sim r^{1.1}$ profile than most other clusters. 
The entropy at $R_{\rm 200}$ measured in all directions of RXJ1159+5531 are consistent with the theoretical expectations for purely gravitational processes operating in galaxy clusters. Its enclosed gas fractions within $R_{\rm vir}$ derived for all directions are consistent with the cosmic value. Compared to other massive clusters observed out to $R_{\rm vir}$ by {\sl Suzaku}, RXJ1159+5531 is less massive and highly evolved. 
The ICM properties of RXJ1159+5531 thus disfavor the ``weakening of accretion shock" explanation (Cavaliere et al.\ 2011) and are consistent with other mechanisms, though they do not require them either.

Deeply related to these mechanisms, 
the entropy behavior in the cluster outskirts has
been frequently associated with the large scale structure. For example, using SDSS, Kawaharada et al.\ (2010) found that in Abell 1689 the direction adjacent to higher galaxy density on the large scale shows larger entropy than in other directions adjacent to smaller galaxy density regions. They propose that the thermalization of the ICM occurs faster along overdense filamentary structures than along low-density void regions. However, Kawaharada et al.\ (2010) employed photometric redshifts when they construct the large scale galaxy density map; the uncertainty of photometric redshifts can be very big. 
Urban et al.\ (2014) studied the Perseus Cluster in eight directions with extensive {\sl Suzaku} observations. They also found that directions along the major axis show higher entropy than other directions. They propose that more frequent major mergers occur along the major directions and such mergers can effectively destroy clumping gas, although we note those directions with high entropy at $R_{200}$ of the Perseus Cluster are associated with high temperatures instead of low density gas. 
However,
we measure the same entropy in each direction at $R_{\rm vir}$ for RXJ1159+5531, while its large scale galaxy distribution is asymmetric (Figure~\ref{fig:env}). Such a lack of connections between gas properties in the outskirt of a cluster and the large scale structures has also been reported for Hydra A (Sato et al.\ 2012) and Abell~1246 (Sato et al.\ 2014).

Finally, it has been suggested that in evolved clusters, the low-density
magnetized ICM will develop a Magnetothermal Instability (MTI), especially in the outer regions (e.g., Parrish et al.\ 2008). The MTI drives turbulent motions and 
deviations from hydrostatic equilibrium in the cluster outskirts. Since our measurements for RXJ1159+5531 indicate that it must be very close to hydrostatic equilibrium,  we can infer that the MTI must not be operating effectively in this system.

\section {\sl Summary}

We revisited the {\sl Chandra} and {\sl Suzaku} observations of the fossil-group/poor-cluster RXJ1159+5531. 
We presented the analysis of the {\sl Suzaku} observations of this system of three new directions. 
We also revised the previous study with deep {\sl Chandra} ACIS-I observations covering out to the virial radii of the north direction of this system. 
Below, we list a few highlights of this paper:

$\bullet$ We have resolved more than half of the CXB component into point sources with deep {\sl Chandra} exposure in the north direction. With the refinement of the {\sl Chandra} data, 
we measured an entropy profile rise all the way to $R_{\rm vir}$ and a baryon fraction within $R_{108}$ consistent with the cosmic value, confirmed our previous study of the this system in the north direction. 

$\bullet$ The gas properties (temperature, density, entropy, and entropy) of all four directions (north, south, east, and west) have very similar values. 
The azimuthal scatter in entropy is much smaller than that found in other galaxy clusters. 
The good agreement of the entropy with that predicted from gravity only simulations at $R_{200}$, 
coupled with the small azimuthal variations found for the ICM spectral properties, especially for the entropy, indicates that the ICM of RXJ1159 is very close to hydrostatic equilibrium and, as such, unusually relaxed compared to other clusters.

$\bullet$ Since the virial entropy in all directions are consistent with the theoretical expectation within uncertainties and the gas fractions derived for all directions are consistent with the cosmic value, there is no need to invoke gas clumpiness or any non-gravitational process in RXJ1159+5531. This result is inconsistent with the weakening of accretion shock explanation given by Cavaliere et al.\ (2011) although their model fits the entropy profiles of most cool-core clusters (Walker et al.\ 2013).

$\bullet$ We inspected the galaxy density distribution centered on RXJ1159+5531 on a scale up to 20 Mpc. The gas properties near $R_{\rm vir}$ seem unrelated to its large structure environment.

\section{Acknowledgments}
We thank the anonymous referee for a careful reading of the manuscript.
We are grateful to Philip Humphrey for early contributions to this project, especially for developing initial versions of software for the data analysis and mass modeling. 
D.A.B. and Y.S.  gratefully acknowledge 
partial support from the National
Aeronautics and Space Administration under Grant No.\ NNX13AF14G
issued through the Astrophysics Data Analysis Program. Partial support
for this work was also provided by NASA through Chandra Award Number
GO2-13159X issued by the Chandra X-ray Observatory Center, which is operated by the Smithsonian Astrophysical Observatory for and on behalf of NASA under contract NAS8-03060. F.G. acknowledges the financial contribution from contract PRIN INAF 2012 (``A unique dataset to address the most compelling open questions about X-ray galaxy clustersÓ) and the contract ASI INAF NuSTAR I/037/12/0.

\begin{table*}
\caption{Azimuthal scatter of gas properties at $R_{200}$ with their systematic budgets}
  \begin{center}
    \leavevmode
 \begin{tabular}{lcccc} \hline \hline 
 \colhead{}&\colhead{Temperature}&\colhead{Density}&\colhead{Entropy}&\colhead{Pressure}\\ 
best-fit & $0.076\pm0.050$&$0.086\pm0.060$&$0.154\pm0.067$&$0.096\pm0.067$\\ \hline
$\Delta$NXB&$\pm0.019$&$\pm0.038$&$\pm0.052$&$\pm0.057$\\
$\Delta$CXB&${\pm0.062}$&${\pm0.015}$&${\pm0.048}$&${\pm0.063}$\\
$\Delta$CXB-$\Gamma$&${\pm0.106}$&${\pm0.035}$&${\pm0.147}$&${\pm0.074}$\\
$\Delta$SWCX&${\pm0.025}$&${\pm0.020}$&${\pm0.028}$&${\pm0.021}$\\
$\Delta$PSF&${\pm0.094}$&${\pm0.033}$&${\pm0.154}$&${\pm0.088}$\\
$\Delta$model&${\pm0.024}$&${\pm0.022}$&${\pm0.042}$&${\pm0.007}$\\
$\Delta$solar&${\pm0.061}, {\pm0.013}$&${\pm0.040}, \pm0.024$&${\pm0.073}, {\pm0.044}$&${\pm0.028}, {\pm0.050}$\\
$\Delta$abun&${\pm0.066}, \pm{0.024}$&${\pm0.058}, {\pm0.018}$&${\pm0.073}, {\pm0.020}$&${\pm0.072},{\pm0.077}$\\
$\Delta$nH&${\pm0.061}$&${\pm0.031}$&${\pm0.050}$&${\pm0.105}$\\
$\Delta$Distance&${\pm0.052}$&${\pm0.007}$&${\pm0.015}$&${\pm0.021}$\\
$\Delta$FI-BI&${\pm0.047}$&${\pm0.018}$&${\pm0.026}$&${\pm0.063}$\\
\hline
Relaxed&$0.091\pm0.097$&$0.081\pm0.061$&$$&$$\\
Perturbed&$0.166\pm0.277$&$0.128\pm0.071$&$$&$$\\ \hline
    \end{tabular}
  \end{center}
Systematic error definitions are the same as for Table~3. 
The 1$\sigma$ statistical error on the best fit is obtained from the standard deviation of measurements from 10000 sets of Monte Carlo simulations of the best-fitting spectral models.
Predictions from numerical simulation of azimuthal scatters in temperature and density at $R_{200}$ in relaxed and perturbed clusters are also listed (Vazza et al.\ 2011).
\end{table*}

\end{document}